\documentclass[twoside, a4paper, english, 11pt]{article}

\usepackage{amssymb,amsmath,amsthm}
\usepackage{graphicx}
\usepackage[margin=2cm]{geometry}
\usepackage[toc,page]{appendix}
\usepackage{float}
\usepackage{caption}
\DeclareCaptionFont{10pt}{\fontsize{10pt}{12pt}\selectfont}
\usepackage{afterpage}
\usepackage{listings}
\usepackage{fancyhdr}
\usepackage{color}
\usepackage{hyperref}
\usepackage{babel,blindtext}
\usepackage[normalem]{ulem}
\usepackage{multirow}
\usepackage{lscape}
\usepackage{graphicx}
\usepackage{subfig}
\usepackage{mathtools}
\usepackage{indentfirst}
\usepackage{algorithm}
\usepackage{bm}
\usepackage{booktabs}
\usepackage{natbib}
\usepackage{eurosym}

\DeclareMathOperator{\logit}{logit}
\DeclareMathOperator{\BG}{BG}
\DeclareMathOperator{\diag}{diag}

\begin{document}

\title{Bivariate Gamma Mixture of Experts Models for Joint Insurance Claims Modeling}

\author{Sen Hu, T. Brendan Murphy, Adrian O'Hagan \\ School of Mathematics and Statistics, University College Dublin, Ireland \\  Insight Centre for Data Analytics, Dublin, Ireland}



\maketitle

\abstract{
In general insurance, risks from different categories are often modeled independently and their sum is regarded as the total risk the insurer takes on in exchange for a premium. 
The dependence from multiple risks is generally neglected even when correlation could exist, for example a single car accident may result in claims from multiple risk categories.
It is desirable to take the covariance of different categories into consideration in modeling in order to better predict future claims and hence allow greater accuracy in ratemaking.  
In this work multivariate severity models are investigated using mixture of experts models with bivariate gamma distributions, where the dependence structure is modeled directly using a GLM framework, and covariates can be placed in both gating and expert networks. 
Furthermore, parsimonious parameterisations are considered, which leads to a family of bivariate gamma mixture of experts models.  
It can be viewed as a model-based clustering approach that clusters policyholders into sub-groups with different dependencies, and the parameters of the mixture models are dependent on the covariates.
Clustering is shown to be important in separating the data into sub-groupings where strong dependence is often present, even if the overall data set exhibits only weak dependence. 
In doing so, the correlation within different components features prominently in the model. 
It is shown that, by applying to both simulated data and a real-world Irish GI insurer data set, claim predictions can be improved.}


\section{Introduction}

In general insurance (GI), it is a common phenomenon that claims from multiple risk categories are correlated: in personal motor insurance an accident can lead to claims for the policyholder's own vehicle as well as property damage to others and bodily injury. Dependence can also be present across multiple product lines: claim history and personal characteristics from a motor policy may reflect information pertaining to one's likely home policy claims. 
Many current insurance pricing approaches assume independence among multiple categories, focusing on independent modeling, and then the sum of the risks is designated as the total risk the insurer is taking on.

General linear models (GLMs) (\citealp{Nelder1972glm}) have become the industry's standard approach for claim modeling, typically using a univariate Poisson distribution with a log link for the frequency aspect of modeling (how many claims are made from policies) and a univariate gamma distribution with a log link for severity (expected size of loss for an insurer given a claim has been made). 
Therefore, modeling multiple risks simultaneously using multivariate Poisson distributions (with Poisson univariate margins) and multivariate gamma distributions (with gamma univariate margins) within the GLM framework provides a natural and useful extension beyond their univariate counterparts. 
Many attempts have been made to use (finite mixtures of) multivariate Poisson regressions in the literature. For example, 
\cite{Karlis2003} and \cite{Karlis2005} investigated bivariate Poisson regressions, and \cite{Karlis2007} investigated model-based clustering with finite mixtures of multivariate Poisson distributions. In particular, \cite{Bermudez2012} looked at finite mixtures of bivariate Poisson regressions in the application of GI a-priori ratemaking. 
However, the use of bivariate or multivariate gamma distributions has received less attention in the past, especially for GLMs and mixtures of regressions.
Hence it will be the main focus of this article, investigating the question of how multiple claim sizes relate to each other when multiple claims are made simultaneously, with general applications to any positive continuous data.

In recent years many studies have been carried out in the actuarial literature on multivariate models in various insurance contexts for positive continuous data, but mainly for distribution estimation or risk capital analysis. For example, multivariate normal distributions (\citealp{Panjer2002}), multivariate Tweedie distributions (\citealp{Furman2008}; \citealp{Furman2010}), multivariate Pareto distributions (\citealp{Chiragiev2007}; \citealp{Asimit2010}) and multivariate mixed Erlang distributions (\citealp{Lee2012}; \citealp{Willmot2015}). 
Multivariate gamma distributions have also been studied for in the same context, see \cite{Furman2005}, \cite{Furman2010} and \cite{Furman2008mvGamma}. Note that, since there are various definitions of multivariate gamma distributions, the versions used in the literature sometimes have been defined differently. 
However, none of them are studied in a regression context or a mixture model context for a-priori insurance ratemaking. 
In recent years, the use of copulas has become a popular strategy for multivariate modeling in insurance and statistics, including ratemaking, see \cite{Frees1998} and \cite{Frees2016}. 
An advantage of the copula approach is that it uses a two stage procedure that models the marginal distributions and a copula function (which captures the dependence structure) separately, so it can make use of the rich resources of univariate modeling. However, this particular feature also implies the shortcomings of copulas including estimation issues. For more detailed discussions, see \cite{Mikosch2006}, \cite{Lee2012} and \cite{Grazian2015}.

In most univariate GI ratemaking models, frequency and severity are assumed to be independent. Therefore, multivariate severity models serve an important purpose for multivariate ratemaking. 
For example, a more serious car crash incident may lead the policyholder to make higher claim sizes in multiple risk categories, given claims have been made on multiple categories. 
Furthermore, some policyholders are more prone to making multiple claims but the claim sizes are usually small, compared to other policyholders who represent higher risks, which could suggest heterogeneity in the policy portfolio. The heterogeneity of risk could also be caused by different driving behaviours and attitudes of the policyholders. 
By identifying high risk customers it is also possible to develop further cross-selling approaches in insurance marketing. 
One main issue caused by heterogeneity is that the bivariate claims data are usually very dispersed, hence data may only present a very weak correlation overall, which could be used as justification for implementing univariate modeling without considering dependencies among risks. It could be expected that by segregating the policyholders' data, i.e. considering the heterogeneity, the dependence structures within sub-groups are amplified. 
Generally the unobserved heterogeneity cannot be measured directly by actuaries. Mixture models provide a natural approach for this problem. 
In the standard mixture model framework, model-based clustering is implemented only on the observed independent variables - no covariates are considered in the process.  
In the GI ratemaking case, such characteristics of the policy/policyholder/insured object need to be taken into account in the model. Finite mixtures of bivariate generalized linear regressions, or more generally the mixture of experts approach (MoE) (\citealp{Jacobs1991}), provides such a modeling framework, which models the parameters of the mixture models as functions of the covariates. It can be viewed as the bivariate extension of the univariate finite mixture of generalized linear regression models (\citealp{Grun2008}). 

The structure of the paper proceeds as follows: 
Section~\ref{section:mvGamma} introduces the chosen bivariate gamma distributions and the extension to higher-dimensions;
Section~\ref{section:MoE} discusses the MoE model family, its usage with bivariate gamma distributions and inference via the EM algorithm;
simulation studies are provided in Section~\ref{section:simulationI} and Section~\ref{section:simulationII}, and a real-world data analysis using an Irish GI motor policy data set is studied in Section~\ref{section:result}. The article concludes in Section~\ref{section:conclusion}.

\section{Bivariate gamma distribution}
\label{section:mvGamma}

There are various definitions of bivariate and multivariate gamma distributions - for a detailed review see \cite{Kotz2000} and \cite{Balakrishna2009}. 
In this work we consider the bivariate and multivariate gamma definitions provided by \cite{Cheriyan1941} and \cite{Ramabhadran1951}, and detailed in \cite{Mathai1991}. 
The advantages of this distribution definition are, as pointed out in \cite{Joe1997}, that an ideal multivariate parametric model should incorporate the following features: (1) the margins are of univariate gamma distributions, belonging to the same parametric family; (2) it is easily interpreted since different parameters are responsible for the dependence structures and the marginals; (3) there is a flexible and wide range of dependence structures depending on applications, and it can be generalized to the n-variate case easily; (4) the densities are computationally feasible and relatively easy to work with (although it may depend on the complexity of the dependence structure).  
Note that a bivariate case of the multivariate gamma distribution (BG) is primarily illustrated in this work, but the extension to higher dimensions could be constructed similarly.

Let $X_1, X_2, X_3$ be independent gamma random variables, where $X_i \sim Gamma(\alpha_i, \beta)$ for $i=1, 2, 3$, with different shape parameters $\alpha_i > 0$ and a common rate parameter $\beta > 0$. Then vector $\boldsymbol{Y}$ is defined as:
\begin{equation}
\boldsymbol{Y} = \begin{bmatrix} Y_1 \\ Y_2  \end{bmatrix}  =  \begin{bmatrix} X_1+X_3 \\ X_2+ X_3  \end{bmatrix} \sim \BG(\alpha_1, \alpha_2, \alpha_3, \beta) .
\label{eq:bivgamma_def}
\end{equation} 
It follows from (\ref{eq:bivgamma_def}) that it has density (using the trivariate reduction technique)
\begin{equation} 
\begin{split}
f_{Y_1, Y_2}(y_1,y_2) = \frac{\beta^{\alpha_1+\alpha_2+\alpha_3}e^{-\beta(y_1+y_2)}}{\Gamma(\alpha_1)\Gamma(\alpha_2)\Gamma(\alpha_3)} \int_{x_3=0}^{\min(y_1, y_2)} e^{\beta x_3} x_3^{\alpha_3 - 1} (y_1 - x_3)^{\alpha_1 -1} (y_2 - x_3)^{\alpha_2 -1} dx_3 .
\end{split}
\end{equation}
Since the density is not of complete analytical form, numerical integration is needed to calculate the density. 
This definition has the benefit that the marginals are also gamma distributions such that $Y_1 \sim Gamma(\alpha_1+\alpha_3,\beta)$,
$Y_2 \sim Gamma(\alpha_2+\alpha_3, \beta)$. Alternatively, the distribution can be defined as $\boldsymbol{Y}=\boldsymbol{AX}$, where 
\begin{equation*}
\boldsymbol{A}=\begin{bmatrix}
1 & 0 & 1 \\
0 & 1 & 0 \\
\end{bmatrix},
\hskip .3cm
\boldsymbol{X}= \begin{bmatrix} X_1 \\ X_2 \\ X_3 \end{bmatrix}.
\end{equation*}
The conditional expectation and variance-covariance matrix of $\boldsymbol{Y}$ are
\begin{equation}
\begin{split}
\mathbb{E}[\boldsymbol{Y}] &= \boldsymbol{A\theta} = \begin{bmatrix}
\frac{\alpha_1+\alpha_3}{\beta} \\
\frac{\alpha_2+\alpha_3}{\beta}
\end{bmatrix} , \\
Var[\boldsymbol{Y}] &= \boldsymbol{A\Sigma A^{\top}} = \begin{bmatrix}
\frac{\alpha_1+\alpha_3}{\beta^2} & \frac{\alpha_3}{\beta^2} \\
\frac{\alpha_3}{\beta^2} & \frac{\alpha_2+\alpha_3}{\beta^2} \\
\end{bmatrix},
\end{split}
\end{equation}
where $\boldsymbol{\theta}=(\alpha_1/\beta,\alpha_2/\beta, \alpha_3/\beta)^{\top}$ and $\boldsymbol{\Sigma} = diag(\alpha_1/\beta^2, \alpha_2/\beta^2, \alpha_3/\beta^2).$
\cite{Jensen1969} has shown that if $\boldsymbol{Y}=(Y_1,Y_2)^{\top}$ has a bivariate gamma distribution then for any $0 \leq c_1 < c_2$, 
\begin{equation*}
P(c_1 \leq Y_1 \leq c_2, c1 \leq Y_2 \leq c_2) \geq P(c_1 \leq Y_1 \leq c_2)P(c_1 \leq Y_2 \leq c_2),
\end{equation*} 
which suggests that using a univariate ratemaking model underestimates the claim probability compared with using a bivariate model. 
It is also worth noting that this distribution can only model positive covariance on its own ($\alpha_0/\beta^2 >0$), which also motivates the use of finite mixtures of bivariate gamma distributions for more flexible modeling of either positive or negative covariance, in addition to addressing the heterogeneity issue (see Appendix~\ref{App:bivGamma_covariance_proof}). 

\begin{figure}[H]
	\centering
\begin{minipage}{.48\textwidth}
	\setlength{\abovecaptionskip}{1pt plus 3pt minus 2pt}
	\includegraphics[width=\linewidth]{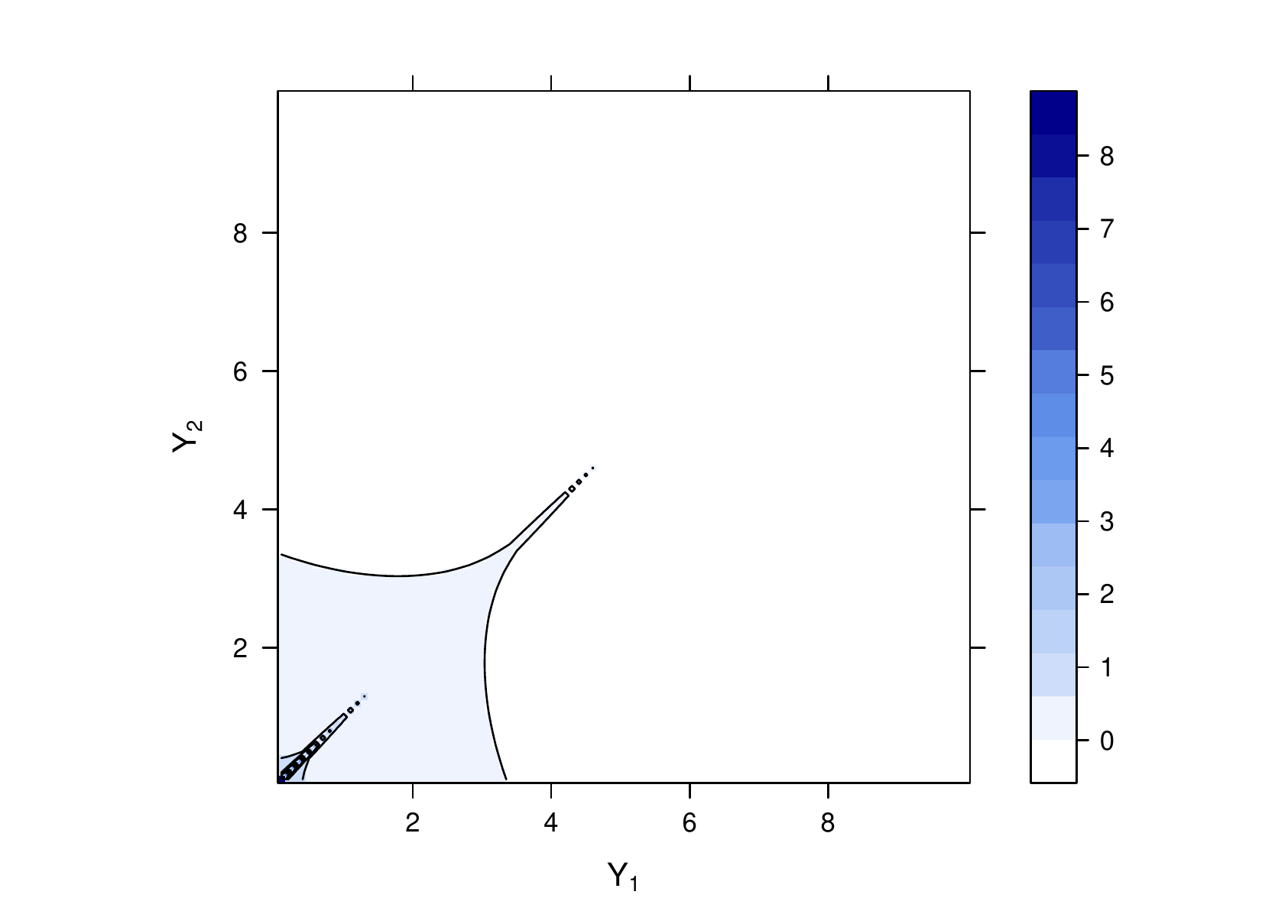}
	\caption*{$\boldsymbol{\alpha}=(0.5,0.5,0.5),\beta=1$}
\end{minipage}	
\begin{minipage}{.48\textwidth}
	\setlength{\abovecaptionskip}{1pt plus 3pt minus 2pt}
	\includegraphics[width=\linewidth]{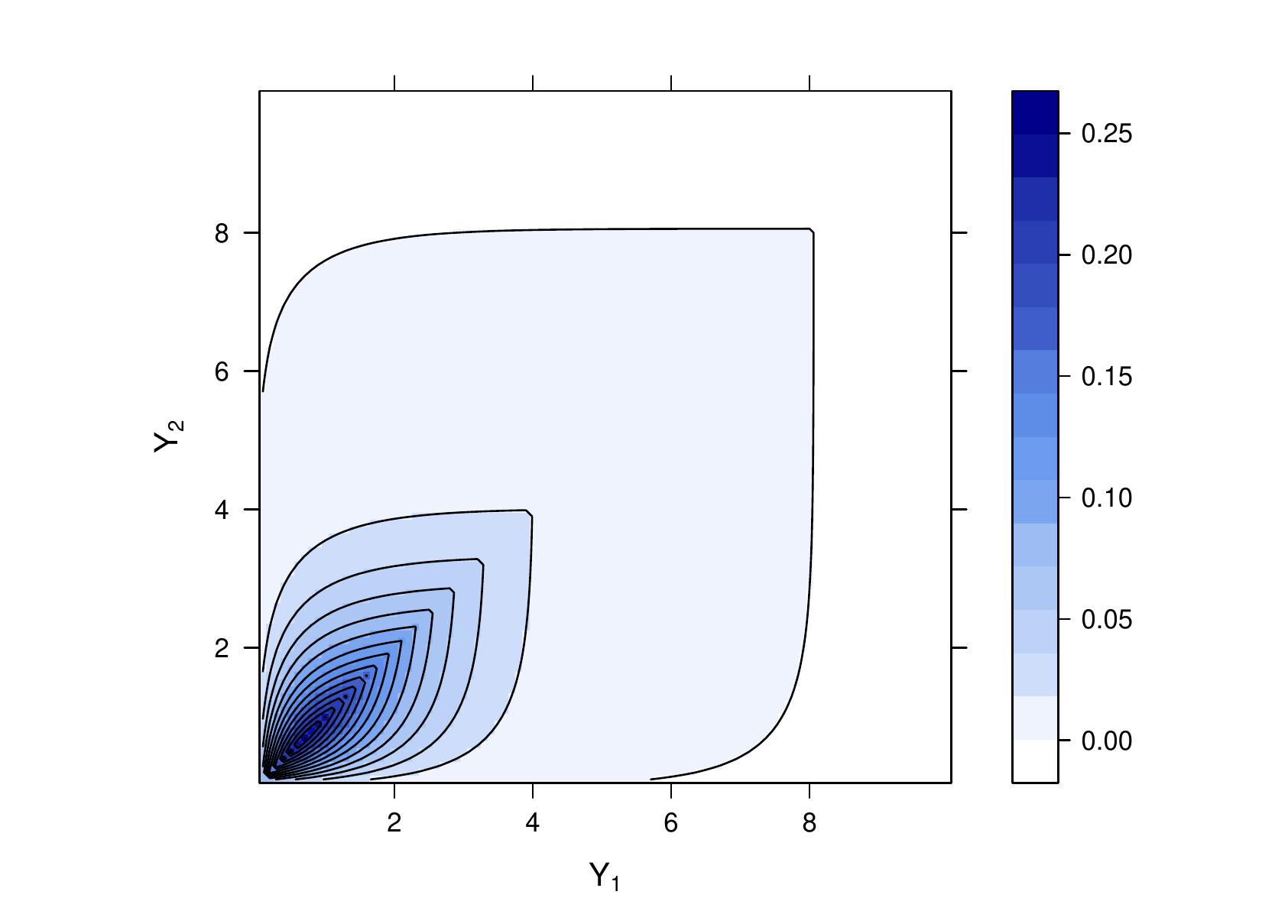}
	\caption*{$\boldsymbol{\alpha}=(1,1,1),\beta=1$}
\end{minipage}	
\begin{minipage}{.48\textwidth}
		\setlength{\abovecaptionskip}{1pt plus 3pt minus 2pt}
	\includegraphics[width=\linewidth]{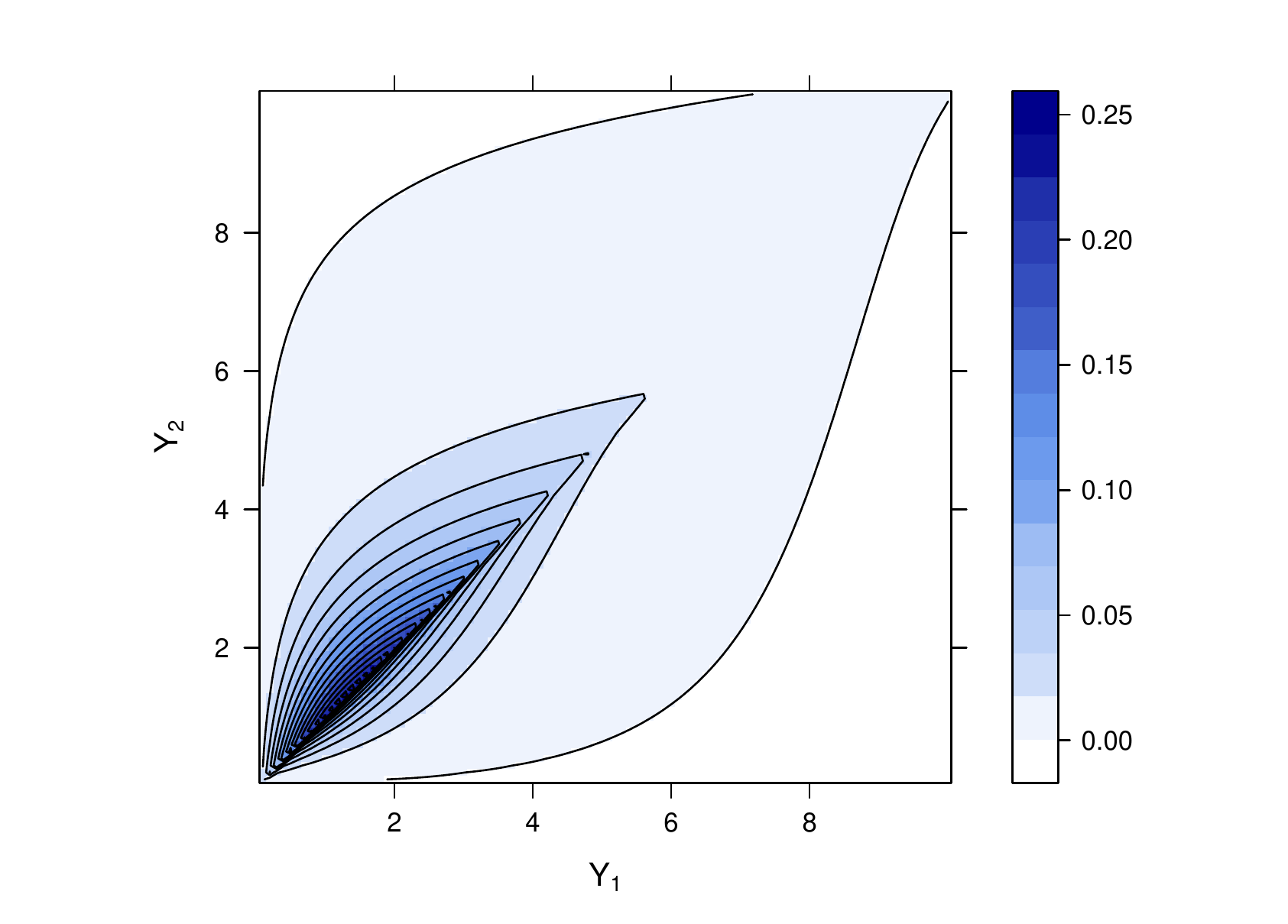}
	\caption*{$\boldsymbol{\alpha}=(0.5,1,2),\beta=1$}
\end{minipage}	
\begin{minipage}{.48\textwidth}
		\setlength{\abovecaptionskip}{1pt plus 3pt minus 2pt}
	\includegraphics[width=\linewidth]{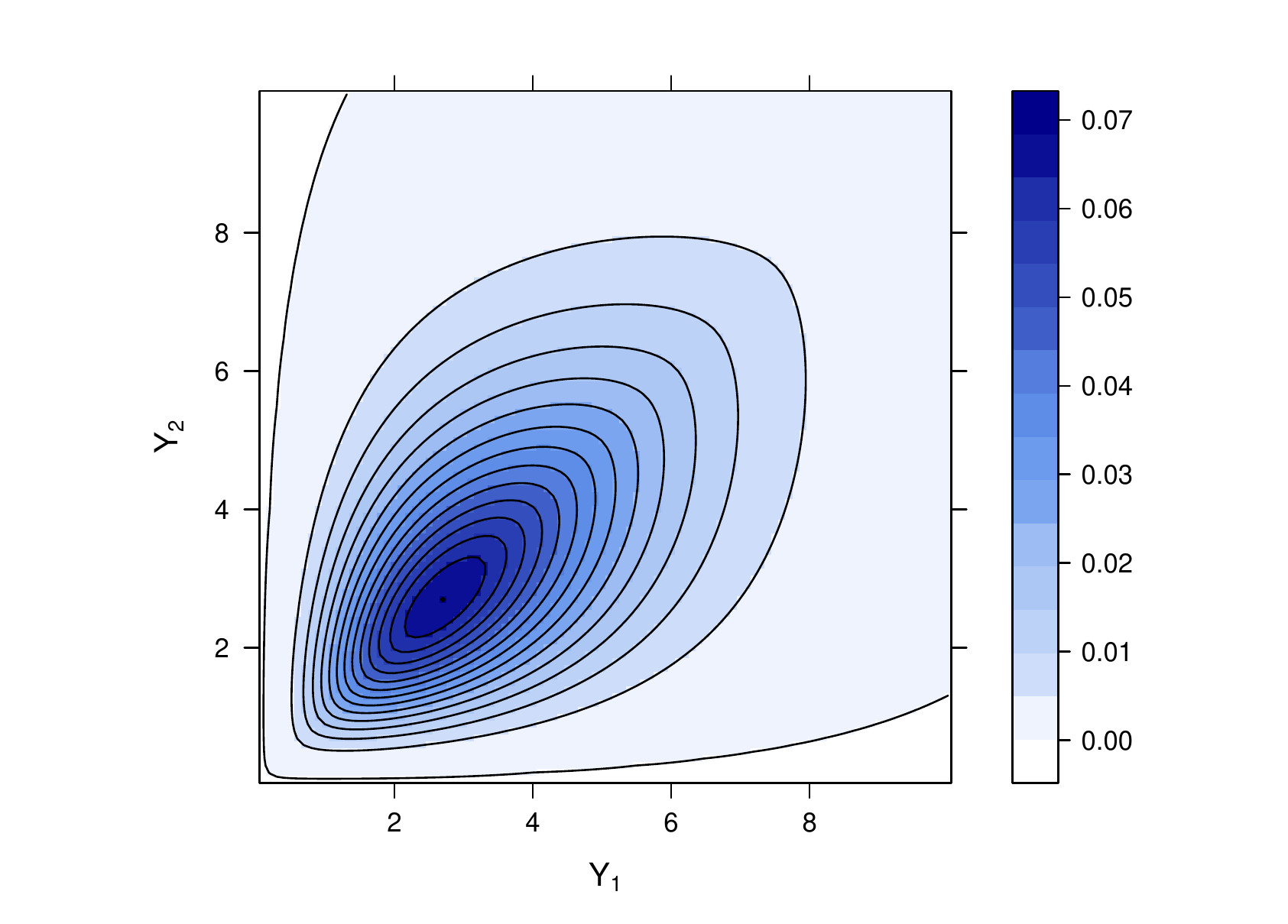}
	\caption*{$\boldsymbol{\alpha}=(2,2,2),\beta=1$}
\end{minipage}
\caption{Plots of densities of bivariate gamma distributions for a range of values of $\boldsymbol{\alpha}=(\alpha_1,\alpha_2,\alpha_3), \beta$.  }
\label{fig:bivgamma_plot}	
\end{figure}

Besides the advantage of having a separate component random variable ($X_3$) accounting for the dependence structure, and having gamma distribution marginals which are consistent with standard univariate claim severity modeling, another advantage of this bivariate gamma distribution is its flexible shape which could capture a wide range of dependence structures. Figure~\ref{fig:bivgamma_plot} shows plots of densities of the $BG(\alpha_1,\alpha_2,\alpha_3,\beta)$ for a range of values of $\boldsymbol{\alpha}, \beta$. 

Extending the bivariate version to higher dimensions can be constructed similarly, with any one or combination of the following cases by \cite{Karlis2005}: 
\begin{itemize}
\item[(1)] Common covariance for all pairs of marginals: 
$$\boldsymbol{Y} = \begin{bmatrix} Y_1 \\ Y_2 \\ Y_3 \end{bmatrix}  =  \begin{bmatrix} X_1 + X_{123} \\ X_2 + X_{123} \\ X_3 + X_{123}  \end{bmatrix} ; $$
\item[(2)] Two-way covariance where the distinct covariance between each pair of marginals is given by 
$$\boldsymbol{Y} = \begin{bmatrix} Y_1 \\ Y_2 \\ Y_3 \end{bmatrix}  =  \begin{bmatrix} X_1 + X_{12} + X_{13} \\ X_2 + X_{12} + X_{23} \\ X_3 + X_{13} + X_{23}  \end{bmatrix} ; $$
\item[(3)] Full covariance trivariate gamma distribution where there is a distinct covariance between each pair of dimensions and a common covariance element across all dimensions
$$\boldsymbol{Y} = \begin{bmatrix} Y_1 \\ Y_2 \\ Y_3 \end{bmatrix}  =  \begin{bmatrix} X_1 + X_{12} + X_{13} + X_{123} \\ X_2 + X_{12} + X_{23} + X_{123} \\ X_3 + X_{13} + X_{23} + X_{123}  \end{bmatrix} ;$$ 
\end{itemize}
each $X_i \sim Gamma(\alpha_i, \beta)$ for $i=1,2,3,12,13,23,123$.

A common method of estimation for the bivariate gamma distribution in (\ref{eq:bivgamma_def}) is the method of moments since its moments can be readily calculated (\citealp{Yue2001}; \citealp{Vaidyanathan2015}). \cite{Tsionas2004} proposed the use of Bayesian Monte Carlo methods for estimating this type of multivariate gamma distribution. However, to the authors' knowledge, there is no existing maximum likelihood estimation procedure for this particular bivariate gamma distribution.
The EM algorithm (\citealp{Dempster1977}) is used for maximum likelihood estimation of parameters of this distribution. Although the MLE under gamma distributions proved to be intractable, we use numerical optimization in the EM algorithm which proves to be stable and efficient. The distribution estimation method is similar to the one used for MoE models below. Full details can be seen in Appendix~\ref{App:DistributionEstimation}.

\section{Clustering and regression with mixture of experts models}
\label{section:MoE}

For insurance claims modeling, covariates need to be included as predictors in regression models for claim predictions; bivariate gamma regressions provide a useful tool for the purpose of ratemaking. Due to the common presence of heterogeneity in the claim severity data, mixture models are a suitable approach for tackling this issue. A standard mixture model clusters outcome variables $\boldsymbol{y_i}$ without considering extra associated information available in the data. Therefore a mixture of bivariate gamma regressions, or in the machine learning terminology, a mixture of experts models (MoE) (\citealp{Jacobs1991}) is considered.
It facilitates flexible modeling, extending the standard mixture models to allow the parameters of the model to depend on concomitant covariates $\boldsymbol{w}_i$. 
Note that higher dimensional cases can be implemented similarly, although the computation may be more complicated.
 
Let $\boldsymbol{y}_1, \boldsymbol{y}_2, \ldots, \boldsymbol{y}_n$ be an identical and independently distributed bivariate sample of outcomes from a population. Suppose the population consists of $G$ components. Each component can be modeled by a bivariate gamma distribution $f(\boldsymbol{y}_i|\theta_g)$ with component-specific parameters $\theta_g = \{ \alpha_{1g}, \alpha_{2g}, \alpha_{3g}, \beta_g \}$, for $g=1,\ldots,G$ and $i=1,\ldots,n$. 
There are also vectors of concomitant covariates $\boldsymbol{w}_1,\boldsymbol{w}_2, \ldots,\boldsymbol{w}_n$ available on which the distribution parameters depend, and which are used to predict future outcome variables. The observed density (conditional on covariates $\boldsymbol{w}_i$) is
\begin{equation}
\label{eq:full_MoE_model}
\centering
\begin{split}
p(\boldsymbol{y}_{i}|\boldsymbol{w}_i) =& \sum_{g=1}^{G} \tau_g(\boldsymbol{w}_i) p(y_{1i}, y_{2i}|\theta_g(\boldsymbol{w}_i)) \\
=& \sum_{g=1}^{G} \tau_g(\boldsymbol{w}_{0i}) p(y_{1i}, y_{2i}| \alpha_{1ig}(\boldsymbol{w}_{1i}), \alpha_{2ig}(\boldsymbol{w}_{2i}), \alpha_{3ig}(\boldsymbol{w}_{3i}), \beta_{i}(\boldsymbol{w}_{4i}))
\end{split}
\end{equation}
where 
\begin{equation}
\centering
\begin{split}
\log(\alpha_{kig}) =& \boldsymbol{\gamma}_{kg}^{\top} \boldsymbol{w}_{ki}, \ \ \ \text{for} \ \ \ k=1,2,3, \\
\log(\beta_{ig}) \  =& \boldsymbol{\gamma}_{4g}^{\top} \boldsymbol{w}_{4i} ,
\end{split}
\end{equation}
and $\tau_g$ is the mixing proportion, i.e. $\sum_{g=1}^{G} \tau_g =1$. 
Different (subsets of) concomitant covariates $\boldsymbol{w_i}$ can go to different parts of the regression models for different parameters since the parameters are independent of each other and hence there exists $\boldsymbol{w}_{0i},\boldsymbol{w}_{1i},\boldsymbol{w}_{2i},\boldsymbol{w}_{3i},\boldsymbol{w}_{4i}$.  
In the machine learning literature, $\tau_g(\boldsymbol{w}_{0i})$ is called the gating network and $p(y_{1i},y_{2i}|\theta_g(\boldsymbol{w}_i))$ is called the expert network. 
When the mixing proportion is regressed on covariates, it is typically modeled using a multinomial logistic regression, with
$\hat{\tau}_g(\boldsymbol{w}_{0i}) = \frac{\exp(\hat{\boldsymbol{\gamma}}_{0g}^{\top} \boldsymbol{w}_{0i})}{\sum_{g^{\prime}=1}^{G}\exp(\hat{\boldsymbol{\gamma}}_{0g^{\prime}}^{\top}\boldsymbol{w}_{0i}) }$ (when $G=2$ it becomes a logistic regression).
The expert network $p(y_{1i},y_{2i}|\theta_g(\boldsymbol{w}_i))$ is typically modeled via a GLM framework with a log link function.
$\boldsymbol{\gamma}_{0g}$, $\boldsymbol{\gamma}_{1g}$, $\boldsymbol{\gamma}_{2g}$, $\boldsymbol{\gamma}_{3g}$, $\boldsymbol{\gamma}_{4g}$ are the regression coefficients for each parameter and component group.
Sometimes MoE is referred to as a conditional mixture model, because for given covariates the distribution of $\boldsymbol{y_i}$ is a mixture model as in Equation~(\ref{eq:full_MoE_model}) (\citealp{Bishop2006}).

For the purpose of insurance claim modeling, the MoE model has the following benefits: (1) both the marginals of risks and their covariance are modeled directly and simultaneously; (2) different heterogeneous groups among policyholders with different claim behaviours can be captured; (3) while individual bivariate gamma distribution can only model positive correlation, MoE is able to model both positive and negative correlations.

In the mixture of bivariate gamma distributions (no concomitant covariates involved), the expectation of $\boldsymbol{Y}$ is
\begin{equation*}
\mathbb{E}[\boldsymbol{Y}] = 
\begin{bmatrix}
\sum_{g} \tau_g  \frac{\alpha_{1g}+\alpha_{3g}}{\beta_g} \\
\sum_{g} \tau_g  \frac{\alpha_{2g}+\alpha_{3g}}{\beta_g} \\
\end{bmatrix},
\end{equation*}
and the variance-covariance matrix is 
\begin{equation*}
Var(\boldsymbol{Y}) = \boldsymbol{A D(\alpha, \beta) A^{\top}} ,
\end{equation*}
where 
\begin{equation*}
\boldsymbol{D}(\boldsymbol{\alpha}, \boldsymbol{\beta}) = \begin{bmatrix}
Var(\frac{\boldsymbol{\alpha}_1}{\boldsymbol{\beta}}) + \mathbb{E}(\frac{\boldsymbol{\alpha}_1}{\boldsymbol{\beta}}) & Cov(\frac{\boldsymbol{\alpha}_1}{\boldsymbol{\beta}}, \frac{\boldsymbol{\alpha}_2}{\boldsymbol{\beta}}) & Cov(\frac{\boldsymbol{\alpha}_1}{\boldsymbol{\beta}}, \frac{\boldsymbol{\alpha}_3}{\boldsymbol{\beta}}) \\
Cov(\frac{\boldsymbol{\alpha}_1}{\boldsymbol{\beta}}, \frac{\boldsymbol{\alpha}_2}{\boldsymbol{\beta}}) & Var(\frac{\boldsymbol{\alpha}_2}{\boldsymbol{\beta}}) + \mathbb{E}(\frac{\boldsymbol{\alpha}_2}{\boldsymbol{\beta}})  & Cov(\frac{\boldsymbol{\alpha}_2}{\boldsymbol{\beta}}, \frac{\boldsymbol{\alpha}_3}{\boldsymbol{\beta}}) \\
Cov(\frac{\boldsymbol{\alpha}_1}{\boldsymbol{\beta}}, \frac{\boldsymbol{\alpha}_3}{\boldsymbol{\beta}}) & Cov(\frac{\boldsymbol{\alpha}_2}{\boldsymbol{\beta}}, \frac{\boldsymbol{\alpha}_3}{\boldsymbol{\beta}}) & Var(\frac{\boldsymbol{\alpha}_3}{\boldsymbol{\beta}}) + \mathbb{E}(\frac{\boldsymbol{\alpha}_3}{\boldsymbol{\beta}})  
\end{bmatrix} .
\end{equation*}
Due to the fact that
\begin{equation*}
Cov(\boldsymbol{Y}_1, \boldsymbol{Y}_2) = Cov(\frac{\boldsymbol{\alpha}_1}{\boldsymbol{\beta}}, \frac{\boldsymbol{\alpha}_2}{\boldsymbol{\beta}}) + Cov(\frac{\boldsymbol{\alpha}_1}{\boldsymbol{\beta}}, \frac{\boldsymbol{\alpha}_3}{\boldsymbol{\beta}}) + Cov(\frac{\boldsymbol{\alpha}_2}{\boldsymbol{\beta}}, \frac{\boldsymbol{\alpha}_3}{\boldsymbol{\beta}}) + Var(\frac{\boldsymbol{\alpha}_3}{\boldsymbol{\beta}}) + \mathbb{E}(\frac{\boldsymbol{\alpha}_3}{\boldsymbol{\beta}}) \ ,
\end{equation*}
negatively correlated $\boldsymbol{\alpha}/\boldsymbol{\beta}$ (i.e. $Cov(\frac{\boldsymbol{\alpha}_i}{\boldsymbol{\beta}}, \frac{\boldsymbol{\alpha}_j}{\boldsymbol{\beta}}) < 0$) can lead to negative correlation of $\boldsymbol{Y}$. A detailed proof of the results for  $\mathbb{E}[\boldsymbol{Y}]$ and $ Var(\boldsymbol{Y})$ above is shown in Appendix~\ref{App:bivGamma_covariance_proof}.

\begin{figure}[ht]
	\centering
	\begin{minipage}{.48\textwidth}
		\centering
		\setlength{\abovecaptionskip}{1pt plus 2pt minus 2pt}
		\includegraphics[width=0.9\linewidth, height=0.5\linewidth]{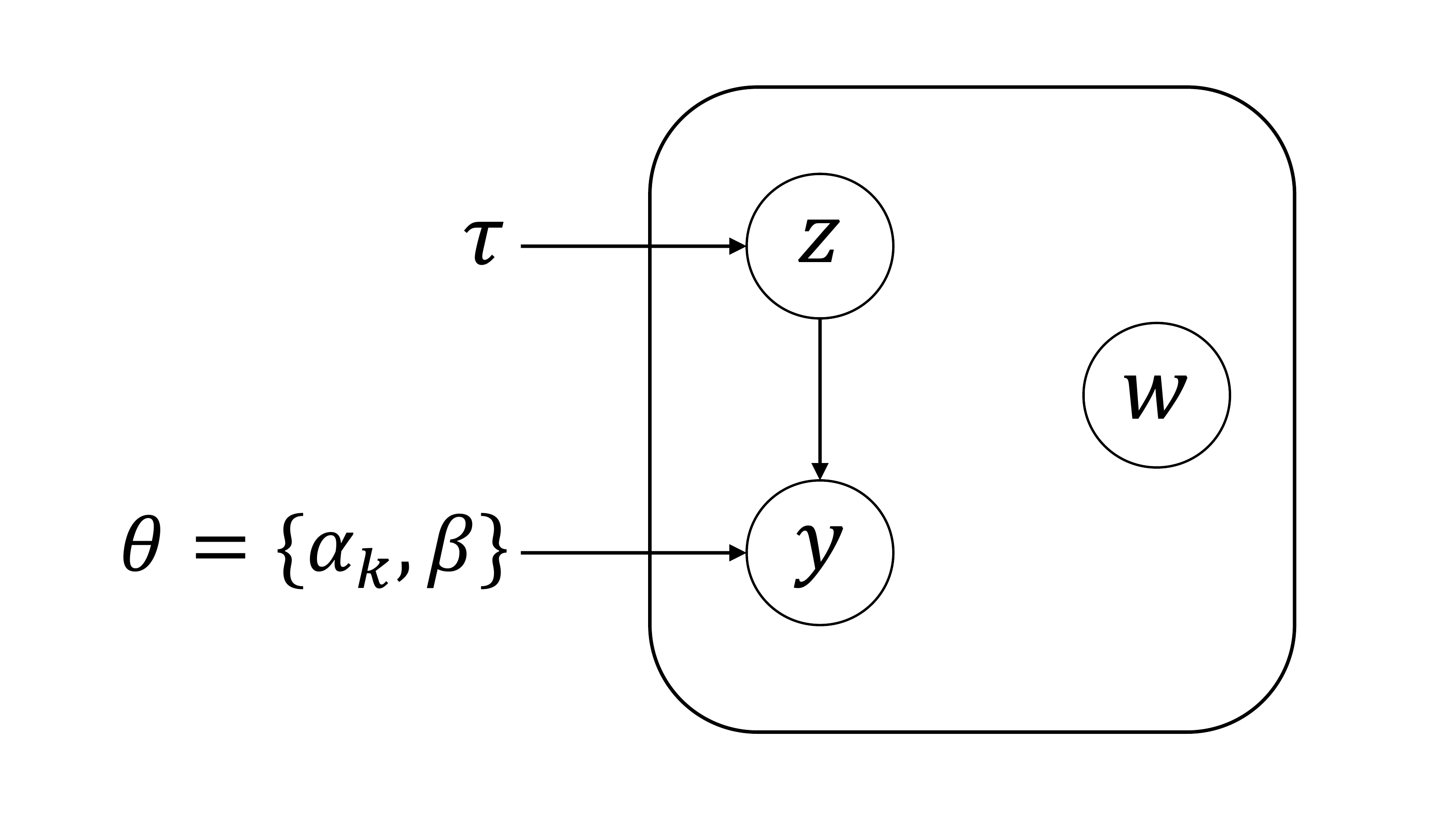}
		\caption*{(1) Mixture model}
	\end{minipage}%
	\begin{minipage}{0.48\textwidth}
		\centering
				\setlength{\abovecaptionskip}{1pt plus 2pt minus 2pt}
		\includegraphics[width=0.9\linewidth, height=0.5\linewidth]{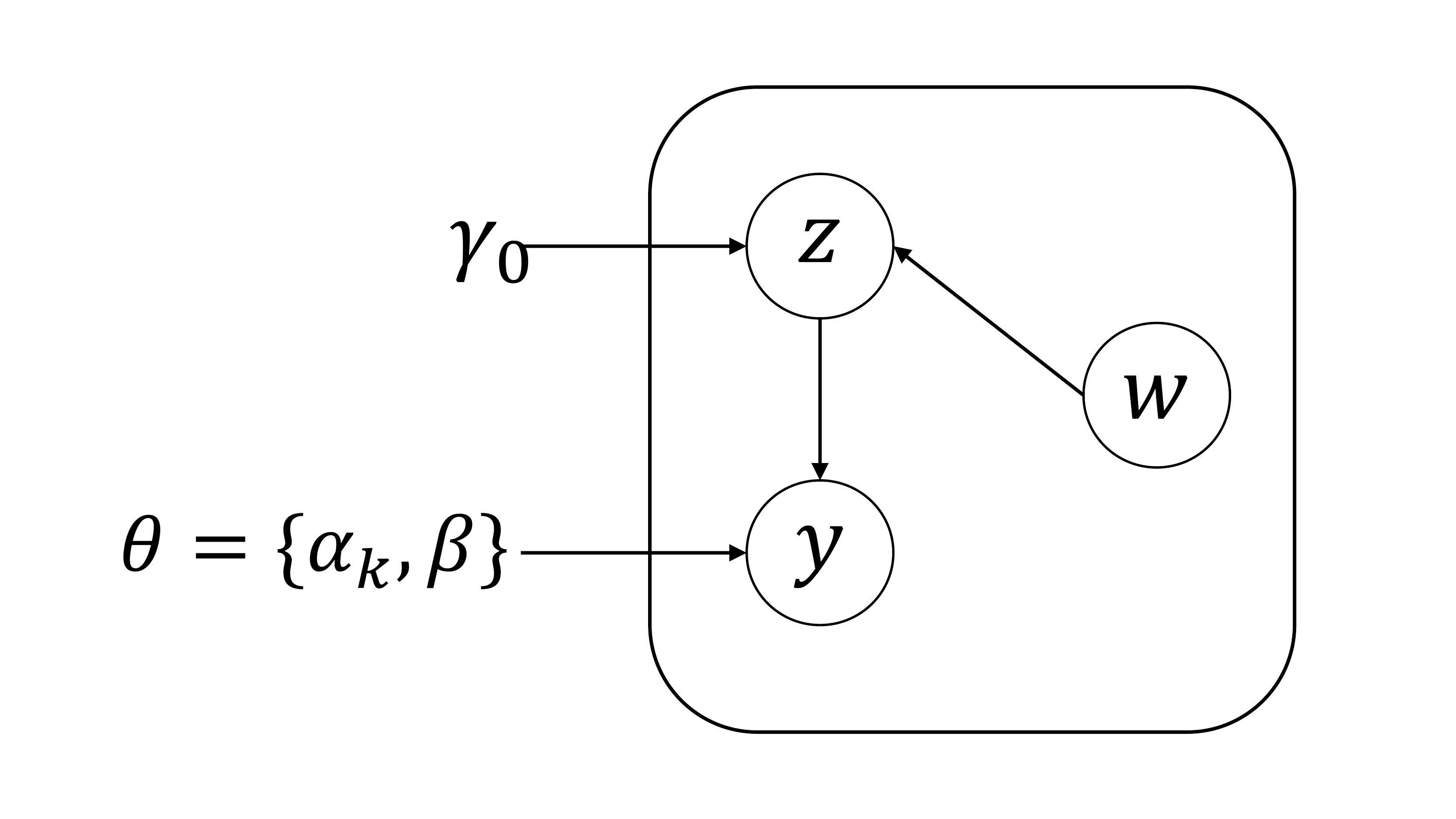}
		\caption*{(2) Gating network MoE}
	\end{minipage}
	\begin{minipage}{.48\textwidth}
		\centering
				\setlength{\abovecaptionskip}{1pt plus 2pt minus 2pt}
		\includegraphics[width=0.9\linewidth, height=0.5\linewidth]{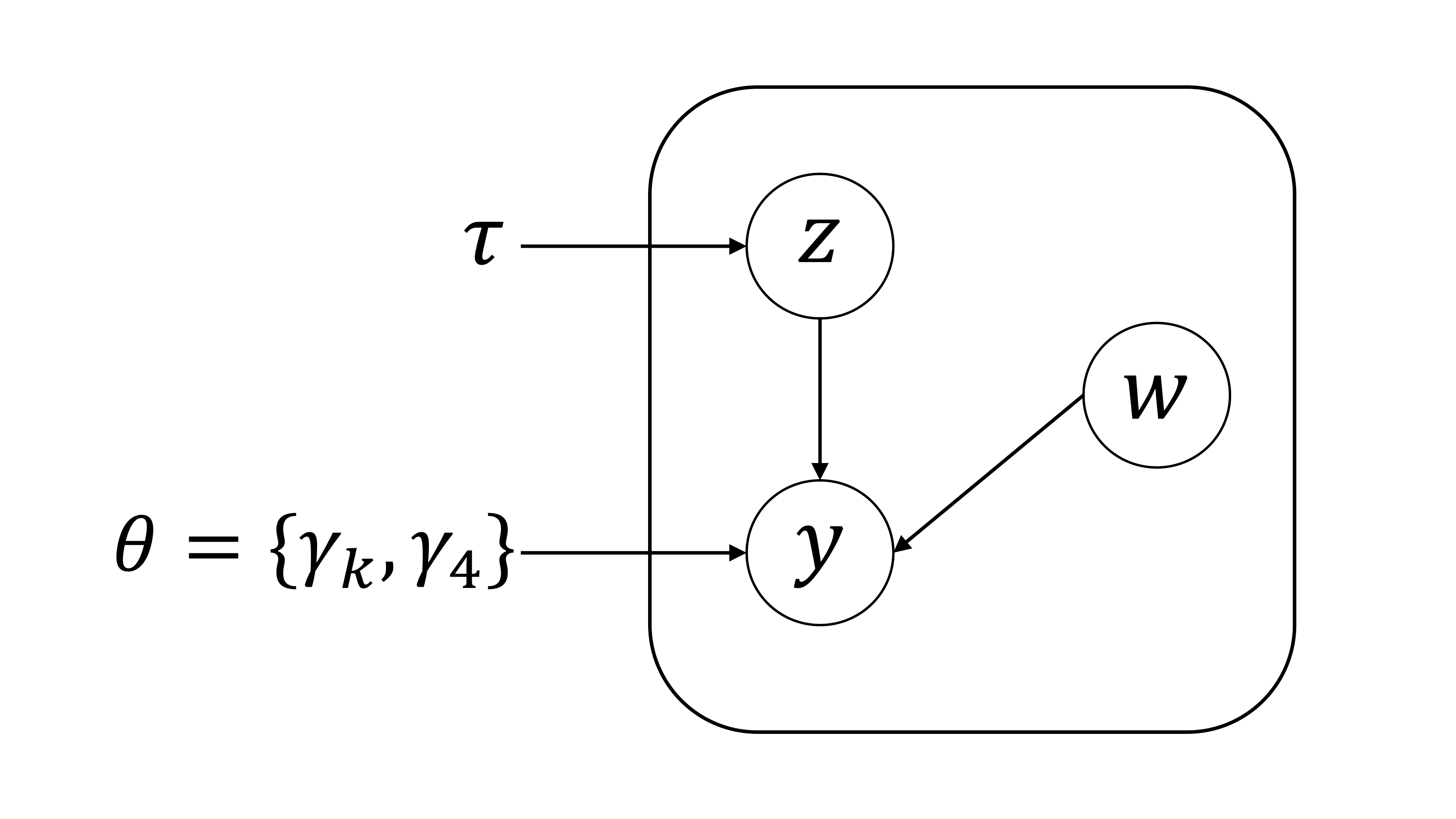}
		\caption*{(3) Expert network MoE}
	\end{minipage}%
	\begin{minipage}{0.48\textwidth}
		\centering
				\setlength{\abovecaptionskip}{1pt plus 2pt minus 2pt}
		\includegraphics[width=0.9\linewidth, height=0.5\linewidth]{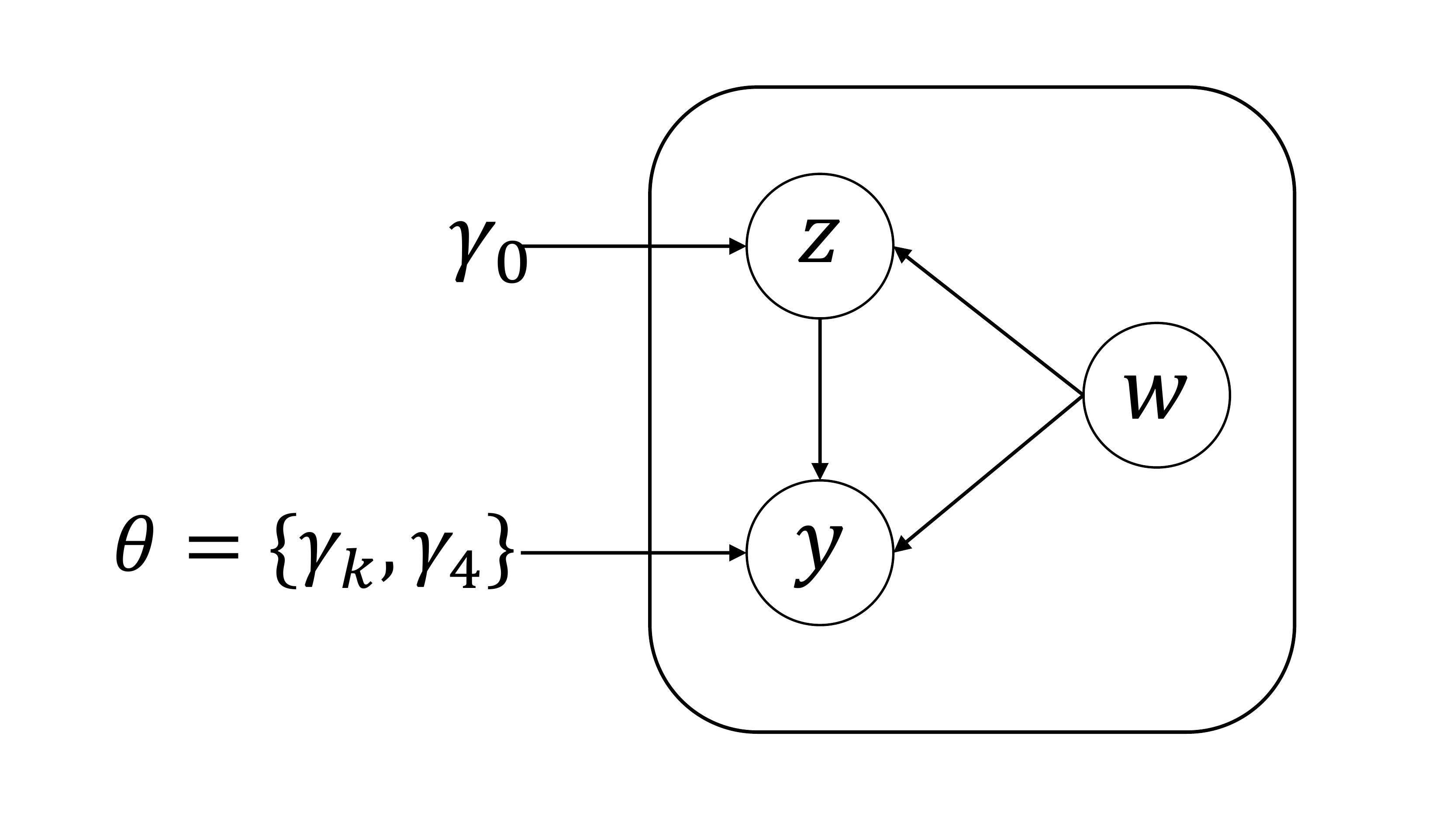}
		\caption*{(4) Full MoE}
	\end{minipage}
\caption{The graphical representation of the four special cases of the MoE model. The differences are due to the presence or absence of directed edges between the covariates $\boldsymbol{w}_i$ and the model parameters. In all plots $k=1,2,3$ corresponding to the parameters $\alpha_1,\alpha_2, \alpha_3$.}
\label{fig:MoE_family}
\end{figure}

For MoE it is possible that none, some or all model parameters could depend on the covariates, which leads to four special cases of MoE models (\citealp{Gormley2011}). It is assumed that the indicator vector $\boldsymbol{z}_{i}=\{z_{i1}, ..., z_{iG} \}$ represents missing group membership, where $z_{ig}=1$ if observation $i$ belongs to group $g$ and $z_{ig}=0$ otherwise:
\begin{enumerate}
	\item \textbf{Standard mixture model:} the distribution of the outcome variable $\boldsymbol{y}_i$ depends on the latent cluster membership variable $\boldsymbol{z}_i$, the model does not depend on any covariates $\boldsymbol{w}_i$:
	\begin{equation*}
	p(\boldsymbol{y}_{i}) = \sum_{g=1}^{G}\tau_g p(y_{1i}, y_{2i}| \alpha_{1g}, \alpha_{2g}, \alpha_{3g}, \beta_g).
	\end{equation*}
	\item \textbf{Gating network MoE:} the distribution of $\boldsymbol{y}_i$ depends on the latent $\boldsymbol{z}_i$ which also depends on $\boldsymbol{w}_i$. The outcome variable $\boldsymbol{y}_i$ does not depend on concomitant covariates $\boldsymbol{w}_i$:
	\begin{equation*}
	p(\boldsymbol{y}_{i}|\boldsymbol{w}_i) = \sum_{g=1}^{G} \tau_g(\boldsymbol{w}_{0i}) p(y_{1i}, y_{2i}| \alpha_{1g}, \alpha_{2g}, \alpha_{3g}, \beta_g) .
	\end{equation*}
	\item \textbf{Expert network MoE:} the distribution of $\boldsymbol{y}_i$ depends on both $\boldsymbol{z}_i$ and $\boldsymbol{w}_i$ while the distribution of $\boldsymbol{z}_i$ does not depend on $\boldsymbol{w}_i$:
	\begin{equation*}
	p(\boldsymbol{y}_{i}|\boldsymbol{w}_i) = \sum_{g=1}^{G} \tau_g p(y_{1i}, y_{2i}|\alpha_{1g}(\boldsymbol{w}_{1i}), \alpha_{2g}(\boldsymbol{w}_{2i}), \alpha_{3g}(\boldsymbol{w}_{3i}), \beta_g(\boldsymbol{w}_{4i})) .
	\end{equation*}
	\item \textbf{Full MoE:} the distribution of $\boldsymbol{y}_i$ depends on $\boldsymbol{z}_i$ and $\boldsymbol{w}_i$ and the distribution of $\boldsymbol{z}_i$ also depends on $\boldsymbol{w}_i$:
	\begin{equation*}
	p(\boldsymbol{y}_{i}|\boldsymbol{w}_i) = \sum_{g=1}^{G} \tau_g(\boldsymbol{w}_{0i}) p(y_{1i}, y_{2i}|\alpha_{1g}(\boldsymbol{w}_{1i}), \alpha_{2g}(\boldsymbol{w}_{2i}), \alpha_{3g}(\boldsymbol{w}_{3i}), \beta_g(\boldsymbol{w}_{4i})) .
	\end{equation*}
\end{enumerate}
Figure~\ref{fig:MoE_family} shows a graphical model representation of the four cases of the MoE model, similarly as in \cite{Gormley2011} and \cite{Murphy2017}. Furthermore, since all parameters $\alpha_{1g}$, $\alpha_{2g}$, $\alpha_{3g}$, $\beta_g$ are independent of one another, and for example knowing the MLEs of $\boldsymbol{\alpha}_{ig}=(\alpha_{1ig},\alpha_{2ig},\alpha_{3ig})$ (i.e. when it is regressed on covariates) can lead to calculating $\boldsymbol{\beta}_{g}$ (i.e. when it is not regressed on covariates) and vice versa, interest lies in allowing more parsimonious parameterisations within the bivariate gamma MoE. These are further developed based on the four special cases of the MoE family, as shown in Table~\ref{tab:parsimonious_parameterisation_MoE}. This yields a family of models, capable of offering additional parsimony in the component densities, as summarised in Table~\ref{tab:bivGamma_MoE_family}. 

\begin{table}[ht]
	\centering
	\caption{Parsimonious parameterisations of the bivariate gamma MoE models. Each of $\boldsymbol{\tau}, \boldsymbol{\alpha}, \boldsymbol{\beta}$ can be modeled via different choices, depending on whether covariates are used in modeling of that parameter.}
	\label{tab:parsimonious_parameterisation_MoE}
	\begin{tabular}{lll}
		\hline\hline
		Gating $\tau$ & Expert $\alpha_k, k=1,2,3$ & Expert $\beta$ \\
		\hline
		C ($\tau_g; \nexists \gamma_{0g}$) & C ($\alpha_{kg}; \nexists \gamma_{kg}$) & C ($\beta_g; \nexists \gamma_{4g}$) \\
		V ($\tau_{ig}; \exists \gamma_{0g}$) & V ($\alpha_{kig};\exists \gamma_{kg}$) & V ($\beta_{ig}; \exists \gamma_{4g}$) \\ 
		E ($\tau_g = 1/G$) & E ($\alpha_{ki.}$; $\exists \gamma_{k}$) & E ($\beta_{i}$; $\exists \gamma_{4}$) \\
		\ & I \ ($\alpha_{k}$; $\nexists \gamma_{k}$) & I \ ($\beta$; $\nexists \gamma_{4}$) \\
		\hline\hline
	\end{tabular}
\end{table}

\begin{table}[ht]
	\centering
	\caption{Bivariate gamma MoE model family based on parsimonious parameterisation: under each model name ``$\ast$" represents the gating network, which can be either ``C" (with covariates), ``V" (no covariates) or ``E" (pre-defined mixing proportion $1/G$). While all models are possible when $G=1$, they are essentially equivalent to one of the four models indicated by ``$\bullet$". It also indicates which models have covariates in the expert networks. In the table, $k=1,2,3$. }
	\label{tab:bivGamma_MoE_family}
	\resizebox{.89\textwidth}{!}{
	\begin{tabular}{l|l|l|c|c}
		\toprule[0.15 em]
		Name & Model & Parameters & G=1 & Covariates in Expert \\
		\hline
		II	 & distribution estimation & $\alpha_k, \beta$ & $\bullet$ & \\
		$\ast$CC  & standard model-based clustering & $\alpha_{kg}, \beta_g$ & & \\
		$\ast$CI  & model-based clustering & $\alpha_{kg}, \beta$ & & \\
		$\ast$IC  & model-based clustering & $\alpha_k, \beta_g$ & & \\
		\hline
		EE   & bivariate gamma regression & $\alpha_{ki}, \beta_i$ & $\bullet$ & $\bullet$ \\
		EI   & bivariate gamma regression over $\alpha_{ki}$& $\alpha_{ki}, \beta$ & $\bullet$ & $\bullet$  \\
		IE   & bivariate gamma regression over $\beta_i$& $\alpha_{k}, \beta_i$ & $\bullet$ & $\bullet$ \\
		\hline
		$\ast$VC  & $\alpha_{kig}$ regressed on covariates& $\alpha_{kig}, \beta_g$ & & $\bullet$  \\
		$\ast$VI  & $\alpha_{kig}$ regressed on covariates& $\alpha_{kig}, \beta$ & & $\bullet$ \\
		$\ast$VV  & $\alpha_{kig}$, $\beta_{ig}$ both regressed on covariates&$\alpha_{kig}, \beta_{ig}$ & & $\bullet$ \\
		$\ast$VE  & $\alpha_{kig}$, $\beta_{i}$ both regressed on covariates &$\alpha_{kig}, \beta_{i}$ & & $\bullet$ \\
		$\ast$CV  & $\beta_{ig}$ regressed on covariates& $\alpha_g, \beta_{ig}$& & $\bullet$  \\
		$\ast$IV  & $\beta_{ig}$ regressed on covariates& $\alpha, \beta_{ig}$& & $\bullet$ \\
		$\ast$EV  & $\alpha_{ki}, \beta_{ig}$ both regressed on covariates&$\alpha_{ki}, \beta_{ig}$&  & $\bullet$  \\
		$\ast$EC  & $\alpha_{ki}$ regressed on covariates& $\alpha_{ki}, \beta_g$ & & $\bullet$ \\
		$\ast$CE  & $\beta_{i}$ regressed on covariates& $\alpha_g, \beta_{i}$ & & $\bullet$ \\
		\bottomrule[0.15 em]
	\end{tabular} }
\end{table}

We focus on maximum likelihood estimation using the EM algorithm (\citealp{Dempster1977}) for model fitting and inference. 
There are two latent variables to be estimated: missing group membership $\boldsymbol{z}_{i}=\{z_{i1}, ..., z_{iG} \}$ where $z_{ig}=1$ if observation $i$ belongs to cluster $g$ and $z_{ig}=0$ otherwise; and the latent variable $X_3$ for each bivariate gamma distribution.
Here only the EM algorithm for the ``VVV" model type from Table~\ref{tab:bivGamma_MoE_family} (i.e. all $\boldsymbol{\tau}, \boldsymbol{\alpha}, \boldsymbol{\beta}$ are regressed on covariates) is shown in detail. The algorithms for other model types in the family can be derived similarly, see Appendix~\ref{App:bivGamma_MoE_family} for details.  
 
For the ``VVV" model type, the complete data likelihood is 
$$\mathcal{L}_c = \prod_{i=1}^{N} \prod_{g=1}^{G} [\tau_g(\boldsymbol{w}_{0i}) p(y_1, y_2, x_3|\theta_g(\boldsymbol{w}_{i}))]^{z_{ig}},$$
where $\theta_g(\boldsymbol{w}_i) =\{\alpha_{1g}(\boldsymbol{w}_{1i}), \alpha_{2g}(\boldsymbol{w}_{2i}), \alpha_{3g}(\boldsymbol{w}_{3i}), \beta_g(\boldsymbol{w}_{4i}) \}= \{\boldsymbol{\gamma}_{1g}, \boldsymbol{\gamma}_{2g}, \boldsymbol{\gamma}_{3g}, \boldsymbol{\gamma}_{4g} \}$.
Because $X_1, X_2, X_3$ are independent, the complete data log-likelihood is:
\begin{equation*}
\begin{split}
\ell_c =& \sum_{i=1}^{N} \sum_{g=1}^{G} z_{ig} \log [\tau_g(\boldsymbol{w}_{0i}) p(y_1, y_2, x_3|\theta_g(\boldsymbol{w}_{i}))] \\
=& \sum_{i=1}^{N}\sum_{g=1}^{G} z_{ig} \log \tau_g(\boldsymbol{w}_{0i}) + \sum_{i=1}^{N}\sum_{g=1}^{G} z_{ig} \log p(y_{1i}-x_{3i}|\alpha_{1ig}(\boldsymbol{w}_{1i}), \beta_{ig}(\boldsymbol{w}_{4i})) \\
&+ \sum_{i=1}^{N}\sum_{g=1}^{G} z_{ig} \log p(y_{2i}-x_{3i}|\alpha_{2ig}(\boldsymbol{w}_{2i}), \beta_{ig}(\boldsymbol{w}_{4i})) + \sum_{i=1}^{N}\sum_{g=1}^{G} z_{ig} \log p(x_{3i}|\alpha_{3ig}(\boldsymbol{w}_{3i}), \beta_{ig}(\boldsymbol{w}_{4i})) .
\end{split}
\end{equation*}
The expectation of this log-likelihood can be obtained in the E-step of the EM algorithm, followed by an M-step that maximises the expectation of the complete data log-likelihood. The estimated parameters, on convergence, achieve at least local maxima of the likelihood of the data. Note that the four parts (including the gating part and the expert part) can be modeled and maximised separately regarding $\boldsymbol{\alpha}$. The full EM algorithm, at the t$^{th}$ iteration, is: \\

\noindent\textbf{\underline{E-step}:}
\begin{equation*}
\begin{split}
\hat{z}_{ig}^{(t+1)} &= \mathbb{E}(z_{ig}|y_{1i},y_{2i}; \hat{\theta}_{g}^{(t)}(\boldsymbol{w}_{i})) = \frac{\hat{\tau}_{g}^{(t)}(\boldsymbol{w}_{0i}) p(y_{1i},y_{2i} ;  \hat{\alpha}_{1ig}^{(t)}, \hat{\alpha}_{2ig}^{(t)}, \hat{\alpha}_{3ig}^{(t)}, \hat{\beta}_{ig}^{(t)})} {\sum_{g^{\prime}=1}^{G} \hat{\tau}_{g^{\prime}}^{(t)}(\boldsymbol{w}_{0i}) p(y_{1i},y_{2i} ; \hat{\alpha}_{1ig^{\prime}}^{(t)}, \hat{\alpha}_{2ig^{\prime}}^{(t)}, \hat{\alpha}_{3ig^{\prime}}^{(t)}, \hat{\beta}_{ig^{\prime}}^{(t)} )} , \\
\hat{x}_{3ig}^{(t+1)} &=\mathbb{E}(X_{3ig}|y_{1i}, y_{2i}; \hat{\theta}_{g}^{(t)}(\boldsymbol{w}_{i})) = \frac{ \frac{\hat{\alpha}_{3ig}^{(t)}}{\hat{\beta}_{ig}^{(t)}} p(y_{1i},y_{2i} ; \hat{\alpha}_{1ig}^{(t)}, \hat{\alpha}_{2ig}^{(t)}, \hat{\alpha}_{3ig}^{(t)}+1, \hat{\beta}_{ig}^{(t)})}{p(y_{1i},y_{2i} ; \hat{\alpha}_{1ig}^{(t)}, \hat{\alpha}_{2ig}^{(t)}, \hat{\alpha}_{3ig}^{(t)}, \hat{\beta}_{ig}^{(t)})} , \\ 
\hat{x}_{1ig}^{(t+1)} &=\mathbb{E}(X_{1ig}|y_{1i}, y_{2i}; \hat{\theta}_{g}^{(t)}(\boldsymbol{w}_{i})) = y_{1i} - \hat{x}_{3ig}^{(t+1)} , \\
\hat{x}_{2ig}^{(t+1)} &=\mathbb{E}(X_{2ig}|y_{1i}, y_{2i}; \hat{\theta}_{g}^{(t)}(\boldsymbol{w}_{i})) = y_{2i} - \hat{x}_{3ig}^{(t+1)} , 
\end{split}
\end{equation*}
\begin{equation*}
\begin{split}
\widehat{\log x}_{3ig}^{(t+1)} &= \mathbb{E}(\log X_{3ig}|y_{1i}, y_{2i}; \hat{\theta}_{g}^{(t)}(\boldsymbol{w}_{i})) = \frac{\int_{0}^{\min(y_{1i},y_{2i})} \log x_{3ig} \ p(x_{3ig},y_{1i},y_{2i};\hat{\theta}_{g}^{(t)}(\boldsymbol{w}_{i})) d x_{3ig} }{p(y_{1i},y_{2i} ; \hat{\alpha}_{1ig}^{(t)}, \hat{\alpha}_{2ig}^{(t)}, \hat{\alpha}_{3ig}^{(t)}, \hat{\beta}_{ig}^{(t)})} , \\
\widehat{\log x}_{1ig}^{(t+1)} &= \mathbb{E}(\log(y_{1i}-X_{3ig}) |y_{1i}, y_{2i}; \hat{\theta}_{g}^{(t)}(\boldsymbol{w}_{i}) = \frac{\int_{0}^{\min(y_{1i},y_{2i})} \log (y_{1i} - x_{3ig}) \ p(x_{3ig},y_{1i},y_{2i};\hat{\theta}_{g}^{(t)}(\boldsymbol{w}_{i})) d x_{3ig} }{p(y_{1i},y_{2i} ; \hat{\alpha}_{1ig}^{(t)}, \hat{\alpha}_{2ig}^{(t)}, \hat{\alpha}_{3ig}^{(t)}, \hat{\beta}_{ig}^{(t)})} , \\
\widehat{\log x}_{2ig}^{(t+1)} &= \mathbb{E}(\log(y_{2i}-X_{3ig})|y_{1i}, y_{2i}; \hat{\theta}_{g}^{(t)}(\boldsymbol{w}_{i}) = \frac{\int_{0}^{\min(y_{1i},y_{2i})} \log (y_{2i}-x_{3ig}) \ p(x_{3ig},y_{1i},y_{2i};\hat{\theta}_{g}^{(t)}(\boldsymbol{w}_{i})) d x_{3ig} }{p(y_{1i},y_{2i} ; \hat{\alpha}_{1ig}^{(t)}, \hat{\alpha}_{2ig}^{(t)}, \hat{\alpha}_{3ig}^{(t)}, \hat{\beta}_{ig}^{(t)})} .
\end{split}
\end{equation*}	

\noindent\textbf{\underline{M-step}:} \\ \\
Update $\hat{\boldsymbol{\gamma}}_{kg}^{(t+1)}$ (and $\hat{\alpha}_{kig}^{(t+1)} = \exp(\hat{\boldsymbol{\gamma}}_{kg}^{(t+1)} \boldsymbol{w_{ki}})$) (for $k=1,2,3$) :
\begin{equation*}
\begin{split}
\underset{\boldsymbol{\gamma}_{kg}}{\arg\max} 
\left(
\sum_{i=1}^{N} \hat{z}_{ig}^{(t+1)} \exp(\boldsymbol{\gamma}_{kg}^{\top}\boldsymbol{w}_{ki}) \log \hat{\beta}_{ig}^{(t)} \right. &- \sum_{i=1}^{N} \hat{z}_{ig}^{(t+1)} \log\Gamma( \exp(\boldsymbol{\gamma}_{kg}^{\top}\boldsymbol{w}_{ki})) \\
&+ \left. \sum_{i=1}^{N} \hat{z}_{ig}^{(t+1)} \exp(\boldsymbol{\gamma}_{kg}^{\top}\boldsymbol{w}_{ki}) \widehat{\log x}_{kig}^{(t+1)} \right) .
\end{split}
\end{equation*}
Update $\hat{\boldsymbol{\gamma}}_{4g}^{(t+1)}$:
\begin{equation*}
\begin{split}
\underset{\boldsymbol{\gamma}_{4g}}{\arg\max} 
&\left(
\sum_{i=1}^{N}  \hat{z}_{ig}^{(t+1)} (\hat{\alpha}_{1ig}^{(t+1)} + \hat{\alpha}_{2ig}^{(t+1)} + \hat{\alpha}_{3ig}^{(t+1)} ) (\boldsymbol{\gamma}_{4g}^{\top} \boldsymbol{w}_{4i}) \right. \\
&\hskip 1cm - \left. \sum_{i=1}^{N} \hat{z}_{ig}^{(t+1)} \exp(\boldsymbol{\gamma}_{4g}^{\top} \boldsymbol{w}_{4i}) (\hat{x}_{1ig}^{(t+1)} + \hat{x}_{2ig}^{(t+1)} + \hat{x}_{3ig}^{(t+1)} ) \right) . 
\end{split}
\end{equation*}
When the mixing proportion is regressed on its covariates (i.e. the gating network), it is modeled using a multinomial logistic regression, with
$\hat{\tau}_g(\boldsymbol{w}_{0i}) = \frac{\exp(\hat{\boldsymbol{\gamma}}_{0g}^{\top} \boldsymbol{w}_{0i})}{\sum_{g^{\prime}=1}^{G}\exp(\hat{\boldsymbol{\gamma}}_{0g^{\prime}}^{\top} \boldsymbol{w}_{0i}) } .$ Otherwise, in the absence of covariates in the gating network, $\hat{\tau}_g = \frac{\sum_{i=1}^{n} \hat{z}_{ig} }{n}$.
Note that there are no complete analytical forms in both the E-step and the M-step, numerical integrations and optimisations are needed respectively which may cause numerical computation complexity.
 
\subsection{EM algorithm set-up}

The use of the EM algorithm has advantages and disadvantages.
First, as in other finite mixture settings, initialization can be done by using a standard clustering algorithm such as \textsf{mclust} (\citealp{Scrucca2017}) or agglomerative hierarchical clustering (\citealp{Everitt2011}), from which initial values for the mixing proportions and classification can be obtained. Initial values of the regression coefficients within components can be obtained by either uniformly sampling the latent $x_{3i}$ from the interval $(0, \min(y_{1i}, y_{2i}))$ or perturbing the $y_{1i}, y_{2i}$ (e.g. half of the $\min(y_{1i}, y_{2i})$).
Since there is a random element used in the initialisation, it may be necessary to run the EM algorithm multiple times from multiple starting points to avoid becoming stuck in local maxima. 

The algorithm is stopped when the change in the log-likelihood is sufficiently small:
\begin{equation*}
\frac{\ell(\boldsymbol{\theta}^{(t+1)}, \boldsymbol{\tau}^{(t+1)}) - \ell(\boldsymbol{\theta}^{(t)}, \boldsymbol{\tau}^{(t)}) }{\ell(\boldsymbol{\theta}^{(t+1)}, \boldsymbol{\tau}^{(t+1)})} < \epsilon .
\end{equation*} 
Suitable terminating conditions should be considered carefully. Since the M-step involves multiple numerical optimizations, towards the end of the algorithm tiny downward likelihood changes could occur occasionally due to the optimization computation complexity for computers. Hence it is recommended that the termination condition is not too small, typically $\epsilon$ should be of the order $1\times 10^{-5}$. 
Alternatively, Aitken’s acceleration criterion (\citealp{Aitken1927}) can be used to assess the convergence (\citealp{Bohning1994}). However, extra caution is needed when the optimisation is complex in this case also. Examples include many regression coefficients needing to be estimated or when the data structure is complex. 

\subsection{Identifiability}

For a finite mixture of bivariate gamma distributions, identifiability is another concern. 
It refers to the existence of a unique characterization for any models in the family. A non-identifiable model may have different sets of parameter values that correspond to the same likelihood. Hence, given the data set one cannot identify the true parameter values using the model (\citealp{Wang2014}). 
As a result, estimation procedures may not be well-defined and not consistent if a model is not identifiable. 
Theoretically, finite mixture models also suffer from the ``label switching" problem. It means the mixture distributions are identical when the component labels are switched. 
It also commonly has the issue of multiple local maxima in the likelihood function. 
Due to the various difficulties, much work on the identifiability of finite mixture models is focused on ``local identifiability", that is, even though there is more than one set of parameter values in the whole parameter space whose likelihoods are the same for the data, there exists identifiable parameter regions within which the mixture of distributions can be identifiable. (\citealp{Wang2014}; \citealp{Kim2015}; \citealp{Rothenberg1971}). 
In practice, Fisher information (or the negative Hessian) for the estimated parameters being positive definite is used to check the existence of local identifiability on the estimated parameter (\citealp{Rothenberg1971}; \citealp{Huang2004}).
 
It follows that, in the proposed EM algorithm for the mixture of bivariate gamma MoEs, the log-likelihood function is maximised numerically, so the returned Hessian matrix needs to be negative-definite. 
Furthermore, since each set of regression coefficients is optimized sequentially and independently, we actually have a block diagonal matrix where each block is the corresponding Hessian matrix. If each block is negative definite, then the whole matrix is negative definite.

Another way to check local identifiability is by re-running the algorithm many times on the same observations. If the algorithm is consistent and repeatedly reaches the same parameter values, it can also be interpreted as identifiable, since estimates from a non-identifiable model will not be consistent. 

\subsection{Model selection}

For the proposed MoE family, different concomitant covariates, if any, can be included in either the gating network, the expert networks of $\alpha_1, \alpha_2, \alpha_3$ and $\beta$, or both. Furthermore, the potential number of components $G$ and the model type from the proposed family are also unknown, which means the number of possible models in the model space is very high. 
Therefore, model selection is important and challenging, especially as the number of covariates increases in the data.

Due to the mentioned issues, a model selection approach is recommended based on forward stepwise selection. At each step it involves either adding a covariate in gating and/or expert networks, or adding a component in $G$, or changing the model type in the model family. The algorithm proceeds as follows:
\begin{enumerate}
	\item To start, fit a $G=1$ model with no concomitant covariates for all allowable model types.
	\item Either add one extra component to the current optimal model, or change the model type, or add one covariate to any (combination) of the expert networks of the current optimal model, based on model selection criterion.
	\item When $G \geq 2$ covariates should also be added to the gating network. 
	\item Stop when there is no improvement in the selection criterion. 
\end{enumerate}

The Akaike information criterion (AIC) (\citealp{Akaike1974}) is a suitable model selection criterion based on penalized likelihood, which is an appropriate technique based on in-sample fit to predict future values. 
From the authors' experience, within the family of bivariate gamma MoE, AIC always selects a better model compared to other criteria. The same general conclusions can also be found in \cite{Vrieze2012} and \cite{Yang2005}. 
It is worth noting that many previous works in the bivariate Poisson model or MoE model settings also support AIC as a selection criterion; see \cite{Karlis2003}; \cite{Karlis2007}; \cite{Bermudez2009}; \cite{Bermudez2012}. 
Other information criteria available in the literature, for example the Bayesian information criterion (BIC) (\citealp{Schwarz1978}) or integrated complete-data likelihood (ICL) (\citealp{Biernacki2000}) are also popular choices for model selection in mixture models and could also be considered.

\section{Simulation study I}
\label{section:simulationI}

To examine the practicality of the proposed method, an artificial data set was simulated from a finite mixture of bivariate gamma distributions as an example. 
In this first simulation study, the primary focus is on the clustering purpose.
The data were simulated based on a gating network MoE, that is, covariates were used only in the gating network to assist with identifying which component the observation was from. 
The purpose is to illustrate that the dependence structure of a given data set can be further enhanced by splitting the data into smaller components, compared to the standard no-clustering case. Therefore, taking the dependence structure into consideration when modeling two perils or risks is better than modeling each peril independently. 

\begin{figure}[htb]
	\centering
	\includegraphics[width=17cm, height=13.5cm]{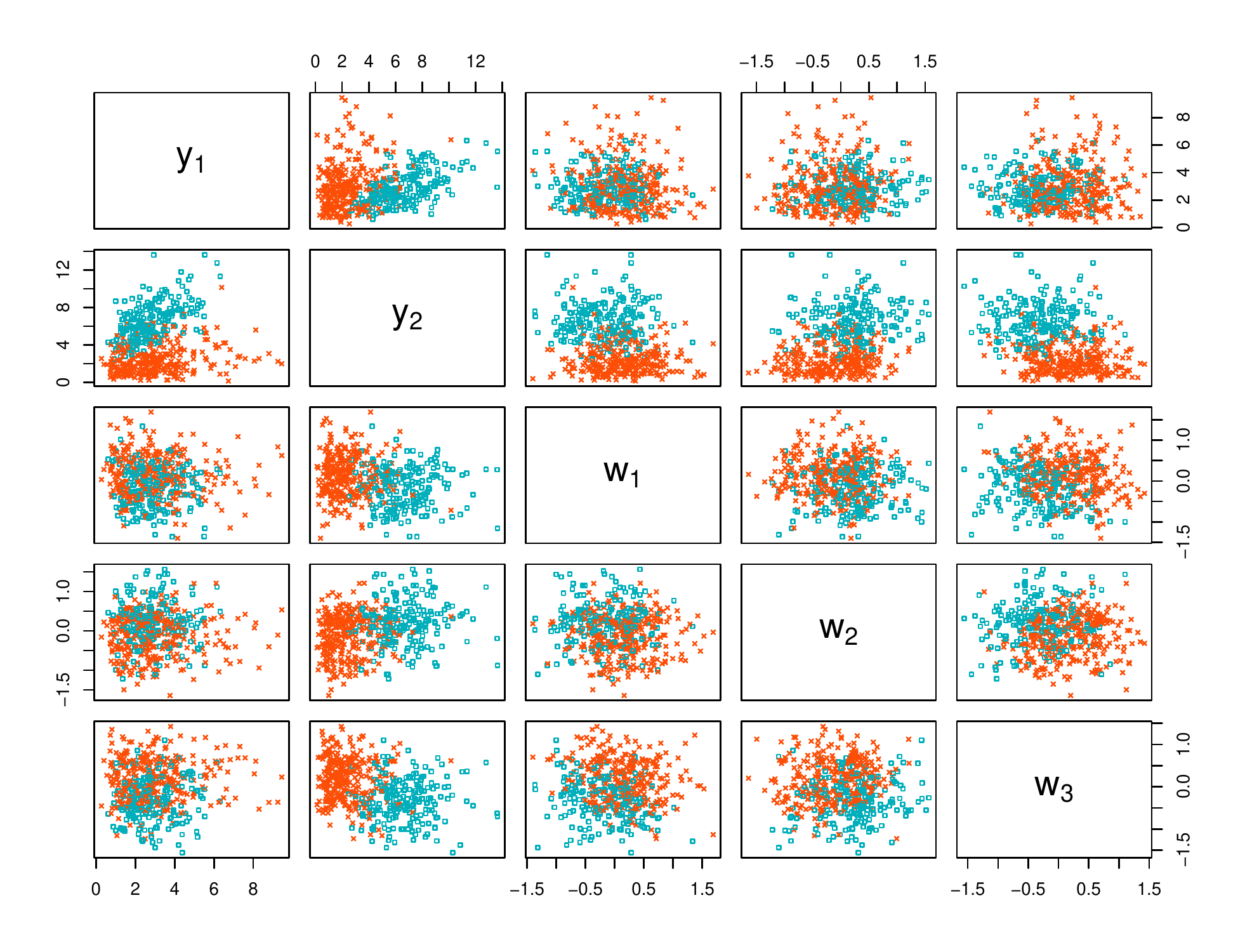}
	\caption{Scatter plot matrix of the simulated data set, colored by component membership: the teal ``$\square$" is component 1 and the red ``$\times$" is component 2. }
	\label{fig:gatingsim_matrix_plot}
\end{figure}

First, a random sample of $n=500$ were generated from the Gaussian distribution as covariates:
\begin{equation*}
\boldsymbol{w} \sim \mathcal{N}\left(\begin{bmatrix}
0 \\
0 \\
0 \\
\end{bmatrix} , \begin{bmatrix}
0.3, 0, 0 \\
0, 0.3, 0 \\
0, 0, 0.3 \\
\end{bmatrix}\right).
\end{equation*} 
The true number of components is $G=2$. The gating is simulated from a logistic regression where the regression coefficients are $\logit(p_{i}) = 1+2w_{1i}-2w_{2i}+3w_{3i}$, based on which a binomial simulation is used. The two components were simulated from two bivariate gamma distributions:
\begin{equation*}
\begin{split}
\boldsymbol{y}^{(g=1)} &\sim \BG(\alpha_1^{(1)}=0.8,\alpha_2^{(1)}=7.9,\alpha_3^{(1)}=5,\beta^{(1)}=1.9), \\
\boldsymbol{y}^{(g=2)} &\sim \BG(\alpha_1^{(2)}=2.6,\alpha_2^{(2)}=2,\alpha_3^{(2)}=0.5,\beta^{(2)}=1). 
\end{split}
\end{equation*}
The simulated data set consists of 205 observations from component 1 and 295 observations from component 2. The scatterplot matrix of the data is shown in Figure~\ref{fig:gatingsim_matrix_plot}, colored by component. In this set-up, $\boldsymbol{w}_1,\boldsymbol{w}_2,\boldsymbol{w}_3$ are not strongly correlated with $\boldsymbol{y}_1,\boldsymbol{y}_2$, but component-wise their correlations are stronger. The two components within $\boldsymbol{w}$ are separating $\boldsymbol{y}$ relatively well, which makes sense since it is known that the data were simulated from a gating network MoE.  

\begin{figure}[H]
	\centering
	\includegraphics[width=11cm, height=7cm]{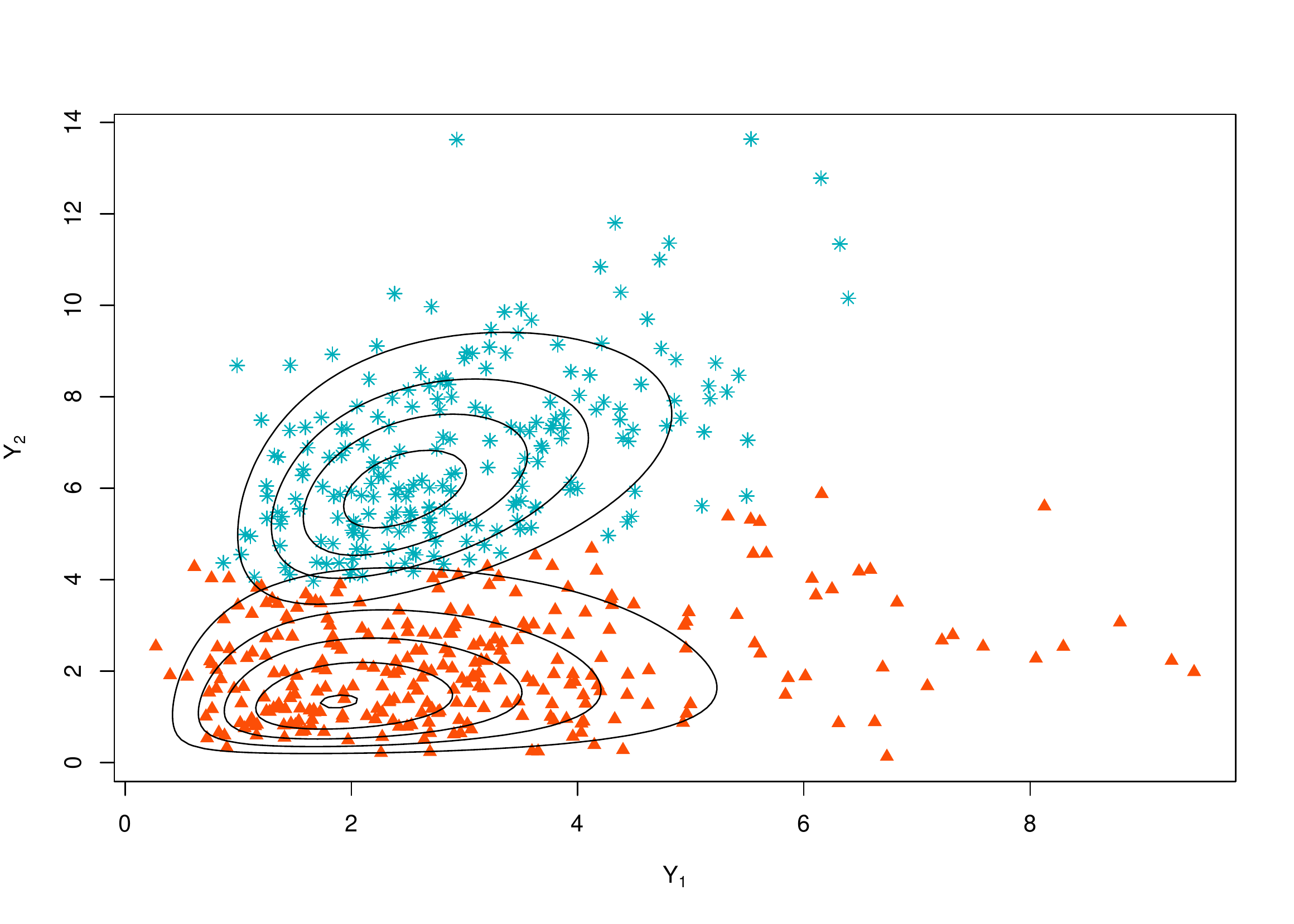}
	\caption{Classification of $\boldsymbol{y}$ without covariates with fitted distribution contours, colored by membership: the teal ``$*$" is component 1, the red ``$\blacktriangle$" is component 2. The optimal model has $G=2$.}
	\label{fig:gatingsim_MBC_classification}
\end{figure}

First, without considering the covariates $\boldsymbol{w}$, clustering over $\boldsymbol{y}$ leads to the optimal number of components for $G=2$, as expected from the true data generating process. It was selected based on AIC when setting $G=1$:5. 
The selection result is presented in Table~\ref{tab:gatingsim_MBC_G_selection}.
Note that BIC gives the same conclusion.  
The parameters estimated are shown in Table~\ref{tab:gatingsim_MBC_parameter_estimation}, together with the true data-generating parameters, which are relatively close to each other.
Figure~\ref{fig:gatingsim_MBC_classification} shows the classification plot when $G=2$, which is relatively close to the true labels. However, it mis-classified many observations into the wrong component since clearly there is some overlap between these two groups. It can be expected that including the covariates may further improve the clustering. 
In fact, in this case the adjusted Rand index is only $0.67$, as in Table~\ref{tab:gatingsim_MBC_G_selection}, and the misclassification rate is 9\%.
As a comparison, model-based clustering with bivariate Gaussian distributions (\textsf{mclust}) (\citealp{Scrucca2017}) on the original data\footnote[1]{Bivariate Gaussian distribution: $G=5$, adjusted Rand index $0.30$} and on its log scale\footnote[2]{Bivariate Gaussian distribution on log data scale: $G=2$, adjusted Rand index $0.64$}, bivariate t distribution\footnote[3]{Bivariate t distribution: $G=3$, adjusted Rand index $0.45$} (\citealp{EMMIXskew}) and skewed t distribution\footnote[4]{Bivariate skewed t distribution: $G=2$, adjusted Rand index $0.50$} (\citealp{EMMIXskew}) all have shown less satisfactory results.  
Note that the clustering process was run many times to ensure the results were stable and accurate. 

\begin{table}[ht]
	\centering
	\caption{Model selection when setting $G=1$:5. The selected best model has $G=2$ (underlined).}
	\label{tab:gatingsim_MBC_G_selection}
	\resizebox{0.6\textwidth}{!}{
	\begin{tabular}{lcccr}
		\toprule[0.15 em]
		\ & log-likelihood & AIC & BIC & Adjusted Rand Index\\
		\midrule
		G=1 & -2079.97 & 4167.94 & 4184.80 & - \\
		G=2 & \underline{-1969.98} & \underline{3957.96} & \underline{3995.89} & \underline{0.67} \\ 
		G=3 & -1968.66 & 3965.32 & 4024.32 & 0.65 \\
		G=4 & -1967.88 & 3973.77 & 4053.84 & 0.43 \\
		G=5 & -1967.59 & 3983.19 & 4084.34 & 0.42 \\
		\bottomrule[0.15 em]
	\end{tabular} }
\end{table}

\begin{table}[ht]
	\centering
	\caption{Parameter estimation when $G=2$ and no covariates are included in the fitted model $\widehat{\boldsymbol{\theta}}$, compared with the true parameter $\boldsymbol{\theta}$. Note that $\bar{\tau}_g$ is the mean of all $\tau_{ig}$ for each $g$.}
	\label{tab:gatingsim_MBC_parameter_estimation}
	\begin{tabular}{c|cc|cc}
		\toprule[0.15 em]
		\ & \multicolumn{2}{|c|}{Component 1} & \multicolumn{2}{|c}{Component 2} \\
		\midrule
		\ & $\boldsymbol{\theta}_{g=1}$ & $\widehat{\boldsymbol{\theta}}_{g=1}$ & $\boldsymbol{\theta}_{g=2}$ & $\widehat{\boldsymbol{\theta}}_{g=2}$ \\
		\midrule
		$\bar{\tau}_g$      & 0.41  & 0.41  & 0.59  & 0.59  \\
		$\alpha_{1g}$ & 0.80  & 2.14  & 2.60  & 2.73  \\
		$\alpha_{2g}$ & 7.90  & 9.66  & 2.00  & 2.05  \\
		$\alpha_{3g}$ & 5.00  & 3.81  & 0.50  & 0.45  \\
		$\beta_{g}$   & 1.90  & 2.00  & 1.00  & 1.08  \\
		\bottomrule[0.15 em]
	\end{tabular}
\end{table}

When taking the covariates $\boldsymbol{w}$ into account, i.e. they are allowed to enter either the gating or expert networks or both, the model selection issue needs to be addressed. 
Choices have to be made for $G$, model types and covariates within each part of expert and gating networks. Stepwise forward selection is used, starting from a null model, which is equivalent to fitting one bivariate gamma distribution to $\boldsymbol{y}$.  
The best four selected models are presented in Table~\ref{tab:gatingsim_model_selection}, selected by AIC (and BIC gives the same conclusion). 
The selected number of components is $G=2$ with no covariates in the expert networks, which coincides with the true simulation process. Table~\ref{tab:gatingsim_bestmodel_regression_coefficient} shows the fitted model paramters compared with the true data-generating parameters. It can be seen that the estimates and the true values are very close.  

\begin{table}[htb]
	\centering
	\caption{The best four bivariate gamma MoE models selected when covariates are considered. The best model is selected via AIC, and is consistent with the true data-generating process.}
	\label{tab:gatingsim_model_selection}
	\resizebox{0.9\textwidth}{!}{
		\begin{tabular}{ccccccccc}
			\toprule[0.15 em]
			Model type & G & $\alpha_1$ expert & $\alpha_2$ expert & $\alpha_3$ expert & $\beta$ expert & gating & AIC & BIC \\
			\midrule
			VCC & 2 & - & - & - & - & $\boldsymbol{w}_1+\boldsymbol{w}_2+\boldsymbol{w}_3$ & 3771.54  & 3822.12  \\
			VCV & 2 & - & -  & - & $\boldsymbol{w}_2$ & $\boldsymbol{w}_1+\boldsymbol{w}_2+\boldsymbol{w}_3$ & 3774.11 & 3833.11 \\
			VCV & 2 & - & -& - & $\boldsymbol{w}_1$ & $\boldsymbol{w}_1+\boldsymbol{w}_2+\boldsymbol{w}_3$ & 3775.16  & 3834.17  \\
			VCV & 2 & - & - & - & $\boldsymbol{w}_3$ & $\boldsymbol{w}_1+\boldsymbol{w}_2+\boldsymbol{w}_3$ & 3775.98  & 3834.98  \\
			\bottomrule[0.15 em]
	\end{tabular} }
\end{table}

\begin{figure}[ht]
	\centering
	\includegraphics[width=11cm, height=7cm]{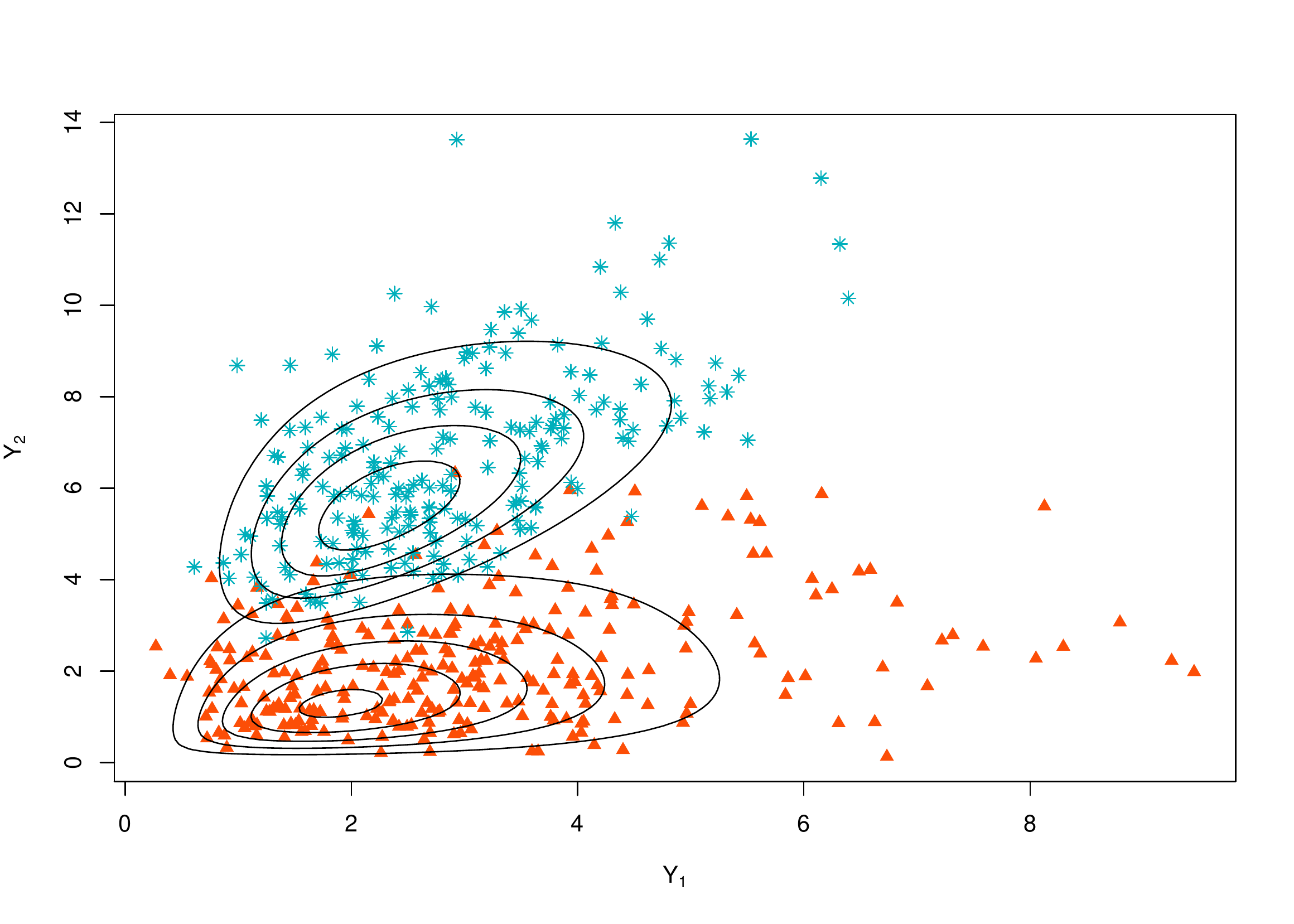}
	\caption{The classification plot of the two components of the best MoE model selected with covariates, with the contour of the fitted bivariate gamma distributions: the teal ``$*$" is component 1 and the red ``$\blacktriangle$" is component 2.}
	\label{fig:gatingsim_bestmodel_classification}
\end{figure}

Figure~\ref{fig:gatingsim_bestmodel_classification} presents the clustering result, which suggests similar component structure versus that of clustering without covariates (as in Figure~\ref{fig:gatingsim_MBC_classification}). However, it is now able to capture the correct membership of the overlapping observations, as in the true data generating process the two components are slightly overlapping. 
The adjusted Rand index is $0.73$ and the misclassification rate is $7\%$, which sees an improvement from the no-covariate case.
Hence, by using covariates in the process of clustering, it further improves the clustering result, which leads to more insightful conclusions on the data and its dependence structure. 
In fact, the overall sample correlation between $Y_1$ and $Y_2$ is $0.17$ in the data, which does not seem to indicate that the two risk perils are strongly correlated, and it appears that it is justified to model them independently as the current common approach. However, by modeling with respect to the dependence structure, the within-component sample correlations are $0.59$ and $0.30$ respectively, which strongly suggests that the dependence structure should be taken into consideration and the two risk perils should be modeled simultaneously.

\begin{table}[ht]
	\centering
	\caption{Estimated parameters of the selected MoE model ($\widehat{\boldsymbol{\gamma}}_{0},\widehat{\boldsymbol{\theta}}_{g}$), compared with the true values ($\boldsymbol{\gamma}_{0}, \boldsymbol{\theta}_{g}$): the top part shows the gating network coefficients; the bottom part shows the estimated $\boldsymbol{\alpha}_{g}, \boldsymbol{\beta}_{g}$ of each component in the expert networks. Note that $\bar{\tau}_g$ is the mean of all $\tau_{ig}$ for each $g$. }
	\label{tab:gatingsim_bestmodel_regression_coefficient}
	\begin{tabular}{c|cc|cc}
		\toprule[0.15 em]
		\multicolumn{5}{c}{\textbf{Gating network}} \\
		\midrule
		\ & intercept & \multicolumn{1}{c}{$\boldsymbol{w}_{1}$} & $\boldsymbol{w}_{2}$ & $\boldsymbol{w}_{3}$ \\
		$\gamma_0$  & 1.00 & \multicolumn{1}{c}{ 2.00 } & -2.00 & 3.00 \\
		$\widehat{\boldsymbol{\gamma}}_{0}$ & 0.57 & \multicolumn{1}{c}{1.77} & -1.71 &  3.05 \\
		\midrule[0.1 em]
		\midrule[0.1 em]
		\multicolumn{5}{c}{\textbf{Expert network}} \\
		\midrule[0.1 em]
		\ & \multicolumn{2}{|c|}{Component 1} & \multicolumn{2}{|c}{Component 2} \\
		\midrule
		\ & $\boldsymbol{\theta}_{g=1}$ & $\widehat{\boldsymbol{\theta}}_{g=1}$ & $\boldsymbol{\theta}_{g=2}$ & $\widehat{\boldsymbol{\theta}}_{g=2}$ \\
		$\bar{\tau}_g$      & 0.41  & 0.43  & 0.59  & 0.57  \\
		$\alpha_{1g}$ & 0.80  & 0.83  & 2.60  & 2.50  \\
		$\alpha_{2g}$ & 7.90  & 7.50  & 2.00  & 1.68  \\
		$\alpha_{3g}$ & 5.00  & 4.32  & 0.50  & 0.75  \\
		$\beta_{g}$   & 1.90  & 1.80  & 1.00  & 1.06  \\
		\bottomrule[0.15 em]
	\end{tabular}
\end{table}

\section{Simulation study II}
\label{section:simulationII}

The first simulation study focused on the clustering perspective. This simulation study mainly focuses on the regression and prediction perspective. 
Similarly, a random sample of size $500$ observations was first generated from the Gaussian distribution as covariates: 
\begin{equation*}
\boldsymbol{w} \sim \mathcal{N}\left(\begin{bmatrix}
0 \\
0 \\
0 \\
\end{bmatrix} , \begin{bmatrix}
0.3, 0, 0 \\
0, 0.3, 0 \\
0, 0, 0.3 \\
\end{bmatrix}\right) .
\end{equation*} 
For the first component, the parameters are generated as 
\begin{equation*}
\begin{split}
\boldsymbol{\alpha}_1^{(g=1)} &= \exp(1 + 0.2\boldsymbol{w}_1 + 0.2\boldsymbol{w}_2), \\
\boldsymbol{\alpha}_2^{(g=1)} &= \exp(0.1 + 0.1\boldsymbol{w}_2 + 0.1\boldsymbol{w}_3),\\
\boldsymbol{\alpha}_3^{(g=1)} &= \exp(0.5 + 0.2\boldsymbol{w}_1 + 0.2\boldsymbol{w}_2 + 0.2\boldsymbol{w}_3),\\
\boldsymbol{\beta}^{(g=1)} &= \exp(0.2 + 0.1\boldsymbol{w}_1 + 0.1\boldsymbol{w}_2 + 0.2\boldsymbol{w}_3).
\end{split}
\end{equation*} 
For the second component, the parameters are generated as
\begin{equation*}
\begin{split}
\boldsymbol{\alpha}_1^{(g=2)} &= \exp(0.1 + 0.1\boldsymbol{w}_1 + 0.1\boldsymbol{w}_2), \\
\boldsymbol{\alpha}_2^{(g=2)} &= \exp(2 + 0.3\boldsymbol{w}_2 + 0.3\boldsymbol{w}_3), \\
\boldsymbol{\alpha}_3^{(g=2)} &= \exp(1.5 + 0.2\boldsymbol{w}_1 + 0.1\boldsymbol{w}_2 + 0.1\boldsymbol{w}_3),\\
\boldsymbol{\beta}^{(g=2)} &= \exp(0.7 + 0.1\boldsymbol{w}_1 + 0.1\boldsymbol{w}_2 + 0.2\boldsymbol{w}_3). \\
\end{split}
\end{equation*}
For each set of parameters $(\alpha_{1i}, \alpha_{2i}, \alpha_{3i}, \beta_i)$ a random sample from this bivariate gamma distribution is simulated, i.e. $\boldsymbol{y}_{i} \sim \BG(\alpha_{1i}, \alpha_{2i}, \alpha_{3i}, \beta_i)$. 
The gating is simulated from a logistic regression model where the regression coefficients are $\logit(p) = 10 + 40\boldsymbol{w}_1 + 30\boldsymbol{w}_2 + 100\boldsymbol{w}_3$, based on which a binomial simulation is used, and the two components are sampled from the simulations above to form the data set.  
The final simulated data set consists of $232$ samples from component 1 and $268$ observations from component 2, with covariates $\boldsymbol{w}_1, \boldsymbol{w}_2, \boldsymbol{w}_3$ and true labels. 

\begin{figure}[htb]
	\centering
	\includegraphics[width=17cm, height=14.5cm]{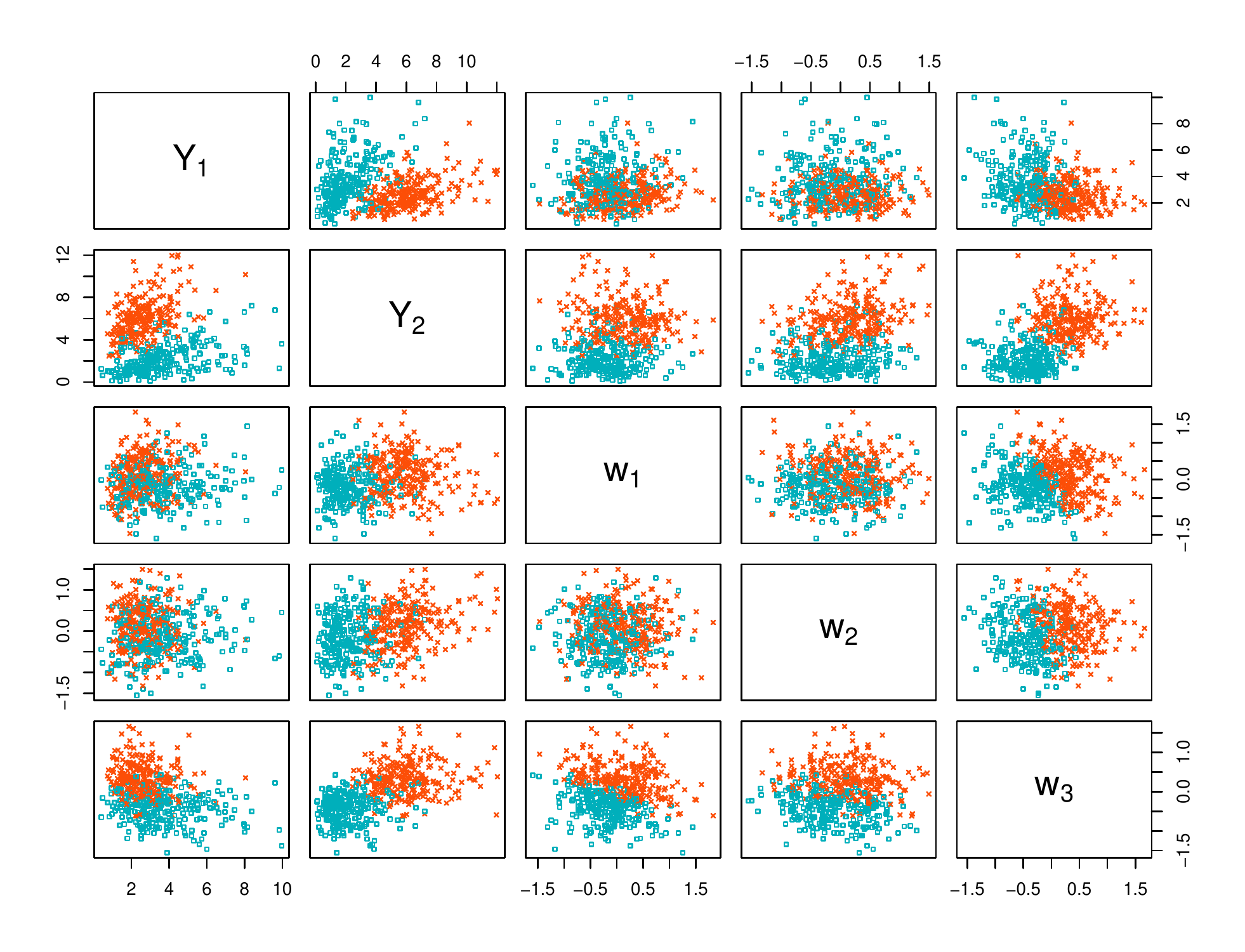}
	\caption{Scatter plot matrix of the simulated data set, colored by component membership: the teal ``$\square$" is component 1 and the red ``$\times$" is component 2.}
	\label{fig:finsim_matrix_plot}
\end{figure}

Figure~\ref{fig:finsim_matrix_plot} shows the scatterplot matrix of the generated data, with different colors and symbols indicating the two components.
Again in this set-up, $\boldsymbol{w}_1, \boldsymbol{w}_2, \boldsymbol{w}_3$ are not strongly correlated with $\boldsymbol{y}$, but component-wise their correlations are much higher, especially regarding $\boldsymbol{y}_2$. 
Therefore, it is expected that $\boldsymbol{w}_1, \boldsymbol{w}_2, \boldsymbol{w}_3$ are significant covariates for the bivariate gamma regressions as well as for the gating network to separate the components. It is also notable that there is slight overlap between the two simulated components. 

\begin{figure}[H]
	\centering
	\includegraphics[width=11cm, height=8cm]{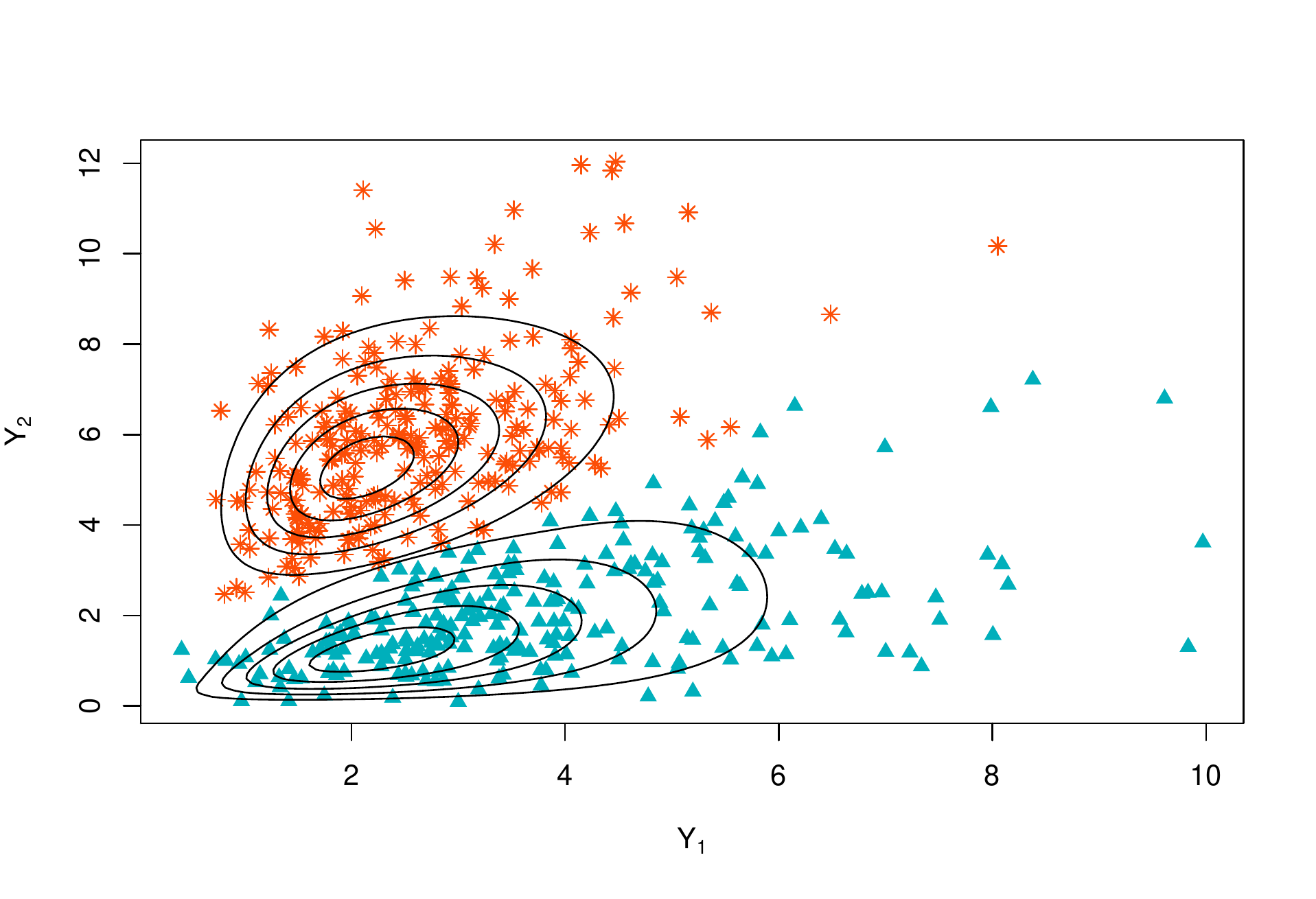}
	\caption{Classification when clustering $\boldsymbol{y}$ without covariates, with fitted distribution contours, colored by membership: the teal ``$\blacktriangle$" is component 1, the red ``$*$" is compoent 2. The optimal model has $G=2$.}
	\label{fig:finsim_MBC_classification}
\end{figure}

First, without considering the covariates $\boldsymbol{w}$, model-based clustering on $\boldsymbol{y}$ leads to the optimal number of components as $G=2$, as expected from the true data generating process. 
This result comes from setting $G=1$:5, the optimal $G$ was selected based on AIC (although BIC gives the same conclusion). The selection result is presented in Table~\ref{tab:finsim_MBC_G_selection}.  
The parameters estimated are shown in Table~\ref{tab:finsim_MBC_parameter_estimation}, together with the means of the true data-generating parameter values, which are relatively close to each other.
Figure~\ref{fig:finsim_MBC_classification} shows the classification plot when $G=2$, which is relatively similar to the true labels. However, it misclassified some observations, so it may be expected that including the covariates may improve the clustering. 
In fact, the no-covariate case gives adjusted Rand index $0.81$ as in Table~\ref{tab:finsim_MBC_G_selection} and the misclassification rate is 5\%.
As a comparison, model-based clustering with bivariate Gaussian distributions (\textsf{mclust}) (\citealp{Scrucca2017}) on the original data\footnote[1]{Bivariate Gaussian distribution: $G=5$, adjusted Rand index $0.39$} and on its log scale\footnote[2]{Bivariate Gaussian distribution on log data scale: $G=2$, adjusted Rand index $0.76$}, bivariate t distributions\footnote[3]{Bivariate t distribution: $G=3$, adjusted Rand index $0.58$} (\citealp{EMMIXskew}) and skewed t distribuitons\footnote[4]{Bivariate skewed t distribution: $G=2$, adjusted Rand index $0.80$} (\citealp{EMMIXskew}), all have shown less satisfactory results.  
Note that the clustering process shown here was run many times to ensure the result was stable and accurate. 

\begin{table}[htb]
	\centering
	\caption{Model selection when setting $G=1$:5. The selected optimal model has $G=2$ (underlined).}
	\label{tab:finsim_MBC_G_selection}
	\resizebox{.6\textwidth}{!}{
	\begin{tabular}{lcccr}
		\toprule[0.15 em]
		\ & log-likelihood & AIC & BIC & Adjusted Rand Index\\
		\midrule
		G=1 & -2077.08 & 4162.17 & 4179.03 & - \\
		G=2 & \underline{-1959.27} & \underline{3936.53} & \underline{3974.46} & \underline{0.81}  \\ 
		G=3 & -1958.69 & 3945.38 & 4004.38 & 0.74  \\
		G=4 & -1954.36 & 3946.71 & 4026.79 & 0.66 \\
		G=5 & -1952.11 & 3952.21 & 4053.36 & 0.47 \\
		\bottomrule[0.15 em]
	\end{tabular} }
\end{table}

\begin{table}[htb]
	\centering
	\caption{Estimated parameters ($\widehat{\boldsymbol{\theta}}_{g}$) when $G=2$ and no covariates included in the fitted model, compared with the mean of the true data-generating parameters $\boldsymbol{\theta}_{g}$.}
	\resizebox{.35\textwidth}{!}{
	\label{tab:finsim_MBC_parameter_estimation}
	\begin{tabular}{c|cc|cc}
		\toprule[0.15 em]
		\ & \multicolumn{2}{|c|}{Component 1} & \multicolumn{2}{|c}{Component 2} \\
		\midrule
		\ & $\boldsymbol{\theta}_{g=1}$ & $\widehat{\boldsymbol{\theta}}_{g=1}$ & $\boldsymbol{\theta}_{g=2}$ & $\hat{\boldsymbol{\theta}}_{g=2}$ \\
		\midrule
		$\bar{\tau}_g$      & 0.50  & 0.49  & 0.50 & 0.51   \\
		$\bar{\alpha}_{1g}$ & 2.74 & 2.49  & 1.11 & 2.14  \\
		$\bar{\alpha}_{2g}$ & 1.10 & 0.81  & 7.48 & 9.37  \\
		$\bar{\alpha}_{3g}$ & 1.65 & 1.53  & 4.48 & 3.51   \\
		$\bar{\beta}_{g}$   & 1.22 & 1.08  & 2.01 & 2.16 \\
		\bottomrule[0.15 em]
	\end{tabular} }
\end{table}

When incorporating the covariates $\boldsymbol{w}_1, \boldsymbol{w}_2, \boldsymbol{w}_3$ in the modeling process using a finite mixture of bivariate gamma regressions, the model selection issue needs to be addressed again. Choices have to be made for $G$, model type and covariates within each part of the expert and gating networks. Stepwise forward selection is used, starting from a null model that is equivalent to fitting one bivariate gamma distribution to $\boldsymbol{y}$.  
The best selected models are presented in Table~\ref{tab:finsim_model_selection}. These models are designated as the best candidates selected by AIC. 
As expected, $G=2$ is selected in all cases, which coincides with the true simulation process. 
Although model 5 in Table~\ref{tab:finsim_model_selection} reflects the true simulation process, the rest of the selected models are slightly better based on AIC, and are only small modifications of model 5, which may be due to the small number of simulated observations. 
This result shows that, although the data are generated using all covariates in all parts of $\boldsymbol{\alpha}$ and $\boldsymbol{\beta}$, a parsimonious parameterisation effectively selects the most efficient parsimonious model.
For now, model 1 is taken as the best model for further investigation, although model 5 represents the true simulation process and the prediction results are very similar among all models.

\begin{table}[htb]
	\centering
	\caption{The best five bivariate gamma MoE models selected according to AIC when covariates are included in the models. Model 5 represents the true simulation process. Note that BIC gives similar conclusion. }
	\label{tab:finsim_model_selection}
	\resizebox{0.9\textwidth}{!}{
	\begin{tabular}{cccccccccc}
		\toprule[0.15 em]
		\ & Model type & G & $\alpha_1$ expert & $\alpha_2$ expert & $\alpha_3$ expert & $\beta$ expert & gating & AIC & BIC \\
		\midrule
		1 & VVC & 2 & $\boldsymbol{w}_1+\boldsymbol{w}_2$ & $\boldsymbol{w}_2+\boldsymbol{w}_3$  & $\boldsymbol{w}_2+\boldsymbol{w}_3$ & - & $\boldsymbol{w}_1+\boldsymbol{w}_2+\boldsymbol{w}_3$ & 3323.35 & 3424.50 \\
		2 & VVC & 2 & $\boldsymbol{w}_1+\boldsymbol{w}_2$ & $\boldsymbol{w}_2+\boldsymbol{w}_3$ & $\boldsymbol{w}_1+\boldsymbol{w}_2+\boldsymbol{w}_3$ & - & $\boldsymbol{w}_1+\boldsymbol{w}_2+\boldsymbol{w}_3$ & 3327.00 & 3436.58 \\
		3 & VVV & 2 & $\boldsymbol{w}_1+\boldsymbol{w}_2$ & $\boldsymbol{w}_2+\boldsymbol{w}_3$ & $\boldsymbol{w}_2+\boldsymbol{w}_3$ & $\boldsymbol{w}_1+\boldsymbol{w}_3$ & $\boldsymbol{w}_1+\boldsymbol{w}_2+\boldsymbol{w}_3$ & 3329.45 & 3447.46 \\
		4 & VVV & 2 & $\boldsymbol{w}_1+\boldsymbol{w}_2$ & $\boldsymbol{w}_2+\boldsymbol{w}_3$ & $\boldsymbol{w}_1+\boldsymbol{w}_2+\boldsymbol{w}_3$ & $\boldsymbol{w}_1+\boldsymbol{w}_3$ & $\boldsymbol{w}_1+\boldsymbol{w}_2+\boldsymbol{w}_3$ & 3331.09 & 3457.52 \\
		5 & VVV & 2 & $\boldsymbol{w}_1+\boldsymbol{w}_2$ & $\boldsymbol{w}_2+\boldsymbol{w}_3$ & $\boldsymbol{w}_1+\boldsymbol{w}_2+\boldsymbol{w}_3$ & $\boldsymbol{w}_1+\boldsymbol{w}_2+\boldsymbol{w}_3$& $\boldsymbol{w}_1+\boldsymbol{w}_2+\boldsymbol{w}_3$ & 3334.23 & 3469.10 \\
		\bottomrule[0.15 em]
	\end{tabular} }
\end{table}

\begin{figure}[ht]
	\centering
\begin{minipage}{.6\textwidth}
	\centering
	\includegraphics[width=1\linewidth, height=0.7\linewidth]{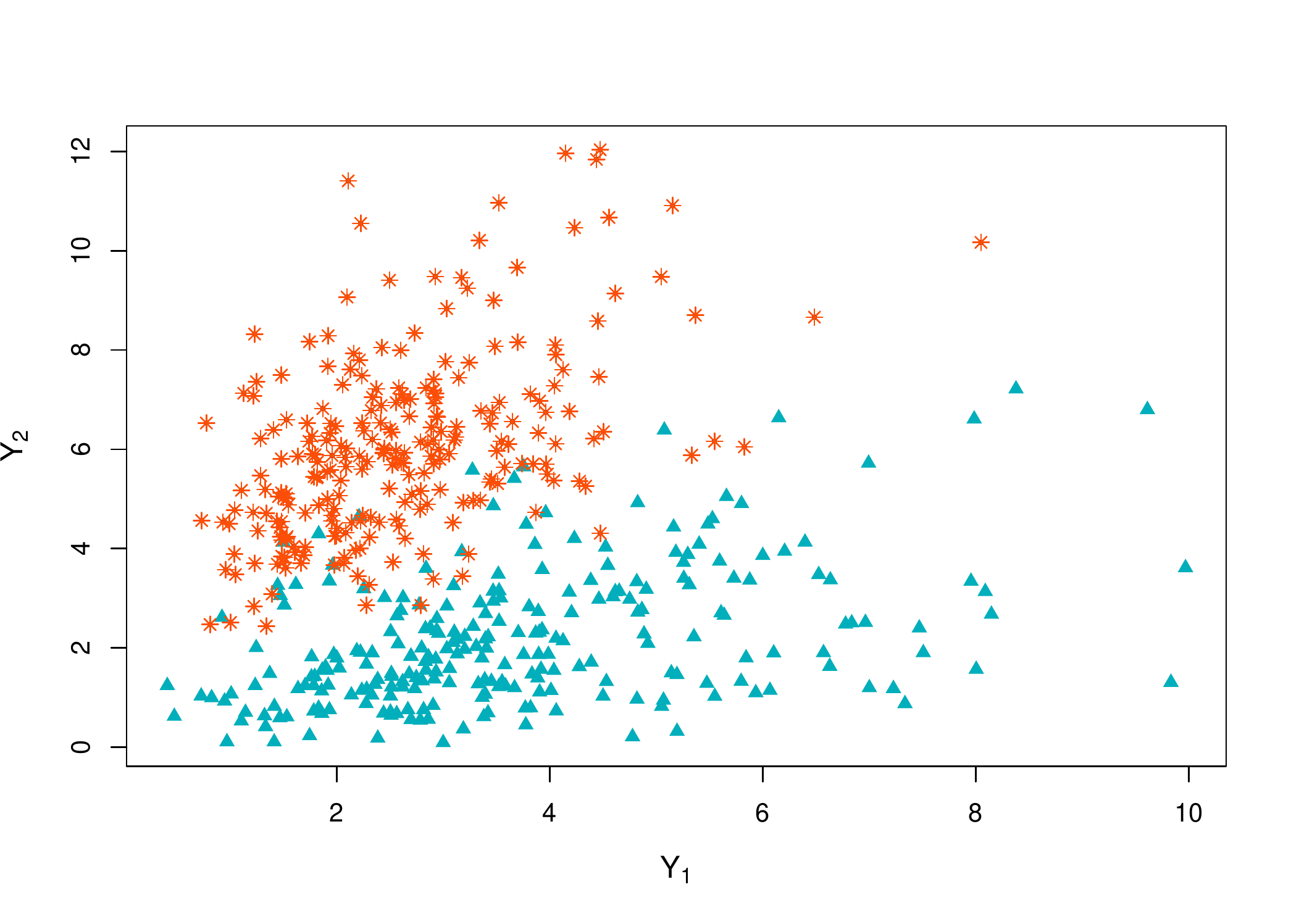}
	\caption*{(a)}
\end{minipage}
\begin{minipage}{.6\textwidth}
	\centering
	\includegraphics[width=1\linewidth, height=0.7\linewidth]{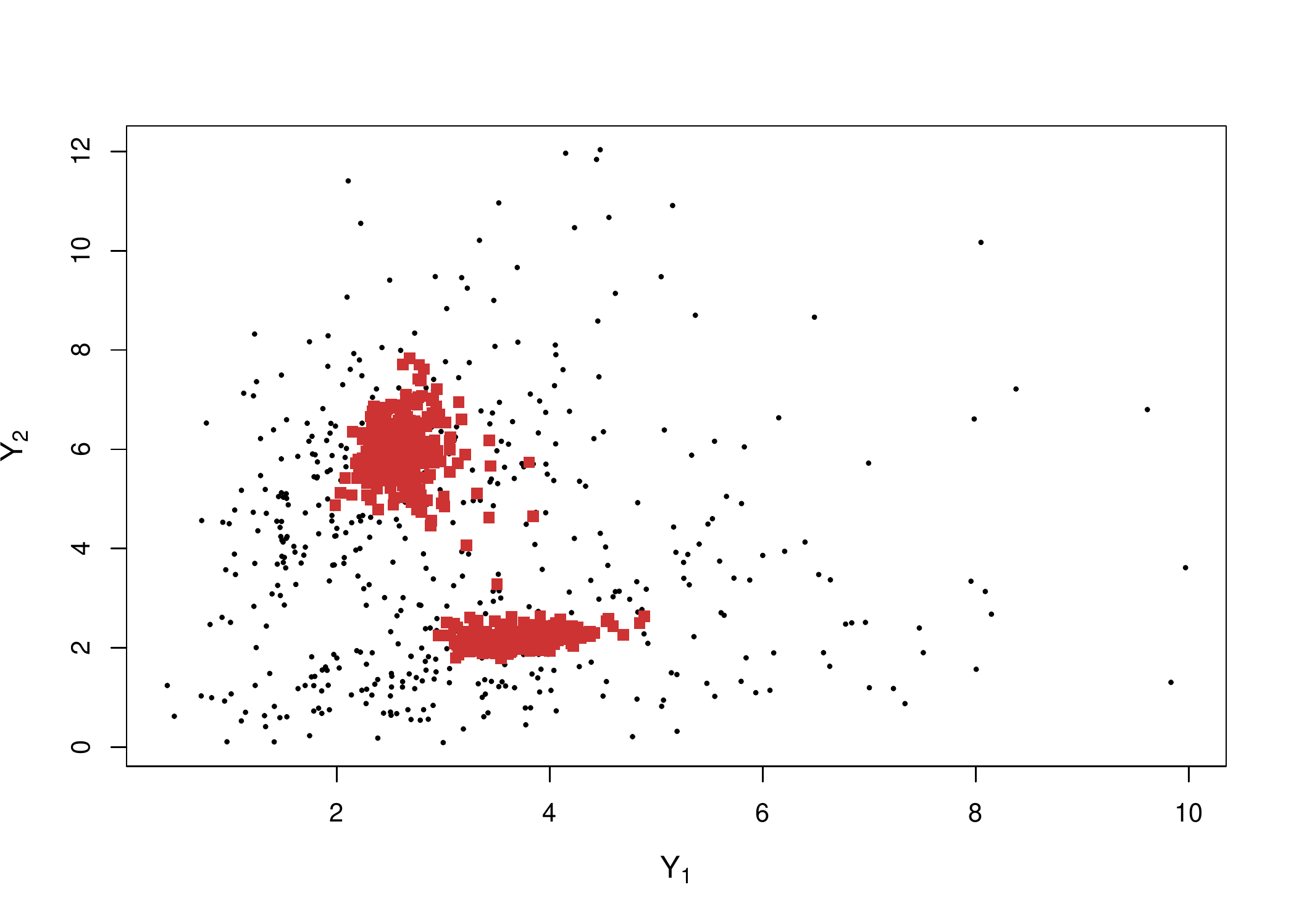}
	\caption*{(b)}
\end{minipage}
\caption{The best MoE model selected: (a) shows the classification plot of $\boldsymbol{y}$, where the teal ``$\blacktriangle$" is component 1, the red ``$*$" is component 2; (b) shows the fitted values of the selected model, where th true values are plotted in small black ``$\centerdot$" and the predictions are plotted in red ``$\blacksquare$".}
\label{fig:simdata_bestmodel_classification_fittedvalues}
\end{figure}

Figure~\ref{fig:simdata_bestmodel_classification_fittedvalues}(a) presents the clustering results for model 1, which suggests similar component structure to clustering without covariates. 
However, it is now able to separate the overlapping parts, which gives an outcome very similar to the true class in Figure~\ref{fig:finsim_matrix_plot}. 
In fact, the adjusted Rand index is $0.98$ with only two observations being misclassified.
Hence, by using covariates in the process of model-based clustering, it further improves the clustering result since extra information is considered, which in turn leads to more insightful conclusions about the data and its dependence structure. 
From a regression perspective, which is the main purpose of this method for predicting claims given covariates while capturing different regression structures, fitted values of this model are presented in  Figure~\ref{fig:simdata_bestmodel_classification_fittedvalues}(b).
Clearly this model is able to capture the two components with two regression lines. 

\begin{figure}[H]
	\centering
	\begin{minipage}{.3\textwidth}
		\centering
		\includegraphics[width=1\linewidth, height=1.8\linewidth]{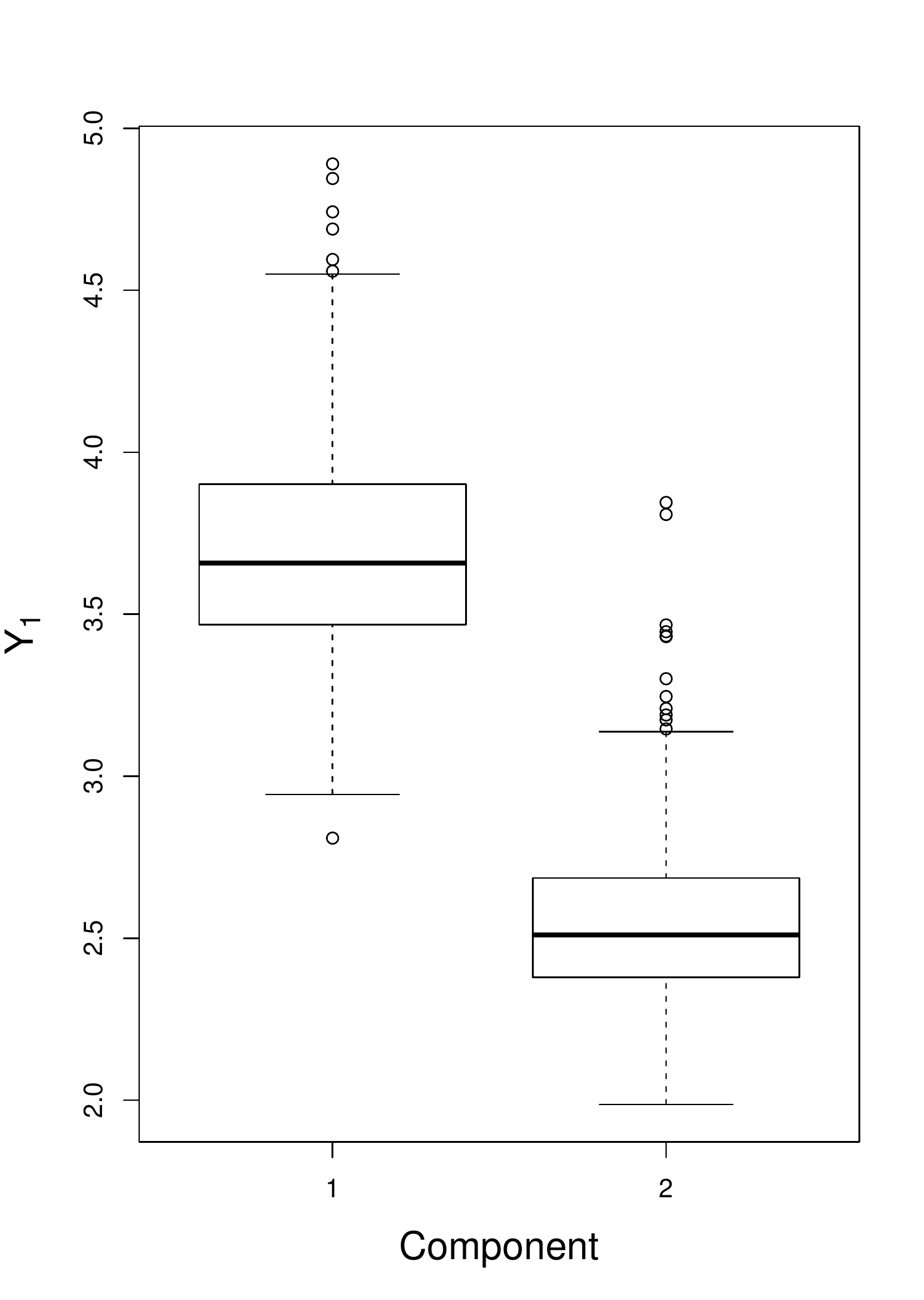}
	\end{minipage}
	\begin{minipage}{0.3\textwidth}
		\centering
		\includegraphics[width=1\linewidth, height=1.8\linewidth]{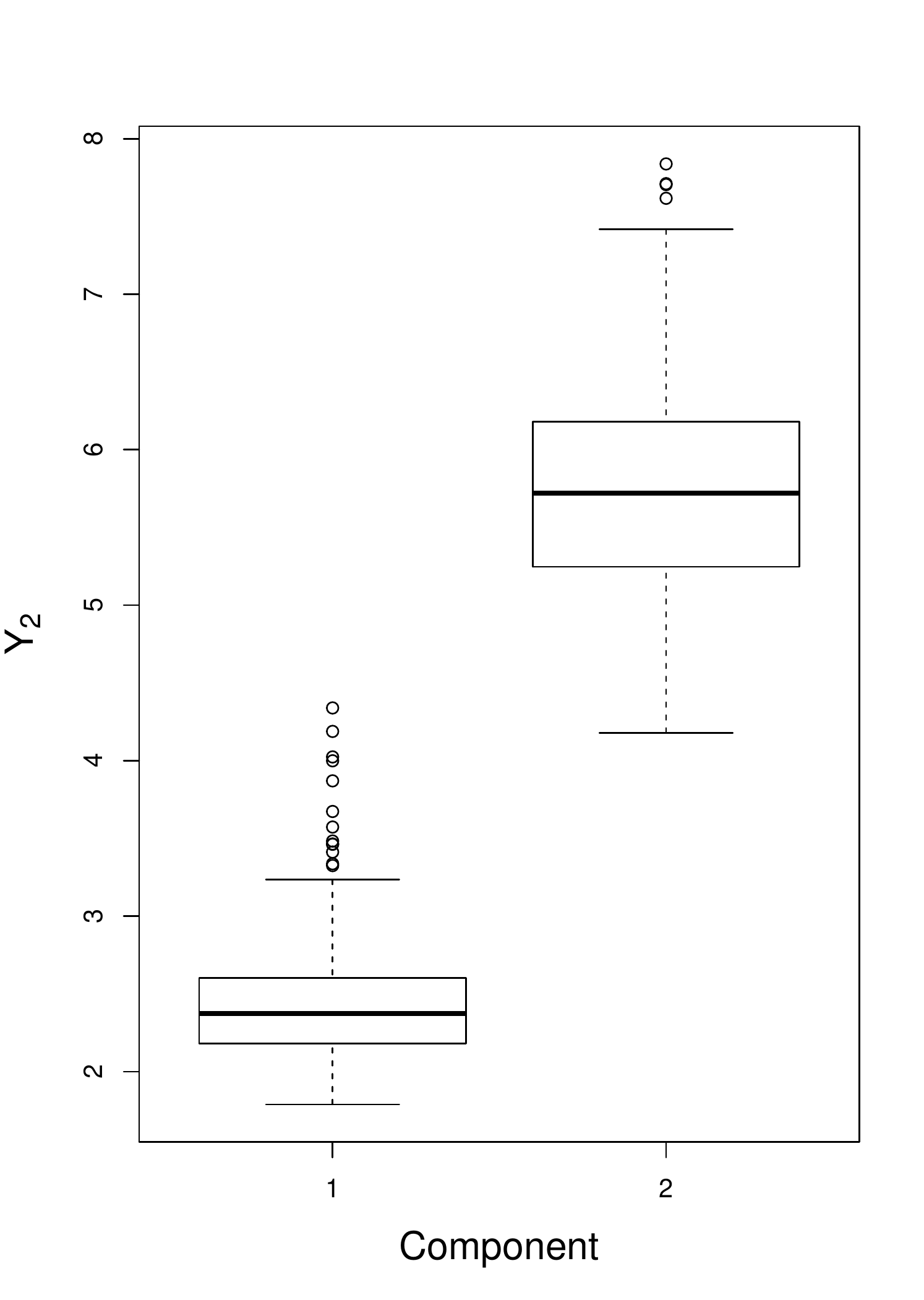}
	\end{minipage}
	\begin{minipage}{0.3\textwidth}
		\centering
		\includegraphics[width=1\linewidth, height=1.8\linewidth]{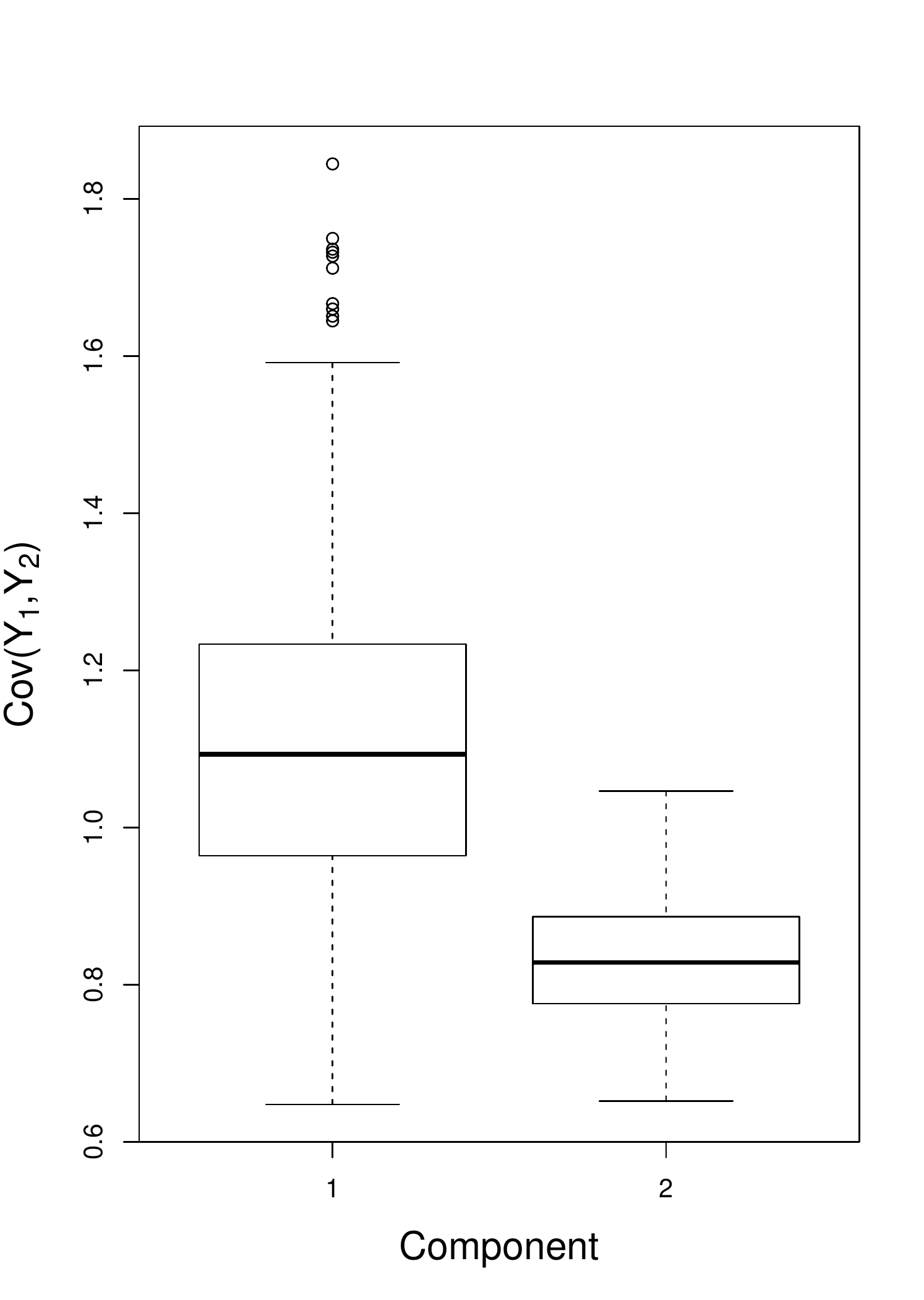}
	\end{minipage}
	\caption{Boxplots of the fitted values within each component for the two responses $Y_1$ and $Y_2$, and the fitted covariance between $Y_1, Y_2$ within each component.}
	\label{fig:finsim_boxplot_fittedvalues_cluster}
\end{figure}

Therefore, when covariates are incorporated the clustering fit is improved compared with the case when no covariates are incorporated. From a regression perspective, using the proposed model improved the (within-sample) fitted values. 
All of $\boldsymbol{w}_1, \boldsymbol{w}_2, \boldsymbol{w}_3$ entered all $\boldsymbol{\alpha}$-expert networks and the gating network, while the parsimonious parameterisation captures the $\boldsymbol{\beta}$ parameter with no covariates.  
The model regression parameters are given in Table~\ref{tab:finsim_bestmodel_regression_coefficient}.
It shows that, in most cases, the regression coefficients are close to the true simulation values.  
Figure~\ref{fig:finsim_boxplot_fittedvalues_cluster} presents the boxplot of the fitted values within each component. It shows that component 1 has mostly higher values of $y_1$ than that of component 2, while $y_2$ values in component 2 are greater than those in component 1, which is consistent with Figure~\ref{fig:simdata_bestmodel_classification_fittedvalues}. Figure~\ref{fig:finsim_boxplot_fittedvalues_cluster} also shows the calculated covariance between $Y_1$ and $Y_2$ from the fitted model, which indicates that the first component represents higher covariance than the second component. In fact, the overall sample correlation between $Y_1$ and $Y_2$ is $-0.01$ in the data. If not considering modeling the dependence structure, it may seem reasonable to model the two risk perils independently. However, with the proposed approach the sample correlations within components are $0.50$ and $0.42$ respectively. This shows that by modeling the dependence structure directly the dependence captured is much stronger than it appears at the outset.  

\begin{table}[t]
	\centering
	\caption{Estimated parameters of the selected MoE model: the top part shows the estimated gating network regression coefficients ($\widehat{\boldsymbol{\gamma}}_0$), compared with the corresponding true values ($\boldsymbol{\gamma}_0$); the bottom part shows the three sets of estimated regression coefficients for the $\alpha_1,\alpha_2, \alpha_3$ expert networks ($\widehat{\boldsymbol{\gamma}}_{1g},\widehat{\boldsymbol{\gamma}}_{1g},\widehat{\boldsymbol{\gamma}}_{1g}$) and the estimated $\widehat{\beta}_g$, compared with their true values $\boldsymbol{\gamma}_{1g}, \boldsymbol{\gamma}_{2g}, \boldsymbol{\gamma}_{3g}$ and the mean of $\beta_{ig}$ (i.e. $\bar{\beta}_g$); }
	\label{tab:finsim_bestmodel_regression_coefficient}
	\resizebox{0.8\textwidth}{!}{
		\begin{tabular}{c||c|cc|cc|cc||cc}
			\toprule[0.15 em]
			\multicolumn{10}{c}{\textbf{Gating network}} \\
			\midrule[0.1 em]	
			\ & \multicolumn{1}{c}{intercept} & $\boldsymbol{w}_1$ & \multicolumn{1}{c}{ $\boldsymbol{w}_2$} & $\boldsymbol{w}_3$ \\
			$\boldsymbol{\gamma}_0$ & \multicolumn{1}{c}{10.00} & 40.00 & \multicolumn{1}{c}{30.00} & 100.00 \\
			$\widehat{\boldsymbol{\gamma}}_0$ & \multicolumn{1}{c}{12.55} & 59.31 & \multicolumn{1}{c}{43.82} & 141.34 \\
			\midrule[0.1 em]
			\midrule[0.1 em]
			\multicolumn{10}{c}{\textbf{Expert network}} \\
			\midrule[0.1 em]
			\ &	\ & $\boldsymbol{\gamma}_{1g}$ & $\widehat{\boldsymbol{\gamma}}_{1g}$ & $\boldsymbol{\gamma}_{2g}$ & $\widehat{\boldsymbol{\gamma}}_{2g}$ & $\boldsymbol{\gamma}_{3g}$ & $\widehat{\boldsymbol{\gamma}}_{3g}$ & $\bar{\beta}_g$ & $\widehat{\beta}_g$ \\
			\multirow{4}{*}{Component $g=1$} 
			& intercept & 1.00   &1.00 & 0.10&0.24 & 0.50&0.31 & \multirow{4}{*}{1.22} & \multirow{4}{*}{1.12} \\
			& $\boldsymbol{w}_1$       & 0.20 &0.10 & 0.00 & 0.00 & 0.20& 0.00 & \\
			& $\boldsymbol{w}_2$       & 0.20 &0.05 & 0.10 &-0.01 & 0.20& 0.24 &  \\
			& $\boldsymbol{w}_3$       & 0.00 &0.00 & 0.10 & 0.68 & 0.20&-0.23 &  \\
			\midrule
			\multirow{4}{*}{Component $g=2$} 
			& intercept & 0.10&0.30 & 2.00  &2.18 & 1.50&1.49& \multirow{4}{*}{2.01} &\multirow{4}{*}{2.32}  \\
			& $\boldsymbol{w}_1$       & 0.10& 0.65& 0.00 &0.00 & 0.20& 0.00 &  \\
			& $\boldsymbol{w}_2$       & 0.10&-0.28& 0.30 &0.21 & 0.10& 0.16 &  \\ 
			& $\boldsymbol{w}_2$       & 0.00& 0.00& 0.30 &0.11 & 0.10&-0.03 &  \\
			\bottomrule[0.15 em]
	\end{tabular} }
\end{table}

\begin{figure}[htb]
\centering
\begin{minipage}{0.6\textwidth}
	\centering
	\includegraphics[width=\linewidth, height=.7\linewidth]{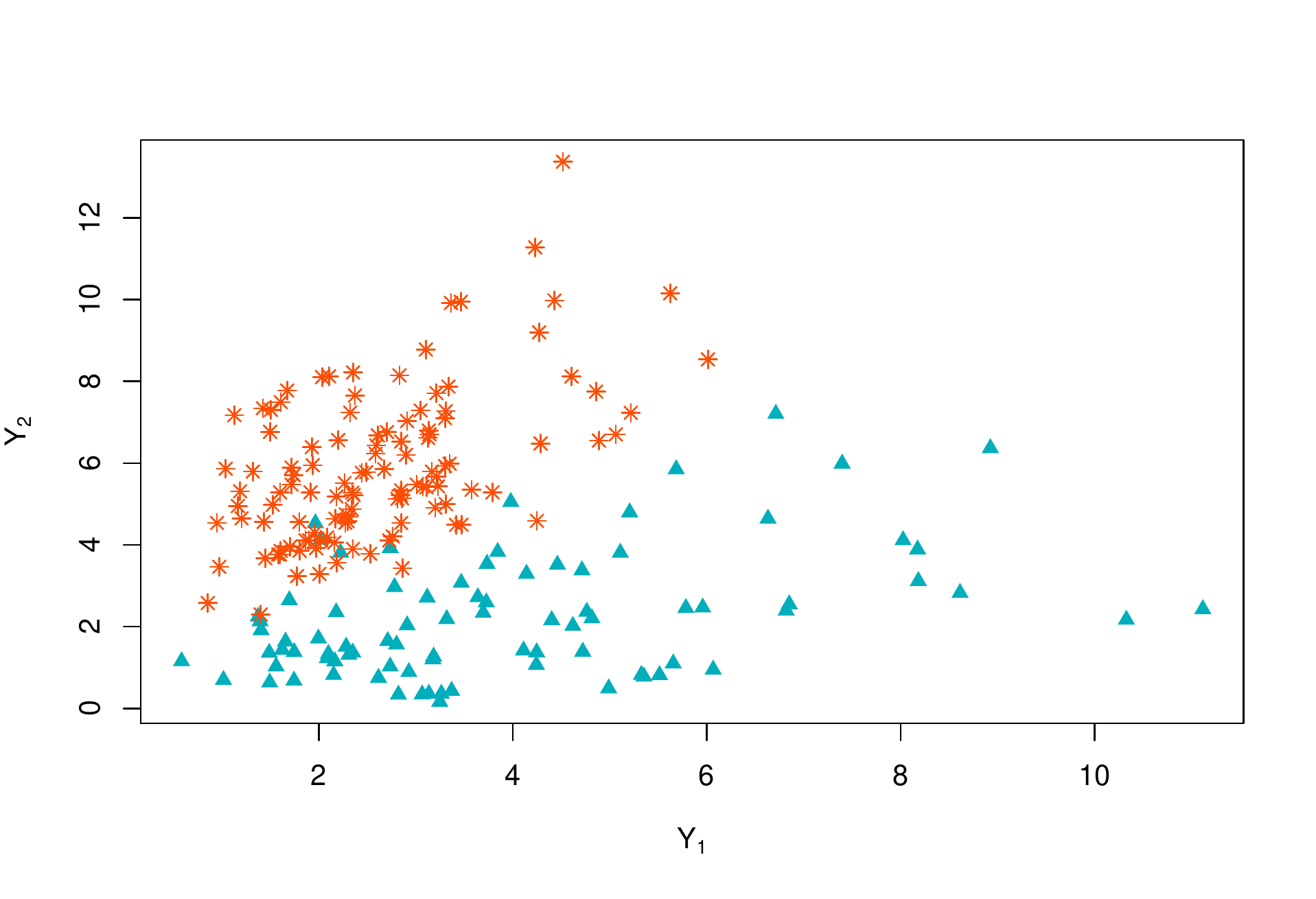}
	\caption*{(a)}
\end{minipage}
\begin{minipage}{0.6\textwidth}
	\centering
	\includegraphics[width=\linewidth, height=.7\linewidth]{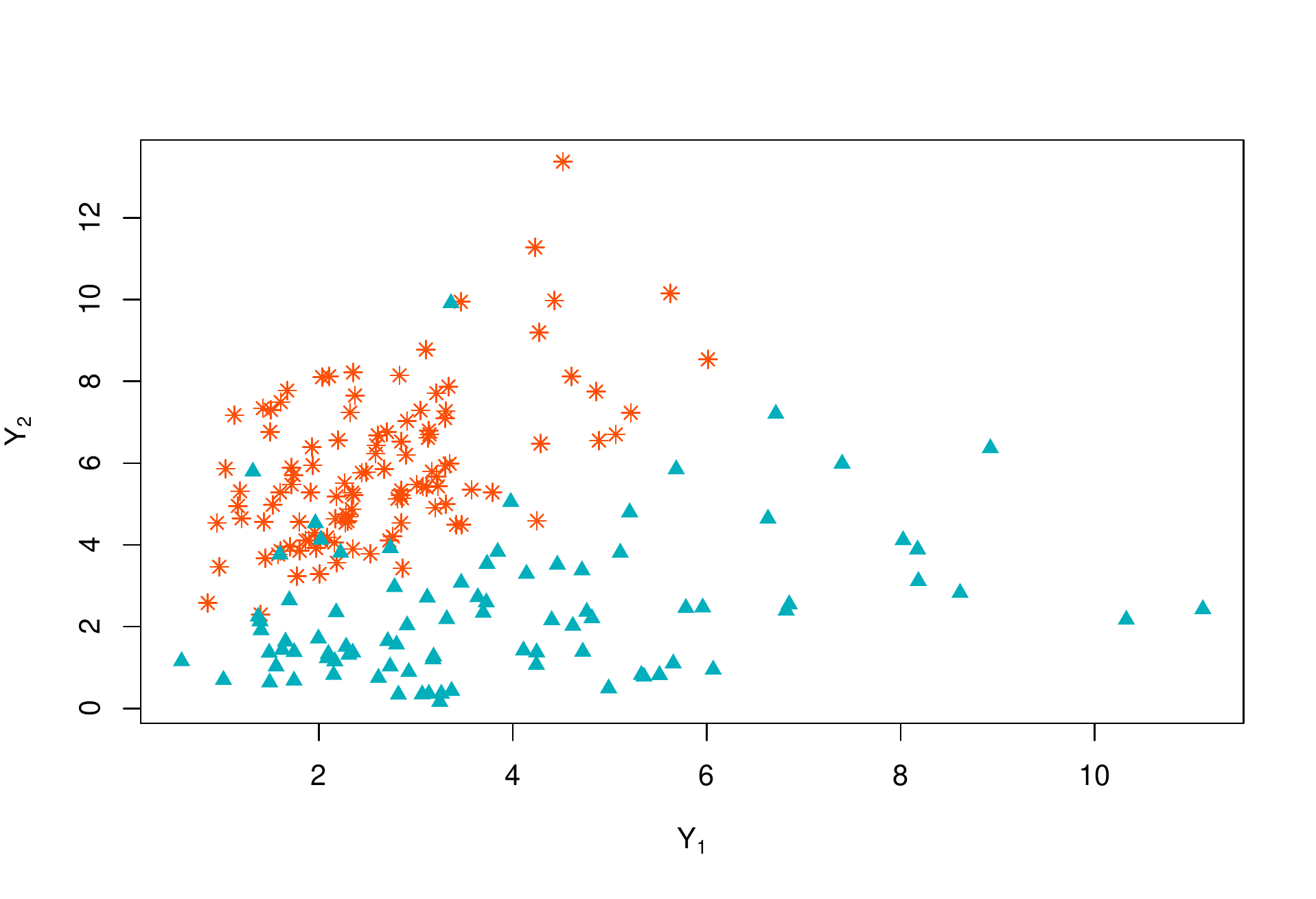}
	\caption*{(b)}
\end{minipage}
\caption{Plot of the response variable $\boldsymbol{y}$ of the test data, generated using the same simulation procedures as the training data: (a) the true cluster labels; (b) the predicted cluster labels using the optimal model, where only three observations are misclassified.}
\label{fig:finsim_testset_plot}
\end{figure}

\begin{figure}[t!]
\centering
\begin{minipage}{0.6\textwidth}
\centering
\includegraphics[width=\linewidth, height=.65\linewidth]{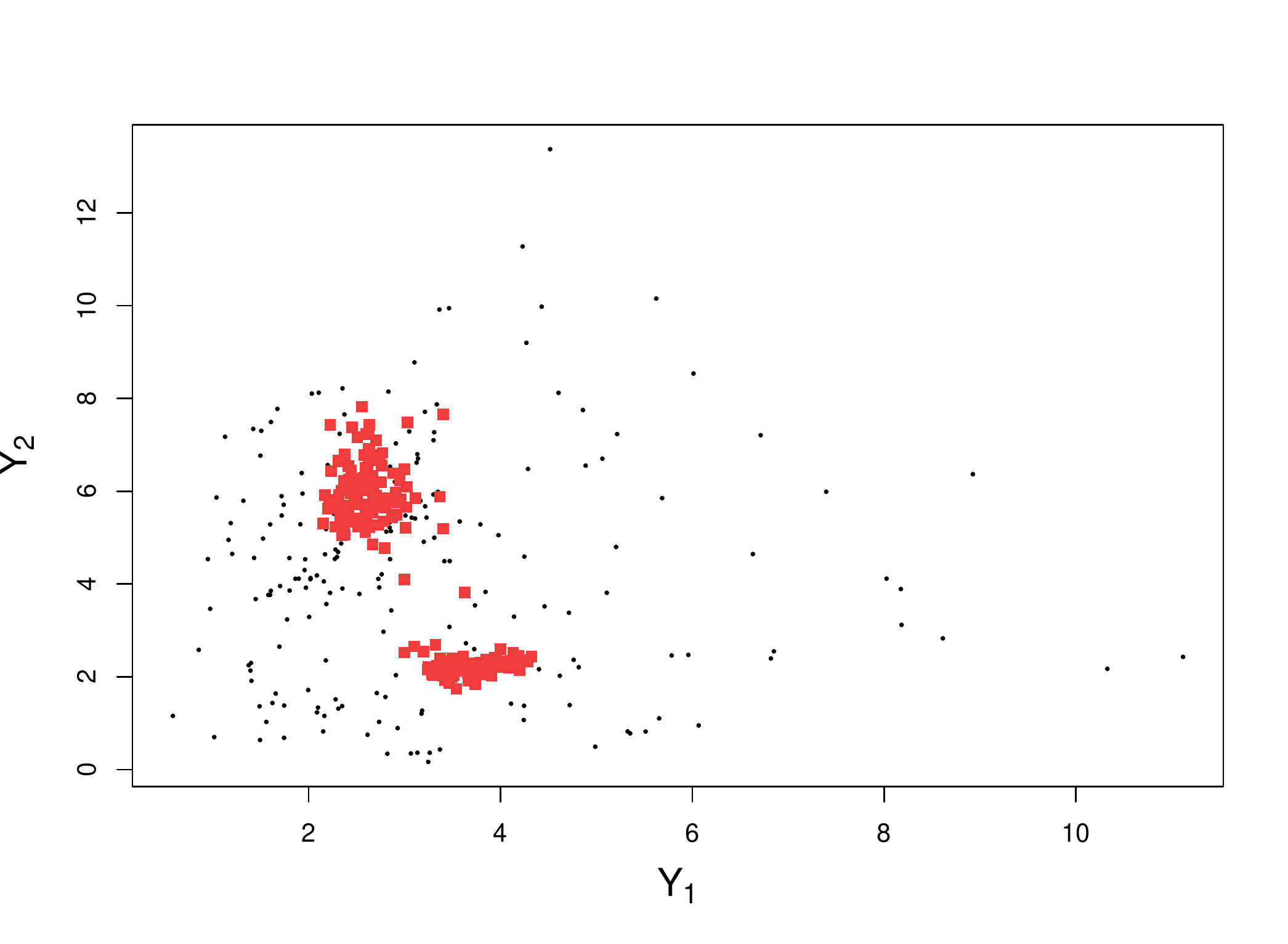}
\caption*{(a)}
\end{minipage}
\begin{minipage}{0.6\textwidth}
\centering
\includegraphics[width=\linewidth, height=.65\linewidth]{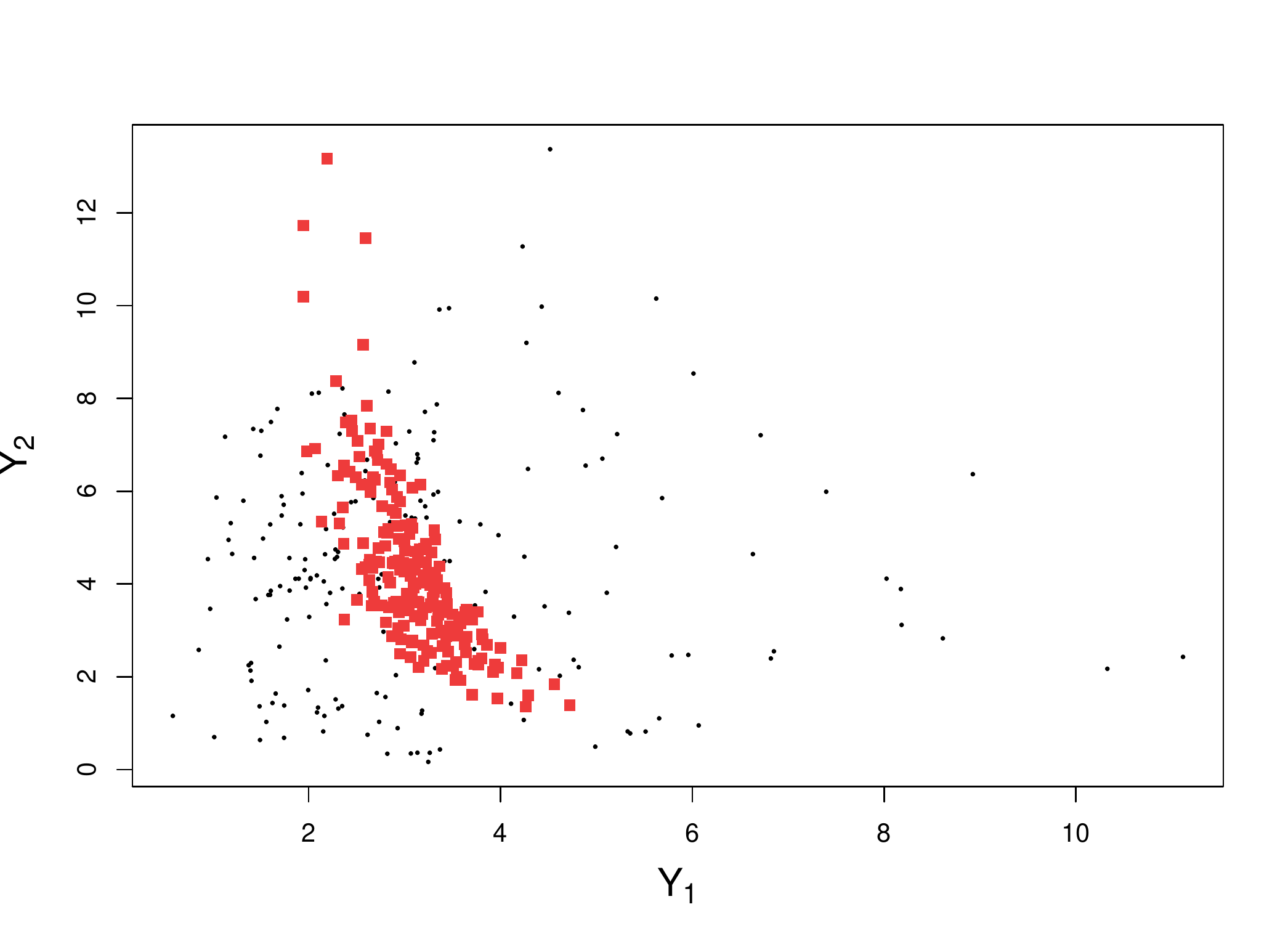}
\caption*{(b)}
\end{minipage}
\caption{Predicted values of $\boldsymbol{y}$ of the test data from (a) the optimal MoE model; (b) two independent univariate GLMs. The true values are plotted in small black ``$\centerdot$", and the predictions are plotted in red ``$\blacksquare$".}
\label{fig:finsim_testset_prediction}
\end{figure}

For testing the predictive power of the model, a new test data set was generated with the same method for $N^{\star}=200$. The response variable $\boldsymbol{y}$ of the test set is presented in Figure~\ref{fig:finsim_testset_plot}(a), colored by true component labels.
The predicted cluster membership is shown in Figure~\ref{fig:finsim_testset_plot}(b), with adjusted Rand index of $0.94$ and only three observations misclassified. 
The predicted values of the selected best model are shown in Figure~\ref{fig:finsim_testset_prediction}(a). For comparison, standard GLMs using the same covariates are also fitted using the simulated data respectively for $\boldsymbol{y}_1$ and $\boldsymbol{y}_2$, and the predicted values using the test data are presented in Figure~\ref{fig:finsim_testset_prediction}(b). It is clear that the univariate model did not capture the structure within the data, and the best MoE model clearly identified the dependence structure within different components. 

To accurately assess the prediction improvement from using MoE models, the out-of-sample predictions are assessed using various metrics, in comparison with the univariate GLM predictions. 
It is important to evaluate the predictive performance of the predictive distributions generated, relative to the respective observations with which the empirical distributions can be calculated, such as predictive CDF versus empirical CDF.  
Key tools used to test different characteristics are: proper scoring rule using the mean of continuous ranked probability score (CRPS) (\citealp{Gschloessl2007}; \citealp{Gneiting2007}); square root mean squared errors (rMSE) (\citealp{Lehmann2006}); Gini index (\citealp{Frees2014}) and Wasserstein distance (\citealp{Vallender1974}). 
The results are shown in Table~\ref{tab:simdata_prediction_comparison}.
Particular interest lies in the predictions of the sum of $Y_1$ and $Y_2$, since it represents the total risk, and the purpose of bivariate modeling is to better predict two risks simultaneously while taking their dependence into account, especially in univariate modelling the sum of models' predictions would be taken as the total risk. It is also expected that predictions within each marginal are improved. 
Table~\ref{tab:simdata_prediction_comparison} shows that the MoE model clearly outperforms the GLMs in all categories with the exception of Wasserstein distance for $Y_2$ which is slightly worse.

\begin{table}[ht]
\centering
\caption{Comparisons between the MoE model and the GLM model for predictive performance using a range of metrics. The underlined values represent the optimal results for each measure.}
\label{tab:simdata_prediction_comparison}
\resizebox{.55\textwidth}{!}{
\begin{tabular}{c|l|rrrr}
\toprule[.15 em]
\ & \ & CRPS & rMSE & Gini & Wasserstein \\
\midrule
\multirow{2}{*}{Sum of $Y_1,Y_2$} & MoE & \underline{1.83} & \underline{2.71} & \underline{0.55} & \underline{1.15} \\
& GLM & 1.86 & 2.95 & 0.54 & 1.26 \\
\midrule
\multirow{2}{*}{$Y_1$} & MoE & \underline{1.02} & \underline{1.61} & \underline{0.56} & \underline{0.76} \\
& GLM & 1.06 & 1.68& 0.55 & 0.89 \\
\midrule
\multirow{2}{*}{$Y_2$} & MoE & \underline{1.48} & \underline{1.64} & \underline{0.62} & 0.80 \\
& GLM & \underline{1.48} & 2.10 & 0.61 & \underline{0.74} \\
\bottomrule[0.15 em]
\end{tabular} }
\end{table}

\section{Irish GI insurer data}
\label{section:result}

A large motor insurance claims data set was obtained from an Irish general insurance company. The data represent the insurer's book of business in Ireland from January 2013 to June 2014, containing 452,266 policies and their characteristics. In Ireland, motor insurance is required for all motorists and is enforced by Irish law, with third party (TP) cover being the minimum requirement (\citealp{MotorLaw}). 
Across the portfolio, there are five distinct categories of claims: accidental damage (AD), third party bodily injury (BI), third party property damage (PD), windscreen (WS) and theft (TF). 
This insurer provides two different types of coverage: third party, fire and theft cover; and comprehensive cover (which includes all five categories). 
Typically, all five categories are modeled independently, then the sum of predicted claims across the five categories is taken as the total risk and the policy price is set based on this value. 
Because all five categories and their pairwise dependencies were investigated, only comprehensive cover was selected.

\begin{figure}[H]
	\centering
	\begin{minipage}{0.65\textwidth}
		\includegraphics[width=\linewidth,height=.77\linewidth]{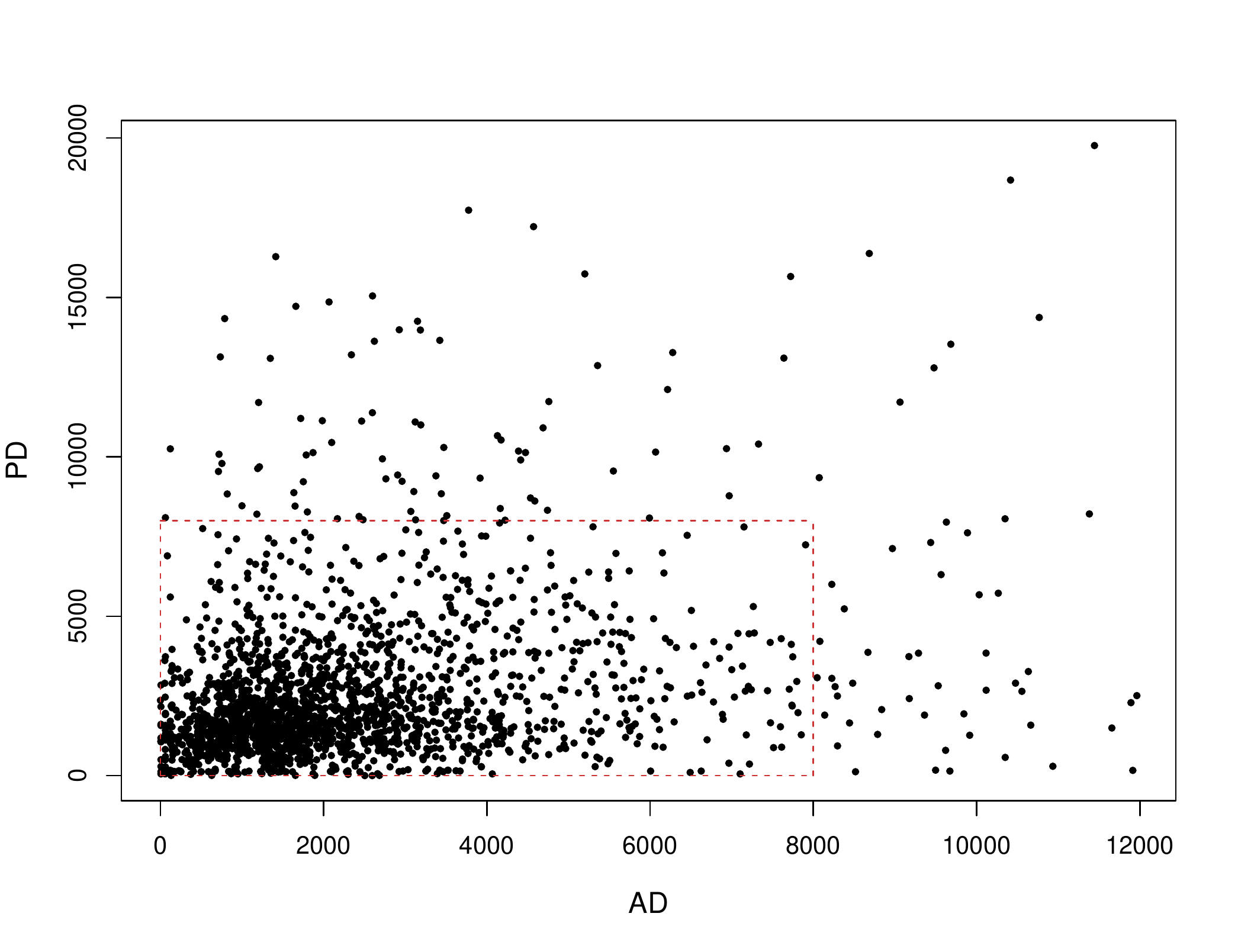}
		\caption*{(a)}
	\end{minipage}
	\begin{minipage}{0.65\textwidth}
		\includegraphics[width=\linewidth,height=.77\linewidth]{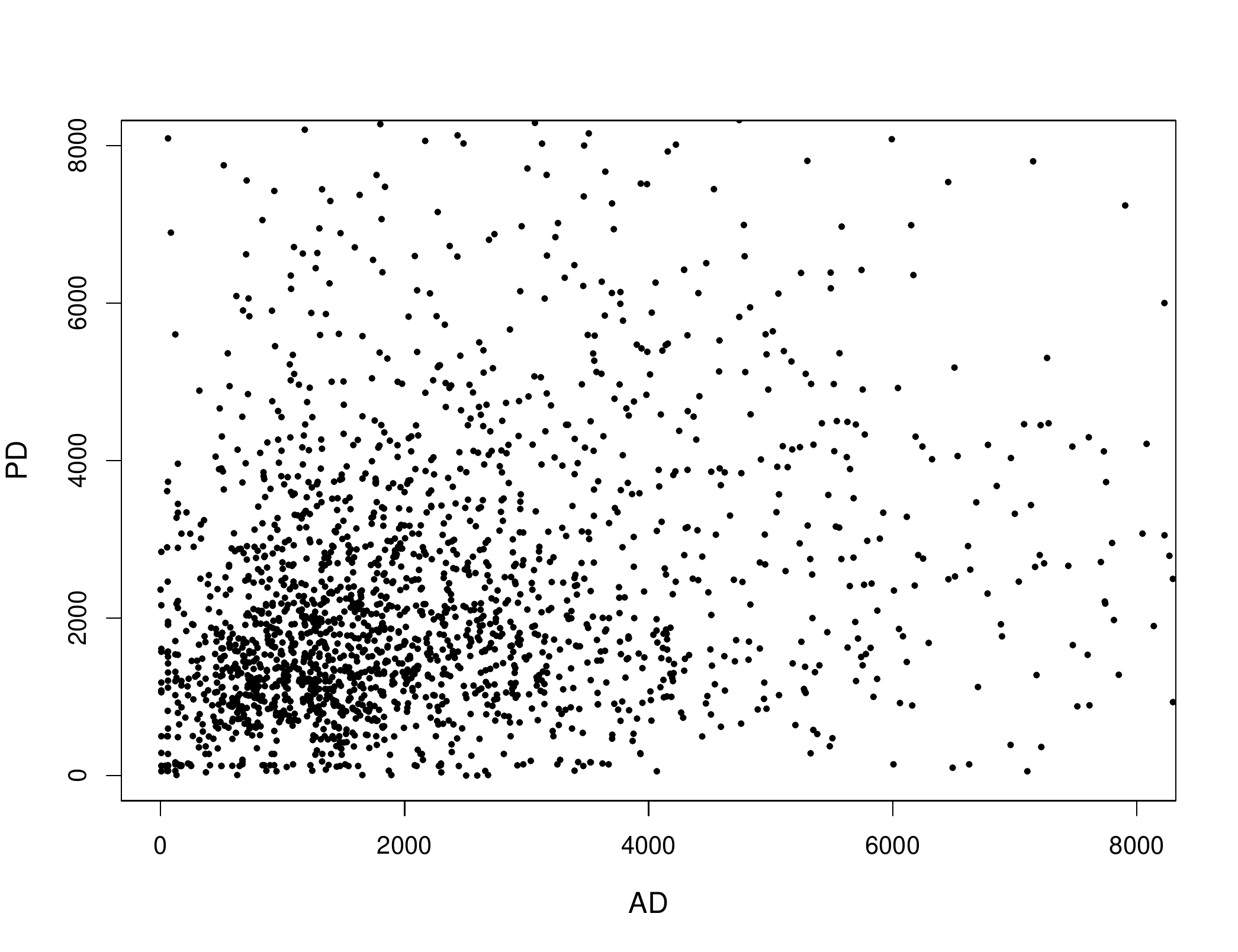}
		\caption*{(b)}
	\end{minipage}
	\caption{The plot of accidental damage (AD) and property damage (PD) claims of the Irish GI data set: (a) shows all the claims; (b) shows the claims smaller than $8,000$ to give a clearer picture of the clustering structure in the region where the majority of claims occur. }
	\label{fig:realdata_response_plot}
\end{figure}

Among the five categories covered per policy, AD and PD represent the highest correlation ($\rho_{(AD,PD)}=0.24$) 
when excluding all policies for which no claims have been made on these two categories. 
Hence only policies having made claims on both AD and PD were selected and their depenence is chosen for investigation using the bivariate gamma distribution. 
Furthermore, in these two categories there exist a few extremely large claims that are much greater than \euro20,000. These observations have been deleted due to their extreme nature and it would be more appropriate to utilise extreme value theory for modeling them. 
This leads to 2,074 policies being present in the final data set. 
The logic of this work on severity is that, when a policyholder makes a claim only in one category and not the other, the frequency models (either multivariate or univariate) will account for this aspect. Bivariate severity models only investigate cases when claims are made on both risks.
It is worth noting that the correlation between AD and PD claims is not very strong, although it represents the strongest correlation in the data. 
One may decide to use independent modeling given this fact. However, it may also indicate that, since the data are very dispersed (see Figure~\ref{fig:realdata_response_plot}), there is underlying heterogeneity. 
By considering a clustering-based approach to segregate the observations, the correlations between AD and PD within different components will be much stronger. \cite{Bermudez2009} concluded that, in the bivariate Poisson distribution case, even when there are small correlations between claims, major differences can still occur in predictions from a bivariate model compared with independent univariate models. It is shown that this same outcome occurs for the bivariate gamma distribution case. 

\begin{table}[ht]
	\centering
	\caption{Descriptions of the selected variables in the Irish GI data set. Note that ``No claim discount" represents the number of years with no claim.}
	\label{tab:realdata_predictor_description}
	\resizebox{9.5cm}{!}{
		\begin{tabular}{ll}
			\toprule[.15 em]
			Variables  & Categories  \\
			\midrule 
			Vehicle fuel type 	  & Diesel; Petrol; Unknown  \\
			Vehicle transmission & Automatic; Manual; Unknown \\
			Annual mileage           & 0-5000; $\ldots$; 45001-50000; 50001+  \\
			Number of registered drivers & 1; 2; 3; 4; 5; 6; 7 \\
			No claim discount & 0; 0.1; 0.2; 0.3; 0.4; 0.5; 0.6\\
			No claim discount protection 	  & No; Yes; Unknown \\
			Main driver license category& B; C; D; F; I; N \\
			\bottomrule[.15 em]
	\end{tabular} }
\end{table}

In the data set, there are about 100 variables presented. 
Due to data confidentiality issues, not all significant predictors could be selected and used in this work.
Seven key covariates are used including policyholder's information (e.g. licence category), insured car information (e.g. fuel type, transmission) and policy information (e.g. no-claim discount).  Table~\ref{tab:realdata_predictor_description} shows the predictors used and their categorical levels. 
It is expected that not all predictors will be significant, since it is acknowledged that in general the severity model requires fewer predictors than the frequency model (\citealp{Coutts1984}; \citealp{Charpentier2014}), and in a previous univariate modeling study using a similar data set, it is also shown that not all the selected predictors provided here are significant (\citealp{Hu2018}).  
Figure~\ref{fig:realdata_response_plot}(a) shows the scatter plot of the AD and PD claim amounts. The claims from the two risk perils are very dispersed, with only a few large claims in both risks. 
The scatters are slightly more skewed towards PD, that is there are more policies having high PD and low AD claim amounts and the number of large PD claim sizes is greater than the number of large AD claims.
There is no clear separation in the scatter plot, suggesting that different components (if any) could be overlapped. There could also be clustering outliers and labelling noise present. 
The vast majority of claims are heavily centred in the interval between $500$ and $3,000$. 
A scatter plot only focusing on the range AD $\leq 8,000$ and PD $\leq 8,000$ is presented in Figure~\ref{fig:realdata_response_plot}(b) for better clarity. 
In this plot, it is easier to see that, while there is a denser cluster in the interval between $500$ and $3000$, there are many claims close to both axes, indicating that many policies have slightly high claim amounts in one peril but very low claim amounts in the other.
The characteristics of the data may explain why the overall correlation is low. 
By using the proposed method focusing on dependence and segregating observations, it is expected that within homogeneous groups the dependence will be much stronger, and therefore important to explicitly include in the model.

\begin{table}[H]
\centering
\caption{Model selection when setting $G=1$:7. The selected optimal model (based on AIC) gives $G=4$ (underlined). Note that BIC gives the same conclusion.}
\label{tab:realdata_MBC_classification_G4}
\begin{tabular}{cccc}
	\toprule[.15 em]
	\  & log-likelihood & BIC & AIC \\
	\midrule
	G=1& -36392.69 & 72815.94 &	72793.39\\
	G=2& -36310.17 & 72689.09 &	72638.36\\
	G=3& -36292.07 & 72691.03 &	72612.11\\
	G=4& \underline{-36245.94} & \underline{72636.99} & \underline{72529.89} \\
	G=5& -36244.47 & 72665.40 &	72530.11\\
	G=6& -36238.72 & 72696.20 &	72532.72\\
	G=7& -36236.37 & 72732.41 & 72540.75 \\
	\bottomrule[.15 em]
\end{tabular}	
\end{table}

\begin{figure}[H]
\centering
\includegraphics[width=13cm, height=9cm]{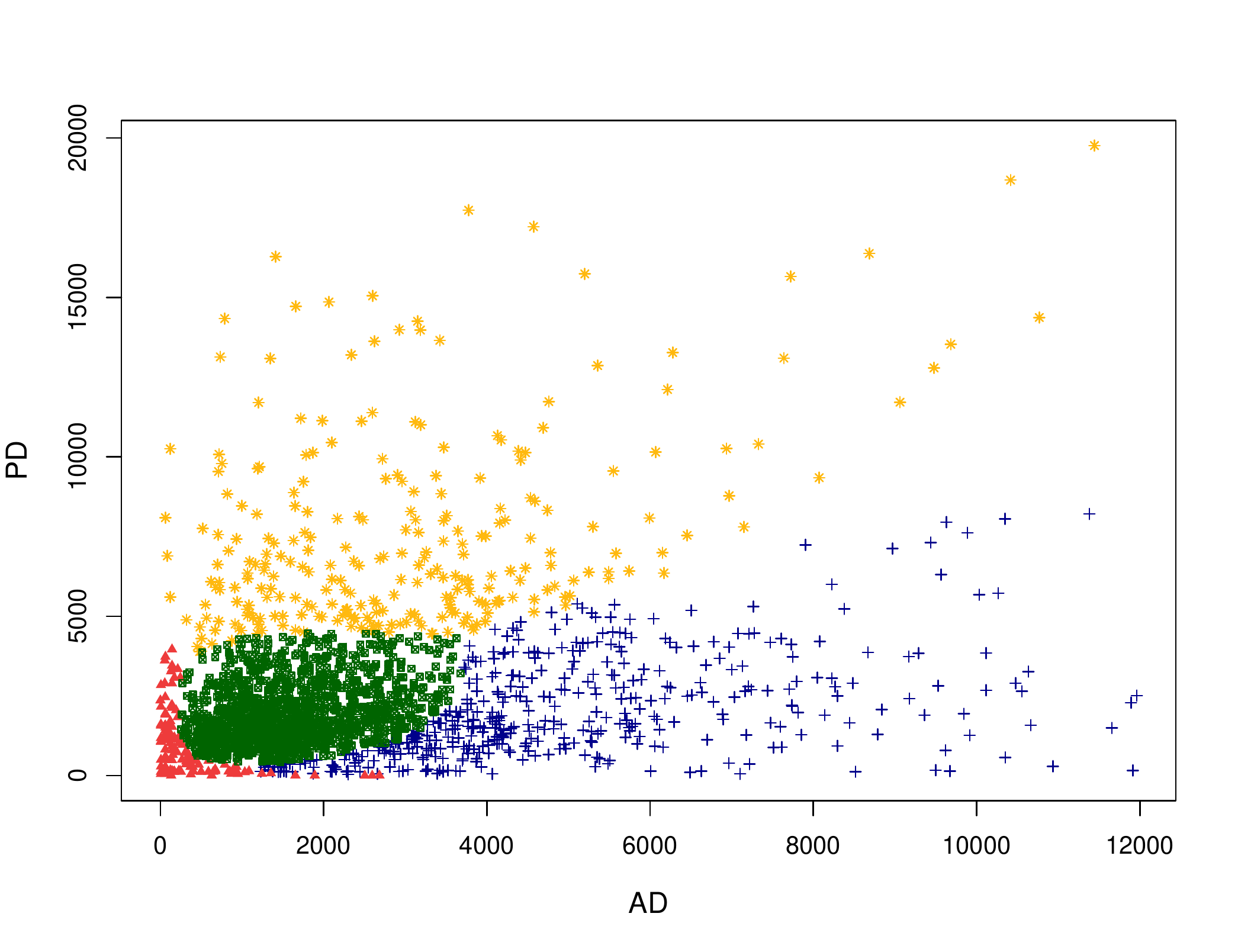}
\caption{Classification plot when clustering accidental damage (AD) and property damage (PD) without covariates. The optimal number of components $G=4$ using AIC.
In the plot the green ``$\boxplus$" represents component 1, the yellow ``$*$" represents component 2, the blue ``$+$" represents component 3 and the red ``$\blacktriangle$" represents component 4.
}
\label{fig:realdata_MBC_classification_G4}
\end{figure}

\begin{figure}[H]
	\centering
	\includegraphics[width=13cm, height=9cm]{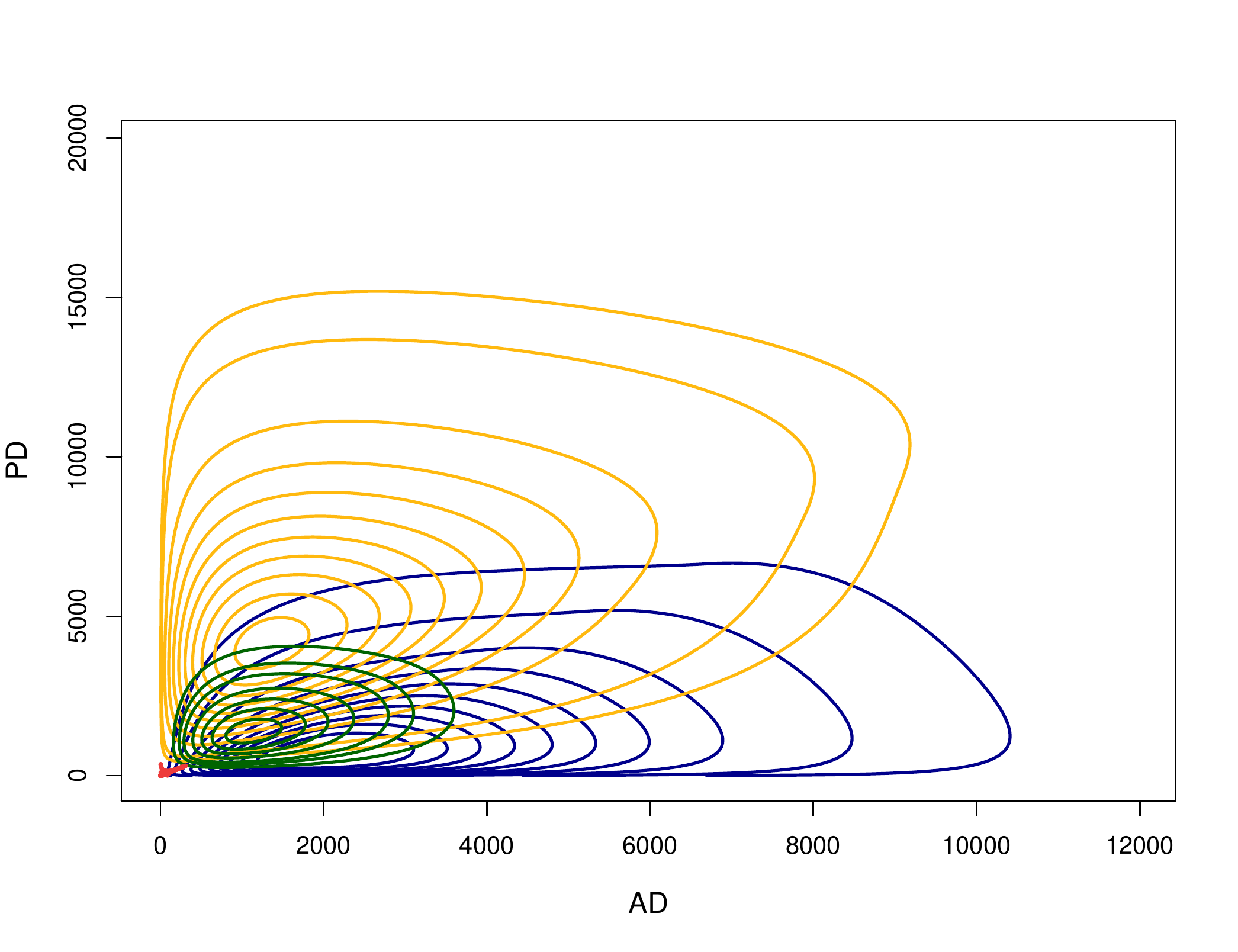}
	\caption{Contour plot based on the fitted optimal number of components $G=4$ when clustering accidental damage (AD) and property damage (PD) without the covariates, corresponding to the classification in Figure~\ref{fig:realdata_MBC_classification_G4}.
	In the plot the green, yellow, blue and red contours represent component 1, 2, 3 and 4 respectively. The plot shows that the four components are highly overlapping.
	}
	\label{fig:realdata_MBC_classification_G4_contour}
\end{figure}

Starting with no covariates in the clustering process using a finite mixture of bivariate gamma distributions, the method selects $G=4$ using AIC, as shown in Table~\ref{tab:realdata_MBC_classification_G4}. Note that selection based on BIC is also shown in Table~\ref{tab:realdata_MBC_classification_G4}, which gives the same conclusion. Figure~\ref{fig:realdata_MBC_classification_G4} shows the classification plot based on the optimal $G=4$, while Figure~\ref{fig:realdata_MBC_classification_G4_contour} shows the corresponding contour plot of the four fitted bivariate gamma distributions, based on the estimated parameters $\boldsymbol{\alpha},\boldsymbol{\beta}$ shown in Table~\ref{tab:realdata_MBC_estimation_G4}. 
As mentioned earlier, there is a dense cluster of AD and PD claim sizes in the interval between $500$ and $3000$, which has been recognized by the model (i.e. component 1). 
Other components include small claim sizes (AD $<1000$ and PD $<4000$) along the axes (component 4), policies having made higher AD claim amounts than PD claims (component 3), and policies having made PD claim amounts greater than or equal to AD claim amounts (component 2). 
Figure~\ref{fig:realdata_MBC_classification_G4_contour} suggests heavy overlapping among the four components, which causes more uncertainty when segregating the policyholders to amplify the dependencies.   
Within the four components, the sample correlations now are $(\rho^{(g=1)}=0.30, \rho^{(g=2)}=0.41, \rho^{(g=3)}=0.50, \rho^{(g=4)}=-0.38)$, which shows that the dependencies are much stronger once heterogeneity is considered, and the amplified dependencies could be incorporated in simutaneous bivariate modeling.  

\begin{table}[H]
\centering
\caption{Estimated parameters when clustering accidental damage (AD) and property damge (PD) without the covariates and $G=4$ groups.}
\label{tab:realdata_MBC_estimation_G4}
\begin{tabular}{c|ccccr}
\toprule[0.15 em]
\ & $\tau_g$ & $\alpha_{1g}$ & $\alpha_{2g}$ & $\alpha_{3g}$ & \multicolumn{1}{c}{$\beta_g$} \\
\midrule
Component g=1 &0.48& 2.32& 2.89& 0.95& 19.27*10$^{-4}$\\ 
Component g=2 &0.14& 0.82& 3.07& 1.02& 6.21*10$^{-4}$ \\
Component g=3 &0.25& 2.16& 0.76& 0.71& 6.99*10$^{-4}$ \\
Component g=4 &0.13& 0.42& 0.60& 0.36& 5.78*10$^{-4}$ \\
\bottomrule[0.15 em]
\end{tabular}
\end{table}

When including covariates in the proposed MoE model, model selection needs to be carefully implemented using forward stepwise selection, and choices need to be made for $G$, model type and covariates in each expert or gating networks (if any). 
It is worth noting that, since there are 7 covariates to select from together with all the model types in the model family, the potential model space is huge. It may be reasonable to expect there are many models with similar predictive power. The optimal model that has been selected based on AIC is presented in Table~\ref{tab:realdata_optimal_model_selected}.
The selected optimal model has model type ``VVV", G=4 and AIC$=72,583.97$. $G$ is consistent with the result with no covariates as before.
Annual mileage and no-claim discount (NCD) are not selected in the final model, while the rest of covariates enter either one or more of the expert or gating networks. Noticeably the gating network and the $\beta$-expert have the most covariates, which makes sense because: (1) it helps to allocate observations into the correct components while within each component the dependence may be simpler to model given the correct allocation; (2) the claim amount range within the data set (from close to $0$ to $20,000$) is large, which may require more covariates to model the rate parameter $\boldsymbol{\beta}$ to separate the smaller and larger claims.  

\begin{table}[ht]
\centering
\caption{Selected covariates in each gating and expert network of the optimal model, which has $G=4$ and model type ``VVV". }
\label{tab:realdata_optimal_model_selected}
\resizebox{.75\textwidth}{!}{
\begin{tabular}{l|ccccccc}
\toprule[.15 em]
	\ & $\alpha_1$ expert & $\alpha_2$ expert & $\alpha_3$ expert & $\beta$ expert & gating \\
\midrule
Vehicle fuel type & $\surd$ & $\surd$ & $\surd$ & $\surd$ & \ \\
Vehicle transmission & $\surd$ & \ & \ & $\surd$ & $\surd$ \\
Annual mileage\\
Number of registered drivers& \ & \ & \ & \ & $\surd$\\
No claim discount (NCD) \\
No claim discount protection& \ & $\surd$ & $\surd$ & $\surd$ & $\surd$ \\
Main driver license category & \ & \ & \ & $\surd$ & $\surd$ \\
\bottomrule[.15 em]
\end{tabular}}
\end{table}

\begin{figure}[ht]
	\centering
	\includegraphics[width=13cm, height=9.5cm]{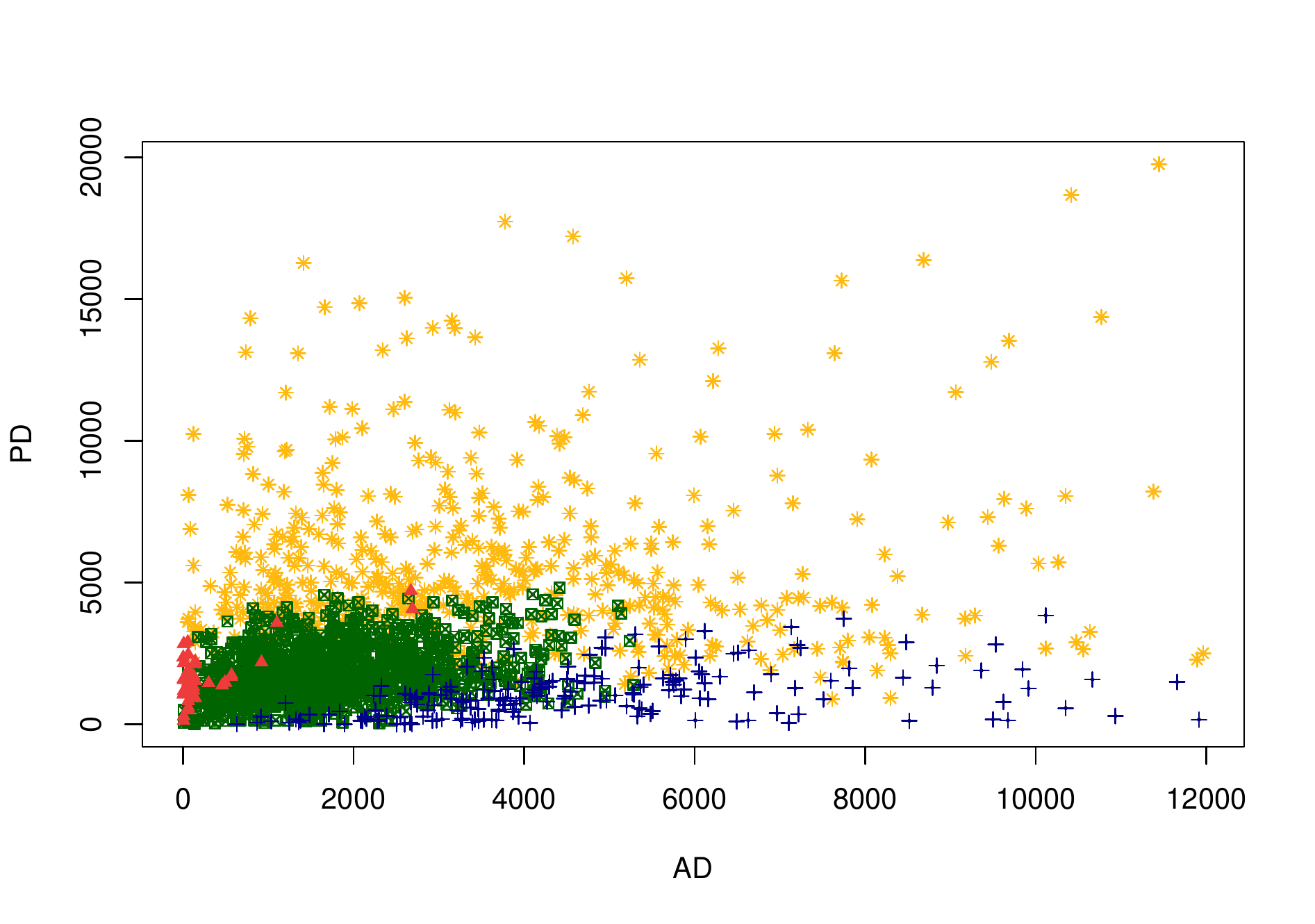}
	\caption{Classification of accidental damage (AD) and property damage (PD) of the selected VVV model with $G=4$, where the green ``$\boxplus$" represents component 1, the yellow ``$*$" represents component 2, the blue ``$+$" represents component 3 and the red ``$\blacktriangle$" represents component 4.}
	\label{fig:realdata_bestmodel_classification}
\end{figure}

The classification plot based on the optimal model is shown in Figure~\ref{fig:realdata_bestmodel_classification}. The general pattern is similar to that in Figure~\ref{fig:realdata_MBC_classification_G4}, 
where component 1 captures the dense claim size cloud of AD and PD in the interval $500-4000$,
component 2 captures large claim amounts particularly when $PD > AD$ ,
component 3 captures AD claims greater than PD 
and component 4 captures small claims of PD greater than AD, which consists of the fewest policies. 
It is useful and important to see that the dependence structures captured here are consistent with the case where no covariates are included, in Figure~\ref{fig:realdata_MBC_classification_G4}.
However, this result makes more sense and has the following differences compared with Figure~\ref{fig:realdata_MBC_classification_G4}: 
(1) there are more policies having higher AD and very low PD claims (i.e. more scatters along the AD-axis from 0), this feature is now better captured by component 3; 
(2) there are some policies having higher PD and very low AD claims, but when PD amounts are larger than $5000$, they also tend to have slightly higher AD amounts, hence component 4  
mostly captures a small proportion of policies having relatively small PD amounts and very small (close to 0) AD amounts; 
(3) component 2 not only captures large PD claims greater than AD, but also overall large claim amounts for both AD and PD; but because of the scarcity of very large claims overall, alternatively it can be interpreted as the large claims noise component.
Note that because this classification is a result of incorporating covariates in regressions, it makes more sense to see that not only there are clear overlaps among all components, but also the component boundaries are more blurred. 
There are a few observations (for example, a few policies labelled red $\blacktriangle$ in component 4 in Figure~\ref{fig:realdata_bestmodel_classification}) set apart from their corresponding components clearly, aside from the covariates' influence this could also potentially be due to outliers or labelling noise in the data. 
 
Based on this classification the sample correlations for the four components are $(\rho^{(g=1)}=0.33, \rho^{(g=2)}=0.05, \rho^{(g=3)}=0.41, \rho^{(g=4)}=0.72)$ respectively. 
In most cases the correlation within component is stronger than the overall correlation ($\rho=0.24$). Component 2 has very low sample correlations, which could be due to the fact that it captures mostly overall large claims, as an overall ``noise" component. 
It again shows that by segregating the data based on the dependence structure directly, within component the dependencies are indeed much stronger than it may first appear.

When different (combinations of) covariates enter different expert networks, it might make model interpretation more difficult. 
In the selected model, vehicle transmission, number of registered drivers, NCD protection and main driver licence category enter the gating networks, which explicitly control which component the policy belongs to, although covariates in each expert also implicitly control the choice of components (\cite{Bermudez2012}). 
Each of the expert networks has different covariates. Covariance is mainly determined by $\alpha_3$ (and $\beta$). It is interesting to see only vehicle fuel type and NCD protection enter this $\alpha_3$ network, indicating vehicle fuel types and NCD protection affect much of the dependence between AD and PD. 
The optimal model is chosen based on AIC. While it penalises extra paramters less than other metrics such as BIC, it is still surprising to see that in most of the expert networks not many covariates are involved. For the proposed MoE model family, since very likely $G \geq 2$, adding a covariate into the model means $G$ times more parameters in the covariate are added. This largely leads to it prefering a simpler model set up.   

\begin{figure}[ht]
	\centering
	\includegraphics[width=11cm, height=7cm]{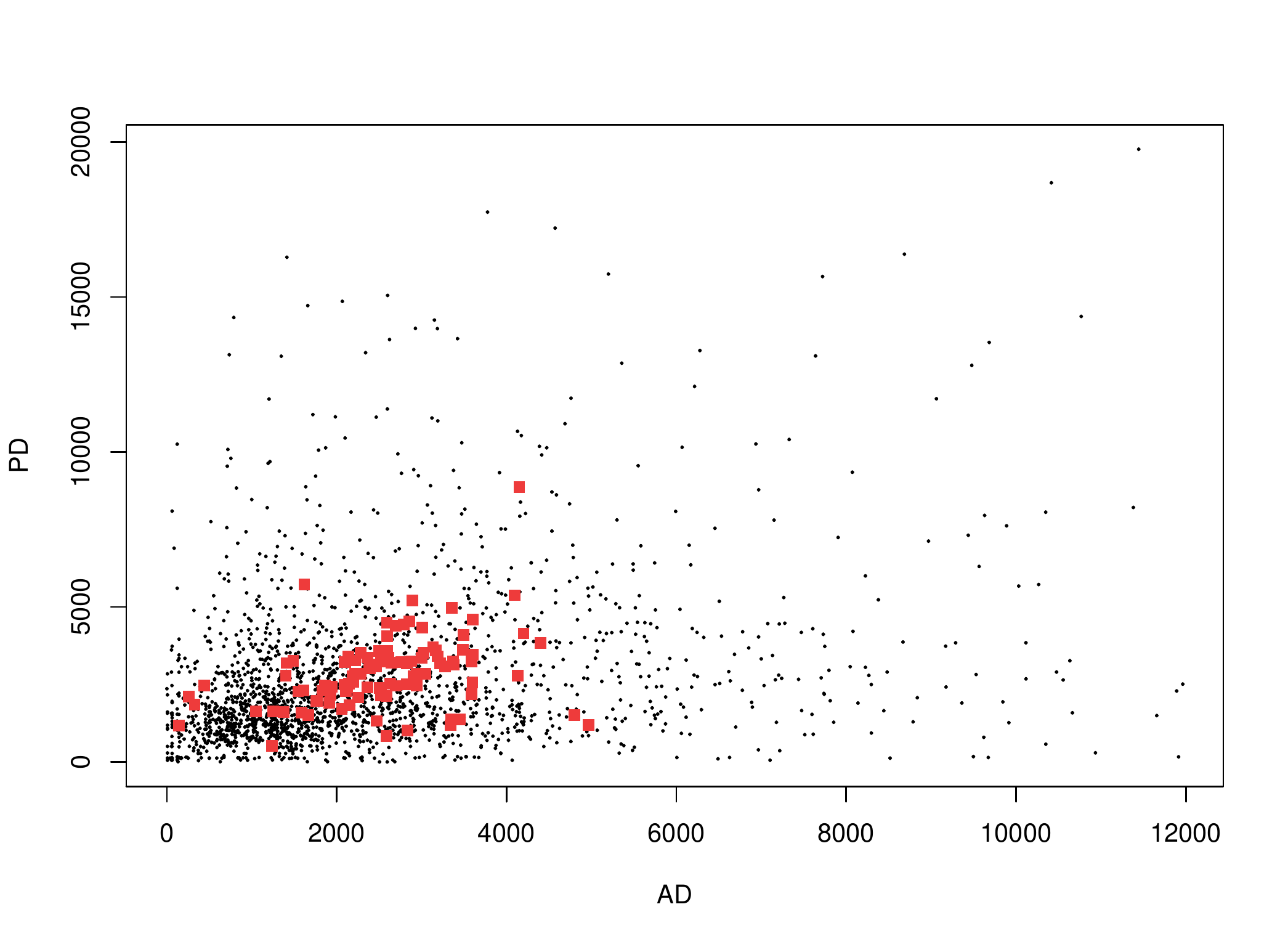}
	\caption{Fitted values of accidental damage (AD) and property damage (PD) using the selected VVV model with $G=4$. The given observations are plotted using a small black ``$\centerdot$", while the predictions are plotted using a red ``$ \blacksquare$" }
	\label{fig:realdata_bestmodel_fittedvalues}
\end{figure}

Figure~\ref{fig:realdata_bestmodel_fittedvalues} shows the final fitted values of the model across all components.
The fitted values cover a wide range of values and capture the very small and large claim amounts. However, they did not capture the very large claims, such as when policies have very large AD claim amounts greater than $6000$ or large PD amounts greater $10,000$.
One potential reason is the limited number of covariates used or available to use. In each of the expert networks only two covariates are used, which are categorical variables with only a limited number of levels. Hence the model might not be able to capture a lot of variation within each component. 
More importantly, it could be because the data are long tailed, with only a few claims having large AD or PD amounts, especially when AD and PD are greater than $8000$.

\begin{figure}[ht]
	\centering
	\begin{minipage}{.47\textwidth}
		\centering
		\includegraphics[width=1\linewidth, height=1\linewidth]{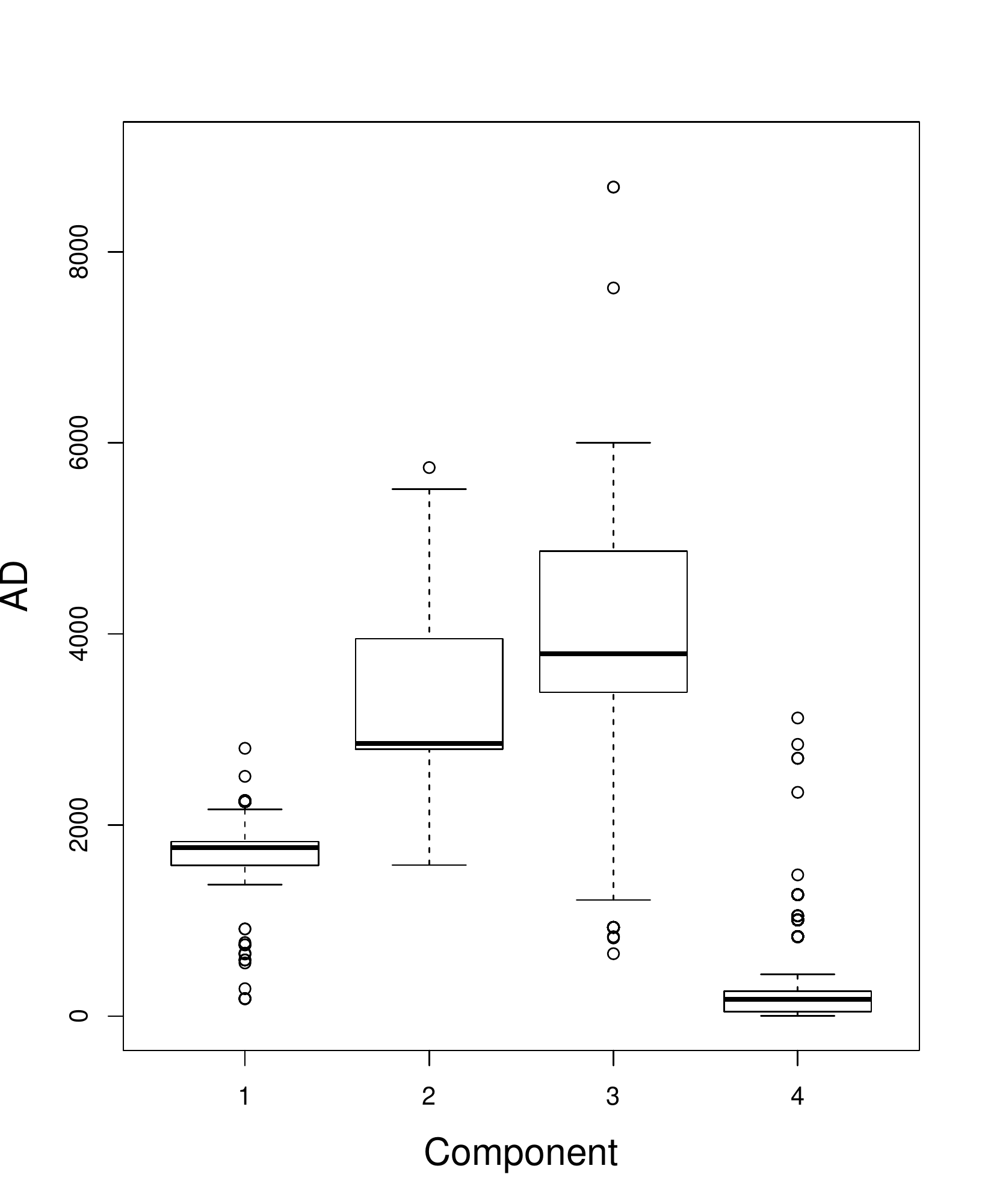}
		\caption*{(a)}
	\end{minipage}
	\begin{minipage}{0.47\textwidth}
		\centering
		\includegraphics[width=1\linewidth, height=1\linewidth]{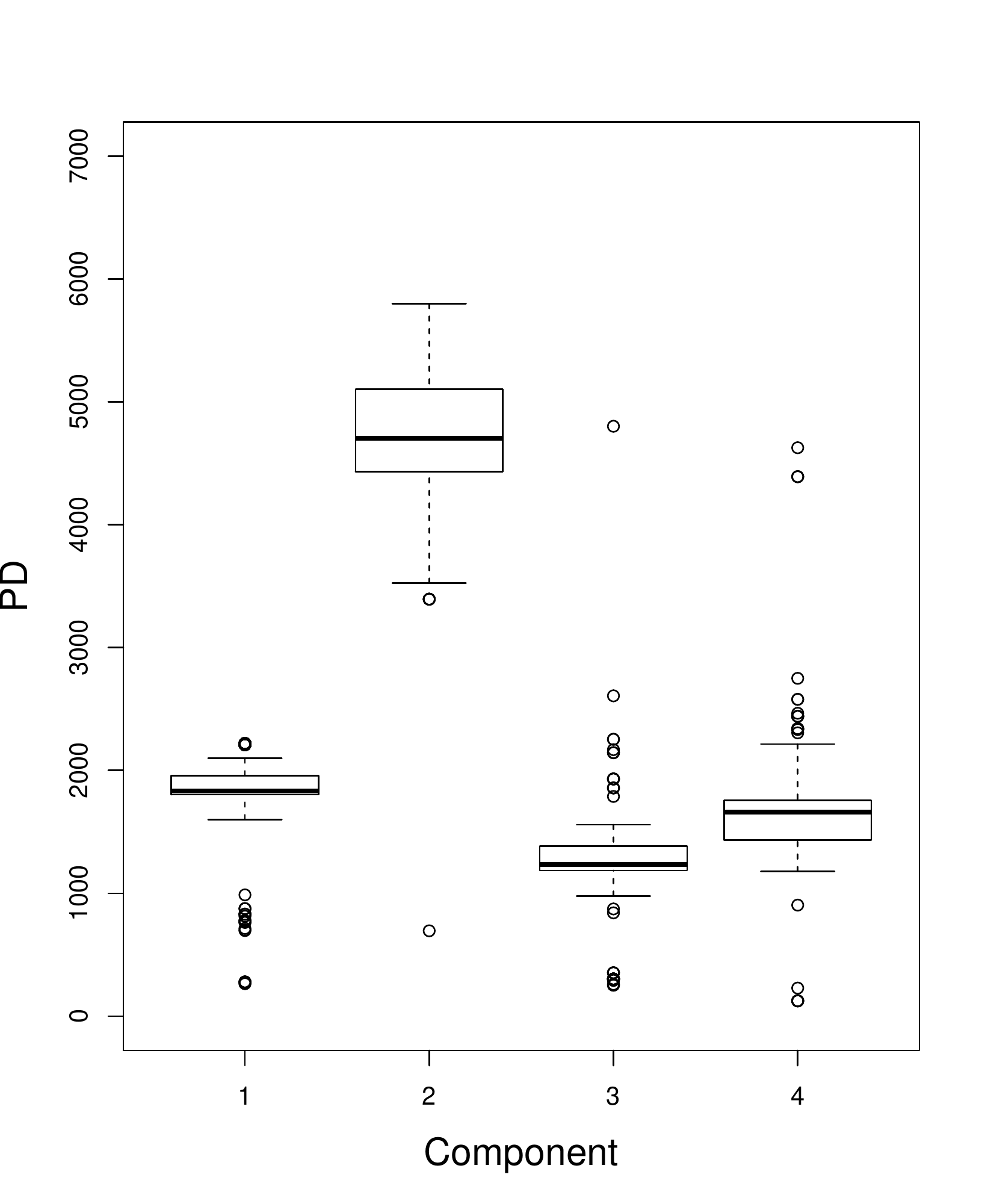}
		\caption*{(b)}
	\end{minipage}
	\caption{Boxplots of the fitted values within each component of the optimal model for the two claim types before mixing: (a) shows accidental damage (AD); (b) shows property damage (PD).}
	\label{fig:realdata_boxplot_fittedvalues_cluster}
\end{figure}

Figure~\ref{fig:realdata_boxplot_fittedvalues_cluster} presents the boxplot of the fitted values within each component before mixing for a clearer picture of comparisons among components. 
It shows that components 2 and 3 have mostly higher values of AD, while component 2 also has mostly high risk on PD. These are consistent with Figure~\ref{fig:realdata_bestmodel_classification} in that component 2 captures mostly large AD and PD claims. Component 1 has medium risk on both AD and PD, while component 4 has very small AD risk but slightly higher PD risk. 
While Figure~\ref{fig:realdata_boxplot_fittedvalues_cluster} shows large variations of predictions among different components,
it is noted that, in the gating network, the fitted mixing probabilities mostly range between 40\% and 70\% to classify observations to each component. Hence, after mixing, the overall predictions are pulled away from the within-component predictions. 
This is mainly due to the overlapping nature of the components, and possibly lack of additional covariates in the data. If the clusters are more separated, or extra information (covariates) is included in the gating network, which may lead to the mixing probabilities being closer to 0 or 1 (i.e. the components are better separated by the MoE model), then predictions could be expected to perform even better.

\begin{figure}[H]
	\centering
	\begin{minipage}{0.6\textwidth}
		\centering
		\includegraphics[width=\linewidth, height=.6\linewidth]{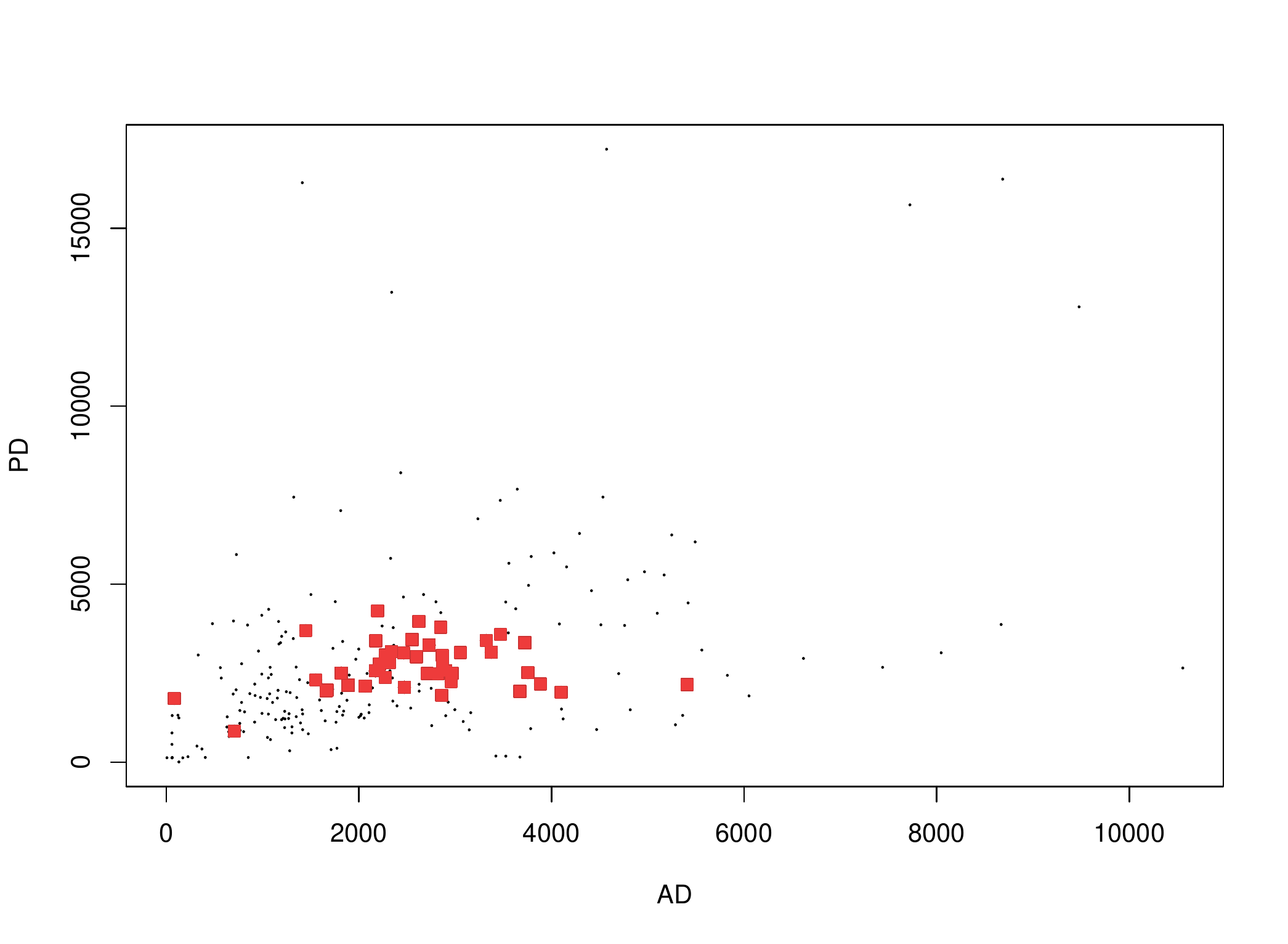}
		\caption*{(a)}
	\end{minipage}
	\begin{minipage}{0.6\textwidth}
		\centering
		\includegraphics[width=\linewidth, height=.6\linewidth]{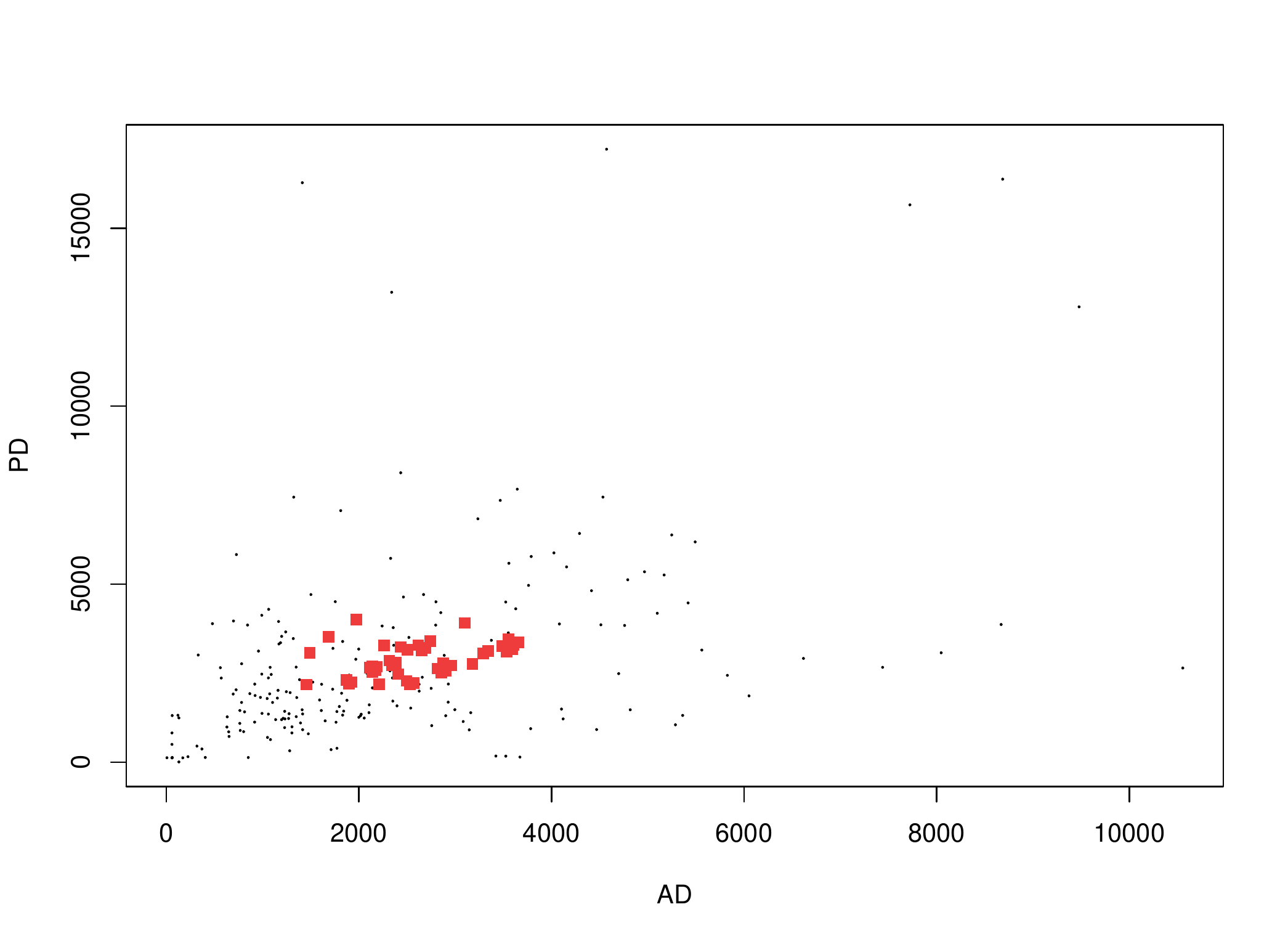}
		\caption*{(b)}
	\end{minipage}
	\caption{Predicted values of accidental damge (AD) and property damage (PD) for the test data set: (a) is the result of the selected optimal MoE model; (b) is the result of the standard univariate GLMs. The true values are plotted in small black ``$\centerdot$", while the predictions are plotted in red ``$\blacksquare$". }
	\label{fig:realdata_testset_prediction}
\end{figure}

For predictions, the data set is split into a training set (N=1874) and a test set (N=200). Figure~\ref{fig:realdata_testset_prediction}(a) shows the predicted values of AD and PD on the test set, while the model is trained using the training set with the previously selected optimal model.
Similarly as in the simulation studies, standard univariate GLMs are fitted for AD and PD respectively based on the training set and predicted on the test set, using the same set of selected covariates. The predictions are shown in Figure~\ref{fig:realdata_testset_prediction}(b). By comparing the two figures, it is clear that the MoE model predictions outperform the GLM predictions, with the MoE predictions spanning a wide range and predicting very small claim amounts and large claim amounts for AD, which can be interpreted as the different dependence structures in the data set being captured. 
To assess the prediction improvements, various metrics are used, as in Section~\ref{section:simulationI} and~\ref{section:simulationII} for the simulation studies. They evaluate the predictive distributions of the model compared with the empirical distributions calculated using the observations. The results are shown in Table~\ref{tab:realdata_prediction_comparison}.
The primary focus for this table is the sum of AD and PD, because it represents the overall total risk of policyholders, and traditionally different perils would be modeled independently, and their sum taken as the total risk. The purpose of the MoE method is that total claim amounts be better predicted when taking the dependence structure into consideration.   
The MoE model outperforms the GLM in all categories in this case, and the improvement is prominent.
It is also expected that, by modeling two risks simultaneously and considering their dependency, the understanding and predictions of each marginal risk are also improved. For each margin, Table~\ref{tab:realdata_prediction_comparison} shows strong improvement, except when considering PD, where rMSE and Gini index are slightly worse, while the rest of the metrics show a clear improvement. It is consistent with Figure~\ref{fig:realdata_testset_prediction}, since in the plot the range of predicted PD claims is similar for both models. This could possibly be due to the fact that PD claims may be affected by other factors not available in the selected data.
Overall it shows that by taking the covariance into account when modeling two risks simultaneously via the bivariate gamma distribution, and taking the heterogeneity into account for enhanced covariance, it leads to better model predictions and improved understanding of the data structure.  

\begin{table}[ht]
	\centering
	\caption{Comparisons between the MoE model and the GLM model for predictive performance regarding accidental damage (AD) and property damage (PD) using the mean of continuous ranked probability score (CRPS), square root mean squared errors (rMSE), Gini index and Wasserstein distance, as discussed before. The underlined values represent the optimal resutls for each measure.}
	\label{tab:realdata_prediction_comparison}
	\resizebox{.65\textwidth}{!}{
	\begin{tabular}{c|l|rrrr}
		\toprule[.15 em]
		\ & \ & CRPS & rMSE & Gini & Wasserstein \\
		\midrule
		\multirow{2}{*}{Sum of AD and PD} & MoE & \underline{2421.38} & \underline{3966.90} & \underline{0.54}  & \underline{2159.26} \\
		& GLM & 2466.63  & 3998.07 & 0.52 & 2216.94 \\
		\midrule
		\multirow{2}{*}{AD} & MoE & \underline{1203.19} & \underline{1854.32} & \underline{0.55} & \underline{1002.13} \\
		& GLM & 1226.90 & 1867.72 & 0.54 & 1062.82 \\
		\midrule
		\multirow{2}{*}{PD} & MoE & \underline{1559.13} & 2786.93 & 0.52 & \underline{1431.44} \\
		& GLM & 1600.67 & \underline{2786.24} & \underline{0.53} & 1481.59 \\
		\bottomrule[0.15 em]
	\end{tabular} }
\end{table}

\section{Conclusion}
\label{section:conclusion}

This work investigated the application of bivariate gamma distributions in the framework of mixture of experts models for insurance claims modeling. More specifically, a finite mixture of bivariate gamma regressions within a GLM framework is investigated, focusing on predicting future claims. 
In doing so, not only is covariance taken into account in regressions for simultaneously modeling of two risks, 
but different claim patterns could be identified with different dependence structures targeting the heterogeneity issue. This leads to amplified within-component dependencies even when the overall dependence of all data is weak.
The bivariate gamma distribution is a natural extension of the univariate gamma GLM when independence between different risk perils is revoked.
It has been shown that claim predictions could be improved for both simulation data sets and the real Irish GI insurer data by using the proposed method. This is done by assessing the sum of the two margins or assessing individual margins independently, since traditionally in GI claim modeling each category is modeled independently and the sum of all categories is taken as the total risk. 

This method could be viewed from two perspectives - claim clustering and claim prediction by regression - both taking the dependence structure into consideration. 
By viewing it from a clustering perspective, given the data are bivariate gamma distributed (which is typically the case for GI claim severity modeling), model-based clustering with the bivariate gamma distributions can segregate the policyholders with different claim-making behaviours into different groups, hence the dependence structure is enhanced and better captured. Furthermore, by including covariates in the clustering process, the result is further improved. 
By viewing it from a regression prediction perspective, the model is essentially equivalent to a mixture of regressions by fitting multiple regressions that model different patterns in the data while combining the different regression lines via a logistic regression at the same time.

Within the bivariate gamma MoE model family, various parameterisations have been proposed to facilitate parsimonious model selection, which allows covariates to enter all or some of the expert networks and the gating network. 
It is also noted that, because covariates entering into different networks may be different, it can lead to interpretability difficulties.
The selection of covariates should also be done carefully, as the effect of the number of covariates on increasing the potential model space is huge, and it will lead to more computational complexity in practice. 
Furthermore, with $G$ increasing, adding one covariates will lead to $G$ times more parameters in the model, hence stepwise selection will often prefer a simpler model.  
Although the simplest option could be that all covariates are included in all parts, this could lead to model parsimony issues. 
We note that variable selection cannot be done by hypothesis testing within each GLM structure involved, hence model selection should be carefully implemented to facilitate easier and better model interpretation. 
If model parsinony is not of primary concern, another possible solution is to utilise logical or expert intuition rather than strictly following a stepwise model selection approach. 
For example, in \cite{Bermudez2012} $G$ is logically chosen based on the characteristics of their insurance claim frequency count data. 

One issue with this bivariate gamma distribution is the computational complexity of its density function, which has prohibited its widespread use in the past. 
When extending to higher dimensional cases as in Section~\ref{section:mvGamma}, the computation becomes even more complex because it requires multiple numerical integrations depending on specification of the dependence structures. 
Further exploration on the higher dimensional cases could constitute future work, in which case the model selection will also involve selecting different dependence structure assumptions. 
Furthermore, since the sufficient statistics of the gamma distribution are $X$ and $\log X$, by implementing the maximum likelihood approach using the EM algorithm, even more computational complexity is introduced, which involves numerical integrations and numerical optimisations in both E and M-steps.
When the tolerance in the EM algorithm is too small, the complexity of numerical computations might sometimes not be able to guarantee the monotonicity of the log-likelihood towards the very end of the algorithm. 
It creates potential obstacles in the implementation of this model including computational stability, accuracy and speed. Parameter tuning should be carefully implemented to ensure the numerical calculations always achieve the best possible results.

In the Irish GI insurer section, the heavily overlapped components cause much of the uncertainty on cluster membership, and there are scarce numbers of very large claims with associated claims data, both affecting the final predictions of the model.
In this work an Expectation/Conditional Maximization algorithm (ECM) (\citealp{Meng1993}) is used for the maximum likelihood approach.
Other variations of the EM algorithm could be adapted such as Classification EM (CEM) (\citealp{Celeux1992}), which may give less fuzzy clustering results and faster convergence.  
Alternatively, if only regarding the proposed method as a clustering approach (i.e. to segregate data for amplified dependence structure), 
after clustering, separate regressions (either bivariate gamma regressions or univariate GLMs) could be fitted within each component independently to improve the model predictions.
In this case, for better clustering results, potentially one can either treat the extreme large claims as outliers and use trimming techniques from the clustering literature, or add an extra component capturing random noise. This could also consitute future work.

If stepwise selection is used, with the number of covariates and $G$ increasing, the potential model space can be huge. It may be reasonable to expect that there are many models with similar goodness-of-fit and predictive power. This could suggest a model averaging approach as future work, similar to the works of \cite{Wei2015}, \cite{Russell2015}, \cite{Hu2018}, where either clustering results, regression predictions or categorical variables' construction can be merged.

\section*{Acknowledgements}
This work was supported by the Science Foundation Ireland funded Insight Research Centre (SFI/12/RC/2289).

\bibliographystyle{wb_stat}
\bibliography{Ref_mvClaims}


\newpage
\appendix
\section*{Appendix}
\section{Proof of $\mathbb{E}(\underline{Y})$, $Var(\underline{Y})$ and $Cov(\underline{Y})$.}
\label{App:bivGamma_covariance_proof}

The bivariate gamma distribution can be defined as $\boldsymbol{Y}=\boldsymbol{AX}$, where $\boldsymbol{X}=(X_1, X_2, X_3)^{\top}$,
$$\boldsymbol{A}=\begin{bmatrix}
1 & 0 & 1 \\
0 & 1 & 1 \\
\end{bmatrix} .
$$
Then the conditional expectation and variance-covariance matrix of $\boldsymbol{Y}$ are
\begin{equation*}
\begin{split}
\mathbb{E}[\boldsymbol{Y}|\boldsymbol{\alpha},\beta] &= \mathbb{E}[\boldsymbol{AX}] = \boldsymbol{A\theta} , \\
Var[\boldsymbol{Y}|\boldsymbol{\alpha},\beta] &= \boldsymbol{A\Sigma A^{\top}},
\end{split}
\end{equation*}
where $\boldsymbol{\theta}=(\alpha_1/\beta,\alpha_2/\beta, \alpha_3/\beta)^{\top}, \boldsymbol{\Sigma} = \diag(\alpha_1/\beta^2, \alpha_2/\beta^2, \alpha_3/\beta^2), \boldsymbol{\alpha}=(\alpha_1,\alpha_2,\alpha_3)^{\top}.$
When $\boldsymbol{Y}$ follows a finite mixture of bivariate gamma distributions, the unconditional expectation of $\boldsymbol{Y}$ is 
\begin{equation*}
\mathbb{E}[\boldsymbol{Y}] =\mathbb{E}[ \mathbb{E}[\boldsymbol{Y}|\boldsymbol{\alpha},\boldsymbol{\beta}]] =  \boldsymbol{A}\mathbb{E}[\boldsymbol{\theta}] = \boldsymbol{A} \begin{bmatrix}
\mathbb{E}[\boldsymbol{\alpha}_{1} /\boldsymbol{\beta}] \\
\mathbb{E}[\boldsymbol{\alpha}_{2} /\boldsymbol{\beta}] \\
\mathbb{E}[\boldsymbol{\alpha}_{3} /\boldsymbol{\beta}] \\
\end{bmatrix} =
\boldsymbol{A} \begin{bmatrix}
\sum_{g} \tau_g  \alpha_{1g} /\beta_g \\
\sum_{g} \tau_g  \alpha_{2g} /\beta_g \\
\sum_{g} \tau_g  \alpha_{3g} /\beta_g \\
\end{bmatrix} =
\begin{bmatrix}
\sum_{g} \tau_g  \frac{\alpha_{1g}+\alpha_{3g}}{\beta_g} \\
\sum_{g} \tau_g  \frac{\alpha_{2g}+\alpha_{3g}}{\beta_g} \\
\end{bmatrix} ,
\end{equation*}
where $\boldsymbol{\alpha}=(\boldsymbol{\alpha}_1, \boldsymbol{\alpha}_2, \boldsymbol{\alpha}_3)^{\top}$, $\boldsymbol{\alpha}_1 = (\alpha_{1,g=1},\ldots,\alpha_{1,g=G})^{\top}$,
$\boldsymbol{\beta} = (\beta_{g=1},\ldots,\beta_{g=G})^{\top}$. 
To calculate the unconditional variance of $\boldsymbol{Y}$, first the second moment of $\boldsymbol{Y}$ conditional on all $\boldsymbol{\alpha}, \boldsymbol{\beta}$ is
\begin{equation*}
\begin{split}
\mathbb{E}[\boldsymbol{YY^{\top}}|\boldsymbol{\alpha}, \boldsymbol{\beta}] &= Var[\boldsymbol{Y}|\boldsymbol{\alpha}, \boldsymbol{\beta}] + \mathbb{E}[\boldsymbol{Y} |\boldsymbol{\alpha}, \boldsymbol{\beta}] [\mathbb{E}[\boldsymbol{Y} |\boldsymbol{\alpha}, \boldsymbol{\beta}]]^{\top} \\
&= \boldsymbol{A\Sigma A^{\top}} + \boldsymbol{A\theta} [\boldsymbol{A\theta}]^{\top} \\
&= \boldsymbol{A} \boldsymbol{B} (\boldsymbol{\alpha}, \boldsymbol{\beta}) \boldsymbol{A}^{\top} ,
\end{split}
\end{equation*}
where 
\begin{equation*}
\boldsymbol{B} (\boldsymbol{\alpha}, \boldsymbol{\beta}) = \boldsymbol{\Sigma} +\boldsymbol{\theta\theta^{\top}} = \begin{bmatrix}
\frac{\boldsymbol{\alpha}_1^2}{\boldsymbol{\beta}^2} + \frac{\boldsymbol{\alpha}_1}{\boldsymbol{\beta}^2} & \frac{\boldsymbol{\alpha}_1 \boldsymbol{\alpha}_2}{\boldsymbol{\beta}^2} & \frac{\boldsymbol{\alpha}_1 \boldsymbol{\alpha}_3}{\boldsymbol{\beta}^2} \\
\frac{\boldsymbol{\alpha}_1 \boldsymbol{\alpha}_2}{\boldsymbol{\beta}^2} & \frac{\boldsymbol{\alpha}_2^2}{\boldsymbol{\beta}^2}+\frac{\boldsymbol{\alpha}_2}{\boldsymbol{\beta}^2} & \frac{\boldsymbol{\alpha}_2 \boldsymbol{\alpha}_3}{\boldsymbol{\beta}^2} \\
\frac{\boldsymbol{\alpha}_1 \boldsymbol{\alpha}_3}{\boldsymbol{\beta}^2} & \frac{\boldsymbol{\alpha}_2 \boldsymbol{\alpha}_3}{\boldsymbol{\beta}^2} & \frac{\boldsymbol{\alpha}_3^2}{\boldsymbol{\beta}^2}+\frac{\boldsymbol{\alpha}_3}{\boldsymbol{\beta}^2}
\end{bmatrix} .
\end{equation*}
This shows that the conditional second moment is polynomial with respect to $\boldsymbol{\alpha}$ and $\boldsymbol{\beta}$. Hence $\mathbb{E}[\boldsymbol{YY^{\top}}] = \mathbb{E}[\mathbb{E}[\boldsymbol{YY^{\top}}|\boldsymbol{\alpha}, \boldsymbol{\beta}]]$ by the law of total expectation. The element-wise expectation of the matrix $\boldsymbol{B}(\boldsymbol{\alpha}, \boldsymbol{\beta})$ is denoted as $\mathbb{E}[\boldsymbol{B}]$, then $\mathbb{E}[\boldsymbol{YY^{\top}}] = \boldsymbol{A} \mathbb{E}[\boldsymbol{B}] \boldsymbol{A}^{\top}$, and the variance-covariance matrix of $\boldsymbol{Y}$ is
\begin{equation*}
\begin{split}
Var(\boldsymbol{Y}) &= \mathbb{E}[\boldsymbol{YY^{\top}}] - \mathbb{E}[\boldsymbol{Y}][\mathbb{E}[\boldsymbol{Y}]]^{\top} \\
&= \boldsymbol{A} \mathbb{E}[\boldsymbol{B}] \boldsymbol{A}^{\top} -  \boldsymbol{A}\mathbb{E}[\boldsymbol{\theta}][\boldsymbol{A}\mathbb{E}[\boldsymbol{\theta}]]^{\top} \\
&= \boldsymbol{A} \left[\mathbb{E}[\boldsymbol{B}] - \mathbb{E}[\boldsymbol{\theta}]\mathbb{E}[\boldsymbol{\theta}]^{\top} \right] \boldsymbol{A}^{\top} \\
&= \boldsymbol{A} \boldsymbol{D}(\boldsymbol{\alpha}, \boldsymbol{\beta}) \boldsymbol{A}^{\top} ,
\end{split}
\end{equation*}
where 
\begin{equation*}
\boldsymbol{D}(\boldsymbol{\alpha}, \boldsymbol{\beta}) = \begin{bmatrix}
Var(\frac{\boldsymbol{\alpha}_1}{\boldsymbol{\beta}}) + \mathbb{E}(\frac{\boldsymbol{\alpha}_1}{\boldsymbol{\beta}}) & Cov(\frac{\boldsymbol{\alpha}_1}{\boldsymbol{\beta}}, \frac{\boldsymbol{\alpha}_2}{\boldsymbol{\beta}}) & Cov(\frac{\boldsymbol{\alpha}_1}{\boldsymbol{\beta}}, \frac{\boldsymbol{\alpha}_3}{\boldsymbol{\beta}}) \\
Cov(\frac{\boldsymbol{\alpha}_1}{\boldsymbol{\beta}}, \frac{\boldsymbol{\alpha}_2}{\boldsymbol{\beta}}) & Var(\frac{\boldsymbol{\alpha}_2}{\boldsymbol{\beta}}) + \mathbb{E}(\frac{\boldsymbol{\alpha}_2}{\boldsymbol{\beta}})  & Cov(\frac{\boldsymbol{\alpha}_2}{\boldsymbol{\beta}}, \frac{\boldsymbol{\alpha}_3}{\boldsymbol{\beta}}) \\
Cov(\frac{\boldsymbol{\alpha}_1}{\boldsymbol{\beta}}, \frac{\boldsymbol{\alpha}_3}{\boldsymbol{\beta}}) & Cov(\frac{\boldsymbol{\alpha}_2}{\boldsymbol{\beta}}, \frac{\boldsymbol{\alpha}_3}{\boldsymbol{\beta}}) & Var(\frac{\boldsymbol{\alpha}_3}{\boldsymbol{\beta}}) + \mathbb{E}(\frac{\boldsymbol{\alpha}_3}{\boldsymbol{\beta}})  
\end{bmatrix} .
\end{equation*}
Hence
\begin{equation*}
\begin{split}
Var(\boldsymbol{Y}_1) &=
Var(\frac{\boldsymbol{\alpha}_1}{\boldsymbol{\beta}}) + \mathbb{E}(\frac{\boldsymbol{\alpha}_1}{\boldsymbol{\beta}}) + Var(\frac{\boldsymbol{\alpha}_3}{\boldsymbol{\beta}}) + \mathbb{E}(\frac{\boldsymbol{\alpha}_3}{\boldsymbol{\beta}}) + 2Cov(\frac{\boldsymbol{\alpha}_1}{\boldsymbol{\beta}}, \frac{\boldsymbol{\alpha}_3}{\boldsymbol{\beta}}) \ , \\
Var(\boldsymbol{Y}_2) &=
Var(\frac{\boldsymbol{\alpha}_2}{\boldsymbol{\beta}}) + \mathbb{E}(\frac{\boldsymbol{\alpha}_2}{\boldsymbol{\beta}}) + Var(\frac{\boldsymbol{\alpha}_3}{\boldsymbol{\beta}}) + \mathbb{E}(\frac{\boldsymbol{\alpha}_3}{\boldsymbol{\beta}}) + 2Cov(\frac{\boldsymbol{\alpha}_2}{\boldsymbol{\beta}}, \frac{\boldsymbol{\alpha}_3}{\boldsymbol{\beta}}) \ , \\
Cov(\boldsymbol{Y}_1, \boldsymbol{Y}_2) &= Cov(\frac{\boldsymbol{\alpha}_1}{\boldsymbol{\beta}}, \frac{\boldsymbol{\alpha}_2}{\boldsymbol{\beta}}) + Cov(\frac{\boldsymbol{\alpha}_1}{\boldsymbol{\beta}}, \frac{\boldsymbol{\alpha}_3}{\boldsymbol{\beta}}) + Cov(\frac{\boldsymbol{\alpha}_2}{\boldsymbol{\beta}}, \frac{\boldsymbol{\alpha}_3}{\boldsymbol{\beta}}) + Var(\frac{\boldsymbol{\alpha}_3}{\boldsymbol{\beta}}) + \mathbb{E}(\frac{\boldsymbol{\alpha}_3}{\boldsymbol{\beta}}) \ .
\end{split}
\end{equation*}

\newpage
\section{Distribution estimation}
\label{App:DistributionEstimation}

This section introduces distribution estimation for the bivariate gamma distribution using the EM algorithm for likelihood maximisation.
Defined as in Equation~(\ref{eq:bivgamma_def}), the distribution has one latent variable $X_3$, while $X_1, X_2, X_3$ are independent. 
Therefore, the complete data log-likelihood is
\begin{equation*}
\ell_c = \sum_{i=1}^{N} \log p(x_3, y_{1i}, y_{2i};\theta) ,
\end{equation*}
and the conditional expectation of the log-likelihood in the E-step at the t$^{th}$ iteration is
\begin{equation*}
Q(\theta^{(t+1)}|\theta^{(t)}) = \mathbb{E}\left[ \sum_{i=1}^{N} \log p(x_3, y_{1i}, y_{2i};\theta^{(t+1)}) \mid y_1, y_2, \theta^{(t)} \right] .
\end{equation*}
The sufficient statistics are $X_3, \log(X_3), \log(y_1 - X_3), \log(y_2 - X_3)$. The EM algorithm is as follows:

\vskip .5cm
\noindent\textbf{\underline{E-step}}
\begin{equation*}
\begin{split}
\hat{x}_{3i}^{(t+1)} &= \mathbb{E}(X_{3i}|y_{1i}, y_{2i}; \hat{\theta}^{(t)}) = \frac{\frac{\hat{\alpha}_{3}^{(t)}}{\hat{\beta}^{(t)}} p(y_{1i}, y_{2i}; \hat{\alpha}_{1}^{(t)}, \hat{\alpha}_{2}^{(t)}, \hat{\alpha}_{3}^{(t)}+1, \hat{\beta}^{(t)})}{p(y_{1i}, y_{2i}; \hat{\alpha}_{1}^{(t)}, \hat{\alpha}_{2}^{(t)}, \hat{\alpha}_{3}^{(t)}, \hat{\beta}^{(t)})} , \\
\hat{x}_{1i}^{(t+1)} &= \mathbb{E}(X_{1i}|y_{1i}, y_{2i}; \hat{\theta}^{(t)}) = \mathbb{E}(y_{1i}-X_{3i}|y_{1i}, y_{2i}; \hat{\theta}^{(t)}) = y_1 - \hat{x}_{3i}^{(t+1)} , \\
\hat{x}_{2i}^{(t+1)} &= \mathbb{E}(X_{2i}|y_{1i}, y_{2i}; \hat{\theta}^{(t)})=\mathbb{E}(y_{2i}-X_{3i}|y_{1i}, y_{2i}; \hat{\theta}^{(t)}) = y_2 - \hat{x}_{3i}^{(t+1)} , \\
\widehat{\log x}_{3i}^{(t+1)} &= \mathbb{E}(\log X_{3i}|y_{1i}, y_{2i}; \hat{\theta}^{(t)}) =\frac{\int_0^{\min(y_{1i},y_{2i})} \log x_{3i} p(x_{3i}, y_{1i}, y_{2i};\hat{\theta}^{(t)})dx_3}{p(y_{1i}, y_{2i}; \hat{\theta}^{(t)})} , \\
\widehat{\log x}_{1i}^{(t+1)} &= \mathbb{E}(\log(y_{1i}-X_{3i})|y_{1i}, y_{2i}; \hat{\theta}^{(t)}) =\frac{\int_0^{\min(y_{1i},y_{2i})} \log(y_{1i}-x_{3i}) p(x_{3i}, y_{1i}, y_{2i};\hat{\theta}^{(t)})dx_3}{p(y_{1i}, y_{2i}; \hat{\theta}^{(t)})} , \\
\widehat{\log x}_{2i}^{(t+1)} &= \mathbb{E}(\log(y_{2i}-X_{3i})|y_{1i}, y_{2i}; \hat{\theta}^{(t)}) = \frac{\int_0^{\min(y_{1i},y_{2i})} \log(y_{2i}-x_{3i}) p(x_{3i}, y_{1i}, y_{2i};\hat{\theta}^{(t)})dx_3}{p(y_{1i}, y_{2i}; \hat{\theta}^{(t)})} . \\
\end{split}
\end{equation*}

\vskip .3cm
\noindent\textbf{\underline{M-step}} \\
\vskip .2cm
For $\hat{\alpha}_{1}^{(t+1)}, \hat{\alpha}_{2}^{(t+1)}, \hat{\alpha}_{3}^{(t+1)}$, numerically solve the following equations respectively:
\begin{equation*}
\begin{split}
N\log \hat{\beta}^{(t)} - N \frac{\Gamma^{\prime} (\alpha_{1})}{\Gamma(\alpha_{1})} + \sum_{1=i}^{N} \widehat{\log x}_{1i}^{(t+1)} &= 0 , \\
N\log \hat{\beta}^{(t)} - N\frac{\Gamma^{\prime}(\alpha_{2})}{\Gamma(\alpha_{2})} + \sum_{1=i}^{N} \widehat{\log x}_{2i}^{(t+1)} &= 0 , \\
N\log \hat{\beta}^{(t)} - N\frac{\Gamma^{\prime} (\alpha_{3})}{\Gamma(\alpha_{3})} + \sum_{1=i}^{N} \widehat{\log x}_{3i}^{(t+1)} &= 0 .
\end{split}
\end{equation*}

\begin{equation*}
\hat{\beta}^{(t+1)} = \frac{N(\hat{\alpha}_{1}^{(t+1)}+\hat{\alpha}_{2}^{(t+1)}+\hat{\alpha}_{3}^{(t+1)})}{\sum_{i=1}^{N}  (\hat{x}_{1i}^{(t+1)}+\hat{x}_{2i}^{(t+1)}+\hat{x}_{3i}^{(t+1)})} .
\end{equation*}

\newpage
\section{EM algorithms for all model types within bivariate gamma MoE model family}
\label{App:bivGamma_MoE_family}

For the EM algorithm for all model types within the bivariate gamma MoE model family as shown in Table~\ref{tab:bivGamma_MoE_family}, the E-step is always the same, except for the parameterisations of $\alpha_{1}$, $\alpha_{2}$, $\alpha_{3}$ and $\beta$. Note that the gating network derivation is the same for all model types as shown in Section~\ref{section:MoE} of the main text, hence is not shown here.
The ``II" model type is estimating one bivarite gamma distribution parameters, which has been shown in Appendix~\ref{App:DistributionEstimation}. The ``$\ast$CC", ``$\ast$CI" ``$\ast$IC" types can be easily derived similarly based on this section and Appendix~\ref{App:DistributionEstimation}. 
At the t$^{th}$ iteration, the E-step and M-step are:\\  

\noindent\textbf{\underline{E-step}}
\begin{equation*}
\begin{split}
\hat{z}_{ig}^{(t+1)} &= \mathbb{E}(z_{ig}|y_{1i},y_{2i}, \hat{\theta}_{g}^{(t)}(\boldsymbol{w}_i)) \\
\hat{x}_{3ig}^{(t+1)} &=\mathbb{E}(X_{3ig}|y_{1i}, y_{2i}; \hat{\theta}_{g}^{(t)}(\boldsymbol{w}_i)) \\
\hat{x}_{1ig}^{(t+1)} &=\mathbb{E}(y_{1i}-X_{3ig}|y_{1i}, y_{2i}; \hat{\theta}_{g}^{(t)}(\boldsymbol{w}_i)) = y_{1i} - \hat{x}_{3ig}^{(t+1)} \\
\hat{x}_{2ig}^{(t+1)} &=\mathbb{E}(y_{2i}-X_{3ig}|y_{1i}, y_{2i}; \hat{\theta}_{g}^{(t)}(\boldsymbol{w}_i)) = y_{2i} - \hat{x}_{3ig}^{(t+1)} \\
\widehat{\log x}_{3ig}^{(t+1)} &= \mathbb{E}(\log X_{3ig}|y_{1i}, y_{2i}; \hat{\theta}_{g}^{(t)}(\boldsymbol{w}_i))\\
\widehat{\log x}_{1ig}^{(t+1)} &= \mathbb{E}(\log(y_{1i}-X_{3i})|y_{1i}, y_{2i}; \hat{\theta}_{g}^{(t)}(\boldsymbol{w}_i)) \\ 
\widehat{\log x}_{2ig}^{(t+1)} &= \mathbb{E}(\log(y_{2i}-X_{3i})|y_{1i}, y_{2i}; \hat{\theta}_{g}^{(t)}(\boldsymbol{w}_i))
\end{split}
\end{equation*}	

\noindent\textbf{\underline{M-step}}
\vskip .3cm 
\begin{itemize}
	\item \textbf{``$\ast$VC" model type:} \\	
	Update $\hat{\boldsymbol{\gamma}}_{kg}^{(t+1)}$ ($k=1,2,3$):
	\begin{equation*}
	\begin{split}
	\hat{\boldsymbol{\gamma}}_{kg}^{(t+1)} = \underset{\boldsymbol{\gamma}_{kg}}{\arg\max} \left(
	\sum_{i=1}^{N} \hat{z}_{ig}^{(t+1)} \exp(\boldsymbol{\gamma}_{kg}^{\top} \boldsymbol{w}_{ki}) \log \hat{\beta}_{g}^{(t)} \right. &- \sum_{i=1}^{N} \hat{z}_{ig}^{(t+1)} \log\Gamma( \exp(\boldsymbol{\gamma}_{kg}^{\top} \boldsymbol{w}_{ki})) \\
	&+ \left. \sum_{i=1}^{N} \hat{z}_{ig}^{(t+1)} \exp(\boldsymbol{\gamma}_{kg}^{\top} \boldsymbol{w}_{ki}) \widehat{\log x}_{kig}^{(t+1)} \right) .
	\end{split}
	\end{equation*} 
	Update $\hat{\beta}_{g}^{(t+1)}$: 
\begin{equation*}
\hat{\beta}_{g}^{(t+1)} = \frac{ \sum_{i=1}^{N} \hat{z}_{ig}^{(t+1)} \left( 
\hat{\alpha}_{1ig}^{(t+1)} + \hat{\alpha}_{2ig}^{(t+1)} + \hat{\alpha}_{3ig}^{(t+1)} \right) }{ \sum_{i=1}^{N} \hat{z}_{ig}^{(t+1)} \left( \hat{x}_{1ig}^{(t+1)} + \hat{x}_{2ig}^{(t+1)} + \hat{x}_{3ig}^{(t+1)} \right) } .
\end{equation*}
\item \textbf{``$\ast$CV" model type:} \\
Update $\hat{\alpha}_{kg}^{(t+1)}$ ($k=1,2,3$) by solving
\begin{equation*}
\sum_{i=1}^{N} \hat{z}_{ig}^{(t+1)} \log \beta_{ig}^{(t)} - \sum_{i=1}^{N} \hat{z}_{ig}^{(t+1)} \frac{\Gamma^{\prime}(\alpha_{kg})}{\Gamma(\alpha_{kg})} + \sum_{i=1}^{N} \hat{z}_{ig}^{(t+1)} \widehat{\log x}_{kig}^{(t+1)} = 0.
\end{equation*}
Update $\hat{\boldsymbol{\gamma}}_{4g}^{(t+1)}$: 
\begin{equation*}
\begin{split}
\hat{\boldsymbol{\gamma}}_{4g}^{(t+1)} = \underset{\boldsymbol{\gamma}_{4g}}{\arg\max} 
& \left(
(\hat{\alpha}_{1g}^{(t+1)} + \hat{\alpha}_{2g}^{(t+1)} + \hat{\alpha}_{3g}^{(t+1)}) \sum_{i=1}^{N}  \hat{z}_{ig}^{(t+1)}  (\boldsymbol{\gamma}_{4g}^{\top} \boldsymbol{w}_{4i}) \right. \\
& \hskip .5cm \left. - \sum_{i=1}^{N} \hat{z}_{ig}^{(t+1)} \exp(\boldsymbol{\gamma}_{4g}^{\top} \boldsymbol{w}_{4i}) (\hat{x}_{1ig}^{(t+1)} + \hat{x}_{2ig}^{(t+1)} + \hat{x}_{3ig}^{(t+1)}) \right) .
\end{split}
\end{equation*}
\item \textbf{``$\ast$VV" model type:} \\
Update $\hat{\boldsymbol{\gamma}}_{kg}^{(t+1)}$ (for $k=1,2,3$):
\begin{equation*}
\begin{split}
\hat{\boldsymbol{\gamma}}_{kg}^{(t+1)} = \underset{\boldsymbol{\gamma}_{kg}}{\arg\max} 
\left(
\sum_{i=1}^{N} \hat{z}_{ig}^{(t+1)} \exp(\boldsymbol{\gamma}_{kg}^{\top} \boldsymbol{w}_{ki}) \log \hat{\beta}_{ig}^{(t)} \right. &- \sum_{i=1}^{N} \hat{z}_{ig}^{(t+1)} \log\Gamma( \exp(\boldsymbol{\gamma}_{kg}^{\top} \boldsymbol{w}_{ki})) \\
&+ \left. \sum_{i=1}^{N} \hat{z}_{ig}^{(t+1)} \exp(\boldsymbol{\gamma}_{kg}^{\top} \boldsymbol{w}_{ki}) \widehat{\log x}_{kig}^{(t+1)}  \right) .
\end{split}
\end{equation*}
Update $\hat{\boldsymbol{\gamma}}_{4g}^{(t+1)}$:
\begin{equation*}
\begin{split}
\hat{\boldsymbol{\gamma}}_{4g}^{(t+1)} = \underset{\boldsymbol{\gamma}_{4g}}{\arg\max} 
& \left(
\sum_{i=1}^{N} \hat{z}_{ig}^{(t+1)} (\hat{\alpha}_{1ig}^{(t+1)} + \hat{\alpha}_{2ig}^{(t+1)} + \hat{\alpha}_{3ig}^{(t+1)}) (\boldsymbol{\gamma}_{4g}^{\top} \boldsymbol{w}_{4i}) \right. \\ 
& \hskip .5cm - \left. \sum_{i=1}^{N} \hat{z}_{ig}^{(t+1)} \exp(\boldsymbol{\gamma}_{4g}^{\top} \boldsymbol{w}_{4i}) (\hat{x}_{1ig}^{(t+1)} + \hat{x}_{2ig}^{(t+1)} + \hat{x}_{3ig}^{(t+1)} ) \right)  .
\end{split}
\end{equation*}
\item \textbf{``$\ast$VI" model type:} \\
Update $\hat{\boldsymbol{\gamma}}_{kg}^{(t+1)}$ ($k=1,2,3$): 
\begin{equation*}
\begin{split}
\hat{\boldsymbol{\gamma}}_{kg}^{(t+1)} = \underset{\boldsymbol{\gamma}_{kg}}{\arg\max} 
\left( 
\sum_{i=1}^{N} \hat{z}_{ig}^{(t+1)} \exp(\boldsymbol{\gamma}_{kg}^{\top} \boldsymbol{w}_{ki}) \log \hat{\beta}^{(t)} \right. &- \sum_{i=1}^{N} \hat{z}_{ig}^{(t+1)} \log\Gamma( \exp(\boldsymbol{\gamma}_{kg}^{\top} \boldsymbol{w}_{ki}))  \\ 
&+ \left. \sum_{i=1}^{N} \hat{z}_{ig}^{(t+1)} \exp(\boldsymbol{\gamma}_{kg}^{\top} \boldsymbol{w}_{ki}) \widehat{\log x}_{kig}^{(t+1)} \right) .
\end{split}
\end{equation*}
Update $\hat{\beta}^{(t+1)}$:
\begin{equation*}
\hat{\beta}^{(t+1)} = \frac{ \sum_{i=1}^{N} \sum_{g=1}^{G} \hat{z}_{ig}^{(t+1)} \left( \hat{\alpha}_{1ig}^{(t+1)} + \hat{\alpha}_{2ig}^{(t+1)} + \hat{\alpha}_{3ig}^{(t+1)} \right) }{ \sum_{i=1}^{N} \sum_{g=1}^{G} \hat{z}_{ig}^{(t+1)} \left( \hat{x}_{1ig}^{(t+1)} + \hat{x}_{2ig}^{(t+1)} + \hat{x}_{3ig}^{(t+1)} \right) } .
\end{equation*}
\item \textbf{``$\ast$IV" model type:} \\
Update $\hat{\alpha}_{k}^{(t+1)}$ ($k=1,2,3$) by solving: 
\begin{equation*}
\sum_{i=1}^{N}\sum_{g=1}^{G} \hat{z}_{ig}^{(t+1)} \log \hat{\beta}_{ig}^{(t)} - \frac{\Gamma^{\prime}(\alpha_{k})}{\Gamma(\alpha_{k})} \sum_{i=1}^{N}\sum_{g=1}^{G} \hat{z}_{ig}^{(t+1)} + \sum_{i=1}^{N}\sum_{g=1}^{G} \hat{z}_{ig}^{(t+1)} \widehat{\log x}_{kig}^{(t+1)} = 0.
\end{equation*}
Update $\hat{\boldsymbol{\gamma}}_{g}^{(t+1)}$:
\begin{equation*}
\begin{split}
\hat{\boldsymbol{\gamma}}_{g}^{(t+1)} = \underset{\boldsymbol{\gamma}_{g}}{\arg\max} 
& \left(
(\hat{\alpha}_{1}^{(t+1)} + \hat{\alpha}_{2}^{(t+1)} + \hat{\alpha}_{3}^{(t+1)}) \sum_{i=1}^{N}  \hat{z}_{ig}^{(t+1)} (\boldsymbol{\gamma}_{g}^{\top} \boldsymbol{w}_{4i}) \right. \\
& \hskip .5cm - \left. \sum_{i=1}^{N} \hat{z}_{ig}^{(t+1)} \exp(\boldsymbol{\gamma}_{g}^{\top} \boldsymbol{w}_{4i}) (\hat{x}_{1ig}^{(t+1)} + \hat{x}_{2ig}^{(t+1)} + \hat{x}_{3ig}^{(t+1)}) \right) .
\end{split}
\end{equation*}
\item \textbf{``$\ast$VE" model type:} \\
Update $\hat{\boldsymbol{\gamma}}_{kg}^{(t+1)}$ ($k=1,2,3$):
\begin{equation*}
\begin{split}
\hat{\boldsymbol{\gamma}}_{kg}^{(t+1)} = \underset{\boldsymbol{\gamma}_{kg}}{\arg\max} \left(
\sum_{i=1}^{N} \hat{z}_{ig}^{(t+1)} \exp(\boldsymbol{\gamma}_{kg}^{\top} \boldsymbol{w}_{ki}) \log \beta_{i}^{(t)} \right. &- \sum_{i=1}^{N} \hat{z}_{ig}^{(t+1)} \log\Gamma( \exp(\boldsymbol{\gamma}_{kg}^{\top} \boldsymbol{w}_{ki}))  \\
&+ \left. \sum_{i=1}^{N} \hat{z}_{ig}^{(t+1)} \exp(\boldsymbol{\gamma}_{kg}^{\top} \boldsymbol{w}_{ki}) \widehat{\log x}_{kig}^{(t+1)} \right).
\end{split}
\end{equation*} 
Update $\hat{\boldsymbol{\gamma}}_{4}^{(t+1)}$:
\begin{equation*}
\begin{split}
\hat{\boldsymbol{\gamma}}_{4}^{(t+1)} = \underset{\boldsymbol{\gamma}_{4}}{\arg\max} 
& \left(
\sum_{i=1}^{N}\sum_{g=1}^{G}  \hat{z}_{ig}^{(t+1)} (\hat{\alpha}_{1ig}^{(t+1)} + \hat{\alpha}_{2ig}^{(t+1)} + \hat{\alpha}_{3ig}^{(t+1)}) (\boldsymbol{\gamma}_{4}^{\top} \boldsymbol{w}_{4i}) \right. \\
& \hskip 2cm - \left. \sum_{i=1}^{N}\sum_{g=1}^{G} \hat{z}_{ig}^{(t+1)} \exp(\boldsymbol{\gamma}_{4}^{\top} \boldsymbol{w}_{4i}) (\hat{x}_{1ig}^{(t+1)} + \hat{x}_{2ig}^{(t+1)} + \hat{x}_{3ig}^{(t+1)}) \right) .
\end{split}
\end{equation*}
\item \textbf{``$\ast$EV" model type:} \\
Update $\hat{\boldsymbol{\gamma}}_{k}^{(t+1)}$ ($k=1,2,3$):
\begin{equation*}
\begin{split}
\hat{\boldsymbol{\gamma}}_{k}^{(t+1)} = \underset{\boldsymbol{\gamma}_{k}}{ \arg\max }
\left(
\sum_{i=1}^{N}\sum_{g=1}^{G} \hat{z}_{ig}^{(t+1)} \exp( \boldsymbol{\gamma}_{k}^{\top} \boldsymbol{w}_{ki}) \log \hat{\beta}_{ig}^{(t)} \right. &- \sum_{i=1}^{N}\sum_{g=1}^{G} \hat{z}_{ig}^{(t+1)}  \log\Gamma(\exp(\boldsymbol{\gamma}_{k}^{\top} \boldsymbol{w}_{ki}))  \\
&+ \left. \sum_{i=1}^{N}\sum_{g=1}^{G} \hat{z}_{ig}^{(t+1)} \exp(\boldsymbol{\gamma}_{k}^{\top} \boldsymbol{w}_{ki}) \widehat{\log x}_{kig}^{(t+1)} \right) .
\end{split}
\end{equation*} 
Update $\hat{\boldsymbol{\gamma}}_{4g}^{(t+1)}$: 
\begin{equation*}
\begin{split}
\hat{\boldsymbol{\gamma}}_{4g}^{(t+1)} = \underset{\boldsymbol{\gamma}_{4g}}{ \arg\max }
& \left(
\sum_{i=1}^{N} \hat{z}_{ig}^{(t+1)} (\hat{\alpha}_{1i}^{(t+1)} + \hat{\alpha}_{2i}^{(t+1)} + \hat{\alpha}_{3i}^{(t+1)}) (\boldsymbol{\gamma}_{4g}^{\top} \boldsymbol{w}_{4i}) \right. \\
& \hskip 1cm - \left. \sum_{i=1}^{N} \hat{z}_{ig}^{(t+1)} \exp(\boldsymbol{\gamma}_{4g}^{\top} \boldsymbol{w}_{4i}) (\hat{x}_{1ig}^{(t+1)} + \hat{x}_{2ig}^{(t+1)} + \hat{x}_{3ig}^{(t+1)}) \right) .
\end{split}
\end{equation*}
\item \textbf{``$\ast$EC" model type:} \\
Update $\hat{\boldsymbol{\gamma}}_{k}^{(t+1)}$ ($k=1,2,3$):
\begin{equation*}
\begin{split}
\hat{\boldsymbol{\gamma}}_{k}^{(t+1)} = \underset{\boldsymbol{\gamma}_{k}}{ \arg\max }
\left(
\sum_{i=1}^{N}\sum_{g=1}^{G} \hat{z}_{ig}^{(t+1)} \exp(\boldsymbol{\gamma}_{k}^{\top} \boldsymbol{w}_{ki}) \log \hat{\beta}_{g}^{(t)} \right. &- \sum_{i=1}^{N}\sum_{g=1}^{G} \hat{z}_{ig}^{(t+1)} \log\Gamma(\exp(\boldsymbol{\gamma}_{k}^{\top} \boldsymbol{w}_{ki})) \\
&+ \left. \sum_{i=1}^{N}\sum_{g=1}^{G} \hat{z}_{ig}^{(t+1)} \exp(\boldsymbol{\gamma}_{k}^{\top} \boldsymbol{w}_{ki}) \widehat{\log x}_{kig}^{(t+1)} \right) .
\end{split}
\end{equation*} 
Update $\hat{\beta}_{g}^{(t+1)}$:
\begin{equation*}
\hat{\beta}_{g}^{(t+1)} = \frac{ \sum_{i=1}^{N} \hat{z}_{ig}^{(t+1)} \left( \hat{\alpha}_{1i}^{(t+1)} + \hat{\alpha}_{2i}^{(t+1)} + \hat{\alpha}_{3i}^{(t+1)} \right) }{ \sum_{i=1}^{N} \hat{z}_{ig}^{(t+1)} \left( \hat{x}_{1ig}^{(t+1)} + \hat{x}_{2ig}^{(t+1)} + \hat{x}_{3ig}^{(t+1)} \right) } .
\end{equation*} 
\item \textbf{``$\ast$CE" model type:} \\
Update $\hat{\alpha}_{kg}^{(t+1)}$ ($k=1,2,3$) by solving
\begin{equation*}
\sum_{i=1}^{N} \hat{z}_{ig}^{(t+1)} \log \hat{\beta}_{i}^{(t)} - \sum_{i=1}^{N} \hat{z}_{ig}^{(t+1)} \frac{\Gamma^{\prime}(\alpha_{kg})}{\Gamma(\alpha_{kg})} + \sum_{i=1}^{N} \hat{z}_{ig}^{(t+1)} \widehat{\log x}_{kig}^{(t+1)} = 0  .
\end{equation*}
Update $\hat{\boldsymbol{\gamma}}_{4}^{(t+1)}$:
\begin{equation*}
\begin{split}
\hat{\boldsymbol{\gamma}}_{4}^{(t+1)} = \underset{\boldsymbol{\gamma}_{4}}{ \arg\max }
& \left(
\sum_{i=1}^{N}\sum_{g=1}^{G}  \hat{z}_{ig}^{(t+1)} (\hat{\alpha}_{1g}^{(t+1)} + \hat{\alpha}_{2g}^{(t+1)} + \hat{\alpha}_{3g}^{(t+1)} ) (\boldsymbol{\gamma}_{4}^{\top} \boldsymbol{w}_{4i}) \right. \\
& \hskip 1cm  - \left. \sum_{i=1}^{N}\sum_{g=1}^{G} \hat{z}_{ig}^{(t+1)} \exp(\boldsymbol{\gamma}_{4}^{\top} \boldsymbol{w}_{4i}) (\hat{x}_{1ig}^{(t+1)} + \hat{x}_{2ig}^{(t+1)} + \hat{x}_{3ig}^{(t+1)})  \right) .
\end{split}
\end{equation*}
\end{itemize}

\end{document}